\documentclass[
floats,
floatfix,
amssymb,
prd,
onecolumn,
superscriptaddress,
nofootinbib,
notitlepage]{revtex4-1}

\usepackage{fontawesome}
\usepackage{blkarray}
\usepackage{graphicx}
\usepackage{amsfonts}
\usepackage{soul}
\usepackage{amssymb}
\usepackage{amsmath}
\usepackage{pifont}
\usepackage{slashed} 
\usepackage{enumerate}
\usepackage{color}
\usepackage{adjustbox}
\usepackage{comment}
\usepackage{bbold}
\usepackage{tikz}
\usepackage{amsmath}
\usepackage{tikz}
\usetikzlibrary{tikzmark}
\usepackage{empheq}
\usepackage{adjustbox}
\usepackage{multirow}
\usepackage[utf8]{inputenc}
\usepackage{adjustbox}

\usepackage{tikz-feynman}
\usepackage{tcolorbox}

\definecolor{oucrimsonred}{rgb}{0.6, 0.0, 0.0}
\definecolor{persianblue}{rgb}{0.11, 0.22, 0.73}
\definecolor{forestgreen}{rgb}{0.13,0.35,0.13}
\definecolor{lightgray}{rgb}{0.83, 0.83, 0.83}
\definecolor{cornellred}{rgb}{0.7, 0.11, 0.11}
\definecolor{navyblue}{rgb}{0.0, 0.0, 0.5}
\definecolor{amethyst}{rgb}{0.6, 0.4, 0.8}
\definecolor{yellow}{rgb}{1.0, 1.0, 0.0}
\definecolor{firebrick}{rgb}{0.7, 0.13, 0.13}
\definecolor{tangerineyellow}{rgb}{1.0, 0.8, 0.0}
\definecolor{deepfuchsia}{rgb}{0.76, 0.33, 0.76}
\definecolor{amber}{rgb}{1.0, 0.75, 0.0}
\definecolor{VioletRed4}{rgb}{0.55, 0.13, .32}
\definecolor{indiagreen}{rgb}{0.07, 0.53, 0.03}
\definecolor{VioletRed4}{rgb}{0.55, 0.13, .32}
\newcommand{\nn}{\nonumber}

\definecolor{oucrimsonred}{rgb}{0.6, 0.0, 0.0}
\newcommand\vertarrowbox[3][6ex]{%
  \begin{array}[t]{@{}c@{}} #2 \\
  \left\uparrow\vcenter{\hrule height #1}\right.\kern-\nulldelimiterspace\\
  \makebox[0pt]{\scriptsize#3}
  \end{array}%
}

\definecolor{mtcolor}{rgb}{.8,.3,.1}

\definecolor{violachiaro}{rgb}{1,0.6,1}
 
\definecolor{gbcolor}{rgb}{.43,.22,.12}
 
\definecolor{gbcolor2}{rgb}{.9,.2,.6}
\definecolor{gbcolor3}{rgb}{.3,.2,.6}

\definecolor{verdechiaro}{rgb}{0.6,1,0.6}
\definecolor{giallochiaro}{rgb}{1,1,0.6}
\definecolor{bluscuro}{rgb}{0.15, 0.2, 0.9}
\definecolor{verdes}{rgb}{0.1, 0.5, 0.1}%
\definecolor{tangerineyellow}{rgb}{1.0, 0.8, 0.0}
\definecolor{americanrose}{rgb}{1.0, 0.01, 0.24}
\definecolor{cobalt}{rgb}{0.0, 0.28, 0.67}
\definecolor{brandeisblue}{rgb}{0.0, 0.44, 1.0}
\definecolor{mycolor}{rgb}{0.0, 0.0, 0.5}
\definecolor{oxfordblue}{rgb}{0.0, 0.13, 0.28}
\definecolor{azure}{rgb}{0.0, 0.5, 1.0}
 
\definecolor{turquoiseblue}{rgb}{0.0, 1.0, 0.94}
\newtcolorbox{mynewbox}[1]{colback=white!5!white,colframe=azure!75!black,fonttitle=\bfseries,title=#1}
\newtcolorbox{mybox}{colback=mycolor!5!white,colframe=azure!75!black}
\newtcolorbox{mynamedbox}[1]{colback=mycolor!5!white,colframe=azure!75!black,title=#1}
\definecolor{venetianred}{rgb}{0.78, 0.03, 0.08}
\newtcolorbox{mynamedbox2}[1]{colback=venetianred!5!white,colframe=venetianred!80!black,title=#1}

\definecolor{myforestgreen}{rgb}{0.13, 0.55, 0.13}
\definecolor{rossocorsa}{rgb}{0.83, 0.0, 0.0}

\makeatletter
\def\l@subsubsection#1#2{}
\makeatother

\definecolor{oucrimsonred}{rgb}{0.6, 0.0, 0.0}
\definecolor{persianblue}{rgb}{0.11, 0.22, 0.73}
\definecolor{forestgreen}{rgb}{0.13,0.35,0.13}
\usepackage[unicode, 
colorlinks=true, 
linkcolor=persianblue, 
citecolor=forestgreen, 
filecolor=persianblue,
urlcolor=oucrimsonred, 
pdfusetitle]{hyperref}

\begin{document}

\title[]{Primordial non-gaussianity up to all orders: \\ theoretical aspects and implications for primordial black hole models} 

\newcommand{\lupm}{LUPM, CNRS, Université Montpellier Place Eugene Bataillon, F-34095 Montpellier, France}
\newcommand{\uniroma}{Dipartimento di Fisica, ``Sapienza'' Universit\`a di Roma, Piazzale Aldo Moro 5, 00185, Roma, Italy}
\newcommand{\infn}{INFN sezione di Roma, Piazzale Aldo Moro 5, 00185, Roma, Italy}

\date{\today}

\author{Giacomo Ferrante}
\email{giacomo.ferrante@umontpellier.fr}
\affiliation{\lupm}
\affiliation{\uniroma}
\author{Gabriele Franciolini}
\email{gabriele.franciolini@uniroma1.it}
\affiliation{\uniroma}
\affiliation{\infn}
\author{Antonio Junior Iovino}
\email{antoniojunior.iovino@uniroma1.it}
\affiliation{\uniroma}
\affiliation{\infn}
\author{Alfredo Urbano}
\email{alfredo.urbano@uniroma1.it}
\affiliation{\uniroma}
\affiliation{\infn}
\vspace{0.5cm}

\begin{abstract}
\noindent  
We develop an exact formalism for the computation of the abundance 
of primordial black holes (PBHs) in the presence of local non-gaussianity (NG) in the curvature perturbation field. 
For the first time, we include NG going beyond the widely used quadratic and cubic approximations, and consider 
a completely generic functional form.
Adopting threshold statistics on the compaction function, we address the computation of the abundance both for narrow and broad power spectra. 
While our formulas are generic, we discuss explicit examples of phenomenological relevance considering the physics case of the curvaton field.
We carefully assess under which conditions the conventional perturbative approach can be trusted. In the case of a narrow power spectrum, this happens only if the perturbative expansion is pushed beyond the quadratic order (with the optimal order of truncation that depends on the  width of the spectrum). Most importantly, we demonstrate that the perturbative approach is intrinsically flawed when considering broad spectra, in which case only the non-perturbative computation captures the correct result. 
Finally, we describe the phenomenological relevance of our results for the connection between the abundance of PBHs and the stochastic gravitational wave (GW) background related to their formation.
As NGs modify the amplitude of perturbations necessary to produce a given PBHs abundance and boost PBHs production at large scales for broad spectra, modelling these effects is crucial to connect the PBH scenario to its signatures at current and future GWs experiments.

 \end{abstract}
\maketitle

{

  \hypersetup{linkcolor=black}
  \tableofcontents
  
}

\section{Introduction}\label{sec:QualiIntro}

Primordial Black Holes (PBHs)\,\cite{Zeldovich:1967lct,Hawking:1974rv,Chapline:1975ojl,Carr:1975qj} have recently attracted ample attention as they may explain some of the LIGO/Virgo/KAGRA observations and significantly contribute to the dark matter in our universe (see e.g. for a review\,\cite{Sasaki:2018dmp}). 
It is known, however, that the precise computation of the PBH abundance at formation is extremely challenging. The standard formation scenario, which we assume throughout this work, 
hypothesises that PBHs form out of the gravitational collapse of large over-densities in the primordial density contrast field\,\cite{Ivanov:1994pa,GarciaBellido:1996qt,Ivanov:1997ia,Blinnikov:2016bxu}. 
Being the latter a random field, the computation of the abundance is statistical in nature, and requires the precise knowledge of the probability density function (PDF) of density fluctuations.

During the recent years, it has become increasingly clear that limiting the analysis to the assumption that the PDF of density fluctuations follows the Gaussian statistics is theoretically flawed. 
The reason is twofold. 
First, density fluctuations in the primordial radiation field originate from curvature perturbations, previously stretched on super-horizon scales during inflation, after their horizon re-entry. 
In the long-wavelength approximation, the equation which relates curvature perturbations to density fluctuations is intrinsically non-linear\,\cite{Harada:2015yda}. For this reason, even under the \emph{assumption} that curvature perturbations follow exact Gaussian statistics (which is in general not true, as we shall discuss next), density fluctuations inherit an unavoidable amount of non-gaussianity from non-linear (NL) corrections\,\cite{DeLuca:2019qsy,Young:2019yug,Germani:2019zez}. 
Second, as anticipated, curvature perturbations do not generically follow a Gaussian statistics.

Let us elaborate more on the second point by considering two different scenarios. 
In single-field inflationary models, 
NG corrections are typically subdominant at scales relevant for CMB observations because suppressed by the tiny values of the slow-roll parameters; however, the formation of PBHs requires a strong violation of the slow-roll paradigm\,\cite{Motohashi:2017kbs}, usually achieved by the presence of an ultra slow-roll (USR) phase that boosts the amplitude of the power spectrum of curvature perturbations at scales relevant for PBH formation\,\cite{Inomata:2016rbd,Garcia-Bellido:2017mdw,Ballesteros:2017fsr,Hertzberg:2017dkh,Kannike:2017bxn,Dalianis:2018frf,Inomata:2018cht,Cheong:2019vzl,Ballesteros:2020qam,Iacconi:2021ltm,Kawai:2021edk}. 
This implies that, at such scales, the suppression of NG, being slow-roll violated, is no longer guaranteed.
Alternatively, one can envision a two-field scenario in which, in addition to the inflaton driving inflation and generating curvature perturbations at CMB scales, there exists a spectator field, dubbed curvaton\,\cite{Lyth:2002my}, 
which is responsible, after its decay into radiation, for the generation of curvature perturbations at scales relevant for PBH formation. 
In this case, it happens again that NG corrections are relevant. As a generic result, the smaller the fraction of total energy density in the curvaton field at the time of its decay, 
the larger NG becomes\,\cite{Malik:2002jb}.

The local NG of the curvature perturbation $\zeta$ is usually parameterized by the expansion
\begin{align}\label{eq:FirstExpansion}
\zeta = \zeta_{\rm G} + \frac{3}{5}f_{\rm NL}\zeta_{\rm G}^2
 + \frac{9}{25}g_{\rm NL}\zeta_{\rm G}^3 + \dots\,,
\end{align}
where $\zeta_{\rm G}$ obeys the Gaussian statistics while the parameters $f_{\rm NL}$, $g_{\rm NL}$, $\dots$ (which,
in full generality, depend on the scale of the perturbation) encode deviations from the Gaussian limit. 
Interestingly, in both models discussed before (that is single-field inflation with an USR phase and two-field curvaton models) it is possible to show, by means of the so-called $\delta N$ formalism, that in the previous expansion 
the parameters $f_{\rm NL}$, $g_{\rm NL}$, $\dots$ are not independent and 
actually give rise to a closed-form resummed expression for $\zeta$ of the schematic form 
$\zeta = F(\zeta_{\rm G})$
 for some appropriate function $F(\zeta_{\rm G})$ of the Gaussian component.

Let us stress that understanding the role of NG in the context of PBH formation is crucial. 
The formation of a PBHs is a rare event; this means that one is interested in the tail of the PDF where small deviations from the Gaussian assumption generically give exponential effects. 
Furthermore, and most importantly, one may be tempted to think that one can always reabsorb NG corrections to the computation of the PBH abundance by means of small retuning of model parameters.
While generically true, this is not harmless, as it would spoil the model predictions which depend on the power spectrum (like the amplitude of the induced gravitational-wave signal) and are directly sensitive to the inclusion of NG corrections. 
Furthermore, with the aim of constraining PBHs with future GW detections, 
missing a precise description of NG corrections may prevent us from setting reliable bounds on their abundance and the fraction of dark matter they can amount for. We will elaborate on this point in the final sections.

Many previous attempts in the literature have tried to address the problem of including NGs in the computation of PBHs abundance, with different levels of approximation. 
Various works exploited the perturbative expansion in  eq.\,\eqref{eq:FirstExpansion} up to a finite number of orders\,\cite{Bugaev:2013vba,Nakama:2016gzw,Byrnes:2012yx,Young:2013oia,Yoo:2018kvb,Kawasaki:2019mbl,Yoo2,Yoo:2019pma,
Riccardi:2021rlf,Taoso:2021uvl,Meng:2022ixx,Escriva:2022pnz}, modelling either the distribution of curvature perturbation or density contrast (where the latter is the physical quantity that should be adopted\,\cite{Young:2014ana}).
Other have tried adopting a path-integral formulation\,\cite{Franciolini:2018vbk}, whose application requires the notion of the full list of n-th order point functions, or non-perturbative approaches\,\cite{Atal:2019cdz,Biagetti:2021eep,Kitajima:2021fpq} focused on specific USR models.
On the other hand, in refs.\,\cite{Shibata:1999zs,Musco:2018rwt} it was suggested that the so-called compaction function $\mathcal{C}$ — defined as twice the
local excess-mass over the co-moving areal radius — is the most suitable parameter to use to determine whether a perturbation collapses to form a PBH. 
This motivated the recent attempt to compute the abundance by applying peak theory to
the full, non-linear, compaction $\mathcal{C}$ in ref.\,\cite{Young:2022phe}.

Focusing on the most recent results, the analysis in ref.\,\cite{Young:2022phe} still suffers from two limitations. First, NGs are taken into account following the spirit of the perturbative  expansion in eq.\,(\ref{eq:FirstExpansion}) with the quadratic and cubic terms that are discussed separately (setting the former to zero when discussing the latter and vice-versa) as if they were independent. As we mentioned, this is seldom the case in concrete models. 
Second, and most importantly, ref.\,\cite{Young:2022phe} makes use of the so-called high-peak approximation. The point is the following. In eq.\,(\ref{eq:FirstExpansion}) $\zeta_{\rm G}$ follows a Gaussian statistics and, as such, it is completely determined by the value of its variance. The latter can be defined as the logarithmic integral over the comoving wavenumber $k$ of the power spectrum of $\zeta_{\rm G}$, that we shall denote as $P_{\zeta}(k)$.  Assuming some functional form for $P_{\zeta}(k)$, therefore, equivalently specifies  the variance of the Gaussian component of the curvature perturbation field. 
The analysis of ref.\,\cite{Young:2022phe} is strictly valid in the limit in which $P_{\zeta}(k)$ takes the form of a Dirac delta function. 
Realistic models, however, tend to generate power spectra that deviate from the above-mentioned limit.

In this work we extend recent literature, and in particular  the analysis proposed in ref.\,\cite{Young:2022phe}, in two important directions.
\begin{itemize}
\item[$\circ$] We consider a generic functional form for the curvature perturbation $\zeta$. 
For definiteness, we shall focus on the curvaton model even though our general approach is applicable to other cases.
\item[$\circ$] We go beyond the high-peak limit and model both narrow and broad power spectra. This is important to match the outcome of realistic inflationary dynamics.
\end{itemize}
Our material is organized as follows. 
In section\,\ref{sec:NGIntro} we discuss in more detail the NG nature of the curvature perturbation field.  
In section\,\ref{sec:NonGauBeta} 
we discuss how to consistently include primordial NG in the computation of the mass fraction of PBHs, that is the fraction of the energy density of the Universe contained in regions overdense enough to form PBHs;
while
in section\,\ref{sec:CompFunc} we discuss the  approach 
based on the compaction function.
In section\,\ref{sec:PBHGW} we discuss the phenomenological implications of our results.
We conclude in section\,\ref{sec:Final}.

\section{Primordial NG up to all orders}\label{sec:NGIntro}

The aim of this paper is to consider the impact of local primordial NG on PBH abundance in the most general way. 
To this end, we will consider the following functional form
\begin{align}
    \zeta(\vec{x}) = F(\zeta_{\rm G}(\vec{x}))\,,\label{eq:MainF}
\end{align}
where $F$ is a generic non-linear function of the Gaussian component $\zeta_{\rm G}$ and the attribute of locality refers to the fact that 
the value of $\zeta$ at the point $\vec{x}$ is fully determined
by the value of $\zeta_{\rm G}$ at the same spatial point.\footnote{In other words, 
the function $F$ in eq.\,(\ref{eq:MainF}) must be independent on spatial derivatives of $\zeta_{\rm G}$ since the latter 
 would make the value of $\zeta$ at $\vec{x}$ 
determined  by the values of 
 $\zeta_{\rm G}$ in a certain neighborhood of $\vec{x}$. 
} 
Whenever unnecessary, we will drop the explicit functional dependence from the spatial coordinates in $\zeta$ and $\zeta_{\rm G}$.

There is a number of relevant cases in the literature of PBH formation that can be traced back to eq.\,(\ref{eq:MainF}), as summarized in the schematic below.
\begin{align}
	\begin{tikzpicture}
	 {\scalebox{1}{
    \node at (0,0) {\scalebox{1}{$\zeta = F(\zeta_{\rm G})$}};
    \draw[->,>=Latex][thick] (-1.5,-0.2)--(-4.0,-1);
    \node at (-7,-0.95) {\scalebox{1}{
    power-series expansion\,\cite{Bugaev:2013vba,Nakama:2016gzw,Byrnes:2012yx,Young:2013oia,Yoo:2018kvb,Kawasaki:2019mbl,Yoo2,
Riccardi:2021rlf,Taoso:2021uvl,Meng:2022ixx,Escriva:2022pnz}
    }};
    \node at (-6.5,-1.5) {\scalebox{1}{
    $\zeta = \zeta_{\rm G} + \frac{3}{5}f_{\rm NL}\zeta_{\rm G}^2
 + \frac{9}{25}g_{\rm NL}\zeta_{\rm G}^3 + \dots$
    }};
    \draw[->,>=Latex][thick] (-0.7,-0.4)--(-3.1,-2.5);
    \node at (-3.25,-3) {\scalebox{1}{
    curvaton\,\cite{Sasaki:2006kq,Pi:2021dft}
    }};    
    \node at (-3.25,-3.55) {\scalebox{1}{
    $\zeta = \log\left[X(r_{\rm dec},\zeta_{\rm G})\right]$
    }};    
    \draw[->,>=Latex][thick] (0.7,-0.4)--(2.5,-2.5);
    \node at (2.5,-2.75) {\scalebox{1}{
    ultra slow-roll\,\cite{Atal:2019cdz}
    }};       
    \node at (2.5,-3.3) {\scalebox{1}{
    $\zeta = -\left(\frac{6}{5}f_{\rm NL}\right)^{-1}\log\left(1-  
    \frac{6}{5}f_{\rm NL}\zeta_{\rm G}\right)$
    }};       
    \draw[->,>=Latex][thick] (1.5,-0.2)--(4.7,-1);
    \node at (6.,-0.9) {\scalebox{1}{
    ultra slow-roll
    }};
    \node at (6.2,-1.25) {\scalebox{1}{
    with an upward step\,\cite{Cai:2022erk}
    }};
    \node at (6.0,-1.8) {\scalebox{1}{
    $\zeta = -\frac{2}{|h|}
    \left[\sqrt{1-|h|\zeta_{\rm G}} - 1\right]$
    }};      
    }}
	\end{tikzpicture}\label{eq:Schema}
\end{align}
In the course of this paper, and in particular in section\,\ref{sec:NonGauBeta}, we will describe the main results of our analysis using the generic functional form $\zeta = F(\zeta_{\rm G})$. 
However, in order to present and discuss explicit examples of phenomenological relevance, we will apply our formulas to the physics-case of the curvaton field.
Let us, therefore, explore in more detail this specific theoretical set-up.

\subsection{Primordial NG in curvaton models}\label{sec:NGIntro}

When presenting results inspired by the curvaton model, we will focus on primordial NG with the following functional form\,\cite{Sasaki:2006kq}
\begin{align}\label{eq:MasterX}
\zeta = \log\big[X(r_{\rm dec},\zeta_{\rm G})\big]\,,
\end{align}
with
\begin{align}\label{eq:XFunction}
X(r_{\rm dec},\zeta_{\rm G}) \equiv& \frac{1}{\sqrt{2 (3+r_{\rm dec})^{1/3}}}
\Bigg\{
\sqrt{
\frac{
-3 + r_{\rm dec}(2+r_{\rm dec}) + [(3+r_{\rm dec})P(r_{\rm dec},\zeta_{\rm G})]^{2/3}
}{
(3+r_{\rm dec})P^{1/3}(r_{\rm dec},\zeta_{\rm G})
}
}  
\nn \\
+ &  
\sqrt{
\frac{(1-r_{\rm dec})}
{P^{1/3}(r_{\rm dec},\zeta_{\rm G})} -
\frac{P^{1/3}(r_{\rm dec},\zeta_{\rm G})}
{(3+r_{\rm dec})^{1/3}} 
+ 
\frac{
(2r_{\rm dec} + 3\zeta_G)^{2}
P^{1/6}(r_{\rm dec},\zeta_{\rm G})
}{r_{\rm dec}
\sqrt{-3 + r_{\rm dec}(2+r_{\rm dec}) +
[(3+r_{\rm dec})P(r_{\rm dec},\zeta_{\rm G})]^{2/3}
}}} 
\Bigg\}\,,
\end{align}
and 
\begin{align}
P(r_{\rm dec},\zeta_{\rm G}) \equiv 
\frac{(2r_{\rm dec} + 3\zeta_{\rm G})^{4}}{16 r_{\rm dec}^2} + \sqrt{
(1-r_{\rm dec})^3(3+r_{\rm dec}) + \frac{(2r_{\rm dec} + 3\zeta_{\rm G})^8}{256 r_{\rm dec}^4}
}\,.\label{eq:PFunction}
\end{align}
This functional form characterizes 
NG of the primordial curvature  perturbation in the curvaton model.
The parameter $r_{\rm dec}$ is the weighted fraction of the curvaton energy density  $\rho_{\phi}$ to the total energy density at the time of curvaton decay, defined by
\begin{equation}
r_{\rm dec} \equiv \left.\frac{3 \rho_{\phi}}{3 \rho_{\phi}+4 \rho_{\gamma}}\right|_{\rm curvaton\,\,decay}\,,
\end{equation}
where $\rho_{\gamma}$ is the energy density stored in radiation after reheating.
In eq.\,(\ref{eq:MasterX}) the gaussian random field $\zeta_{\rm G}$  
corresponds to $\zeta$ in linear approximation.
We remark that eq.\,(\ref{eq:MasterX}) is strictly valid under the approximation that the curvaton potential 
is quadratic (which implies the absence of any non-linear evolution of the curvaton field
between the Hubble exit and the start of curvaton oscillation, see refs.\,\cite{Sasaki:2006kq,Kawasaki:2011pd}).

The above expressions simplify in the limit case $r_{\rm dec}=1$ (that is the case in which the curvaton dominates the energy density of the Universe at the time of its decay) that gives
\begin{align}\label{eq:MasterX2}
\zeta_{\rm log} \equiv \log\big[X(1,\zeta_{\rm G})\big] = \frac{2}{3}\log\bigg(
1+ \frac{3}{2}\zeta_{\rm G}
\bigg)\,.
\end{align}
We note that, if we approximate eq.\,(\ref{eq:MasterX}) at the quadratic order in the gaussian field $\zeta_{\rm G}$, 
we find
\begin{align}\label{eq:ZetaQ}
\zeta_{2} \equiv \zeta_{\rm G} + \frac{3}{5}f_{\rm NL}(r_{\rm dec})\zeta_{\rm G}^2\,,
~~~~~~~{\rm with}~~~~~~~
f_{\rm NL}(r_{\rm dec}) \equiv  \frac{5}{3}\bigg(
\frac{3}{4r_{\rm dec}} - 1 - \frac{r_{\rm dec}}{2}
\bigg)\,.
\end{align}
Similarly, at the cubic order, we find
\begin{align}\label{eq:ZetaC}
\zeta_{3} = 
\zeta_{\rm G}  + \frac{3}{5}f_{\rm NL}(r_{\rm dec})\zeta_{\rm G}^2
 + \frac{9}{25}g_{\rm NL}(r_{\rm dec})\zeta_{\rm G}^3\,,
 ~~~~~~~{\rm with}~~~~~~~
g_{\rm NL}(r_{\rm dec}) \equiv \frac{25}{54}\left(
-\frac{9}{r_{\rm dec}} + \frac{1}{2} + 10r_{\rm dec} + 3r_{\rm dec}^2
\right)\,.
\end{align}
We refer to ref.\,\cite{Sasaki:2006kq} for a comprehensive derivation of eqs.\,(\ref{eq:MasterX},\,\ref{eq:XFunction},\,\ref{eq:ZetaQ},\,\ref{eq:ZetaC}).

The coefficients $f_{\rm NL}(r_{\rm dec})$, $g_{\rm NL}(r_{\rm dec})$, $\dots$ seem to diverge in the limit $r_{\rm dec}\to 0$ (meaning that the smaller the fraction of total energy density in the curvaton field at the time of its decay, 
the larger NG becomes). 
It is important for the rest of the analysis to have clear in mind the actual meaning of this apparent divergence.
Consider the simplest example of curvaton field, that is a massive real scalar field $\phi$ with quadratic potential $V(\phi) = m^2\phi^2/2$.
After the end of inflation the curvaton field  remains approximately constant until the Hubble rate becomes comparable to $m_{\phi}^2$; at this time, the curvaton starts oscillating around the minimum of its potential. The energy density of the oscillating curvaton is 
$\rho_{\phi} = m_{\phi}^2\phi^2$ and can be expanded into a background term $\bar{\rho}_{\phi}$ and 
a perturbation $\delta\rho_{\phi}$ according to
 $\rho_{\phi} = m_{\phi}^2(\bar{\phi}+\delta\phi)^2 = 
m_{\phi}^2\bar{\phi}^2 + m_{\phi}^2(2\bar{\phi}\delta\phi + \delta\phi^2) 
\equiv \bar{\rho}_{\phi} + \delta\rho_{\phi}$. 
Consequently, in the spatially flat gauge, the curvature perturbation associated with the curvaton field takes the form
\begin{align}
\zeta_{\phi} = \frac{2}{3}\frac{\delta\phi}{\bar{\phi}} + 
\frac{1}{3}\left(\frac{\delta\phi}{\bar{\phi}}\right)^2\,,
\end{align}
and it consists of a linear plus a quadratic (hence NG) term. 
The key point is that, after the end of inflation, the {\it total} curvature perturbation is the weighted sum (evaluated at the time of curvaton decay into radiation according to the so-called sudden decay approximation)
\begin{align}
\zeta = \frac{\dot{\rho}_{\gamma}}{\dot{\rho}}\zeta_{\gamma} + 
\frac{\dot{\rho}_{\phi}}{\dot{\rho}}\zeta_{\phi} = \underbrace{\frac{2r_{\rm dec}}{3}\frac{\delta\phi}{\bar{\phi}}}_{\equiv\,\,\zeta_{\rm G}}
+ \frac{3}{4r_{\rm dec}}\left(\frac{2r_{\rm dec}}{3}\frac{\delta\phi}{\bar{\phi}}\right)^2 = 
\zeta_{\rm G} + \frac{3}{4r_{\rm dec}}\zeta_{\rm G}^2\,,\label{eq:ZetaPhi}
\end{align}
where, importantly, we assume  that the curvature perturbation in radiation produced at the end
of inflation is negligible, $\zeta_{\gamma} \simeq 0$.\footnote{More in details, we assume $\zeta_{\gamma} \simeq 0$ at scales relevant for PBH formation. We tacitly work under the assumption that there exists 
some inflationary dynamics at the origin
of curvature perturbation responsible, at much larger length-scales, for the generation of CMB
anisotropies.}  This means that the curvaton contribution to $\zeta$ is not just a  correction on the top of some leading (gaussian) term but it entirely defines $\zeta$.
Formally, in the limit $r_{\rm dec}\to 0$,  eq.\,(\ref{eq:ZetaPhi}) implies that 
$\zeta\to 0$ (so that we do not have any divergence). However, what eq.\,(\ref{eq:ZetaPhi}) is telling us is that, once we define the linear term to be 
$\zeta_{\rm G} \equiv 2/3\,r_{\rm dec}\delta\phi/\bar{\phi}$ and we fix it to some reference value, for decreasing $r_{\rm dec}$ the quadratic correction takes over the linear one, meaning that the level of NG of $\zeta$ increases. 
The above argument only captures the leading quadratic correction in the limit of small $r_{\rm dec}$ (cf. ref.\,\,\cite{Sasaki:2006kq} for the full non-linear relation between $\zeta_{\phi}$ and $\zeta$ that gives rise to the functional form in eq.\,(\ref{eq:MasterX})) but it is sufficient to clarify the meaning of the $r_{\rm dec}\to 0$ limit.

\begin{figure}[!h!]
\begin{center}
\includegraphics[width=1.03\textwidth]{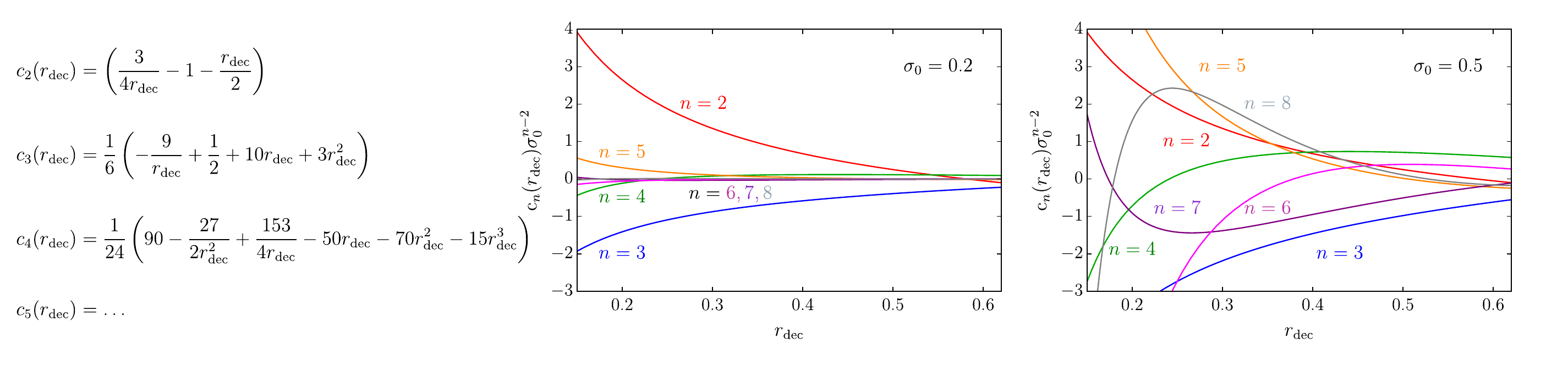}
\caption{\em
First few coefficients of the expansion in eq.\,(\ref{eq:ZetaSeries}) as function of $r_{\rm dec}$ (both analytically and numerically). 
In the figures we plot each coefficient $c_n(r_{\rm dec})$ rescaled by the appropriate power $\sigma_0^{n-2}$ in order to give a more realistic comparison of their relative size.
 }\label{fig:CnCoeff}
\end{center}
\end{figure}

More in general, one can setup the following power-series expansion 
\begin{align}\label{eq:ZetaSeries}
\zeta_N = \sum_{n = 1}^{N} c_n(r_{\rm dec}) \zeta_{\rm G}^{n}\,,~~~~~~~~
{\rm with}~~~~c_1(r_{\rm dec}) = 1\,.
\end{align}
We were not able to find a close recursive expression for the generic coefficient $c_n(r_{\rm dec})$ but
it is not difficult to extract it (analytically or numerically) at any finite order $N$. 
In fig.\,\ref{fig:CnCoeff} we show the first few coefficients of this expansion as function of the parameter $r_{\rm dec}$. 
In order to give a fair idea of their relative size in the expansion $\zeta_N$, 
we plot the quantity $c_n(r_{\rm dec})\sigma_0^{n-2}$ (assuming $\zeta_{\rm G} = O(\sigma_0))$.\footnote{In other words, we rewrite the expansion in eq.\,(\ref{eq:ZetaSeries}) in the form
\begin{align}
\zeta_N = \sum_{n = 1}^{N} c_n(r_{\rm dec}) \zeta_{\rm G}^{n} 
= \zeta_{\rm G} + 
\sigma_0^2\left[
c_2(r_{\rm dec})\left(\frac{\zeta_{\rm G}}{\sigma_0}\right)^2 +
c_3(r_{\rm dec})\sigma_0\left(\frac{\zeta_{\rm G}}{\sigma_0}\right)^3 + 
c_4(r_{\rm dec})\sigma_0^2\left(\frac{\zeta_{\rm G}}{\sigma_0}\right)^4 +\dots + c_n(r_{\rm dec})\sigma_0^{n-2}\left(\frac{\zeta_{\rm G}}{\sigma_0}\right)^n 
+\dots
\right]\,.
\end{align}
If we estimate $\zeta_{\rm G}$ with the variance of its PDF, $\zeta_{\rm G} = O(\sigma_0)$, the previous expansion is controlled by the coefficients $c_n(r)\sigma_0^{n-2}$.
} From this simple plot, we already get the idea that if $\sigma_0 \ll 1$ then the truncated expansion in $\zeta_N$ 
seems an appropriate approximation, and that only the first few orders are enough to capture the full result (cf. fig.\,\ref{fig:CnCoeff} with $\sigma_0 = 0.2$). Furthermore, we also note that the terms with $n=2$ and $n=3$ (that are, respectively, $f_{\rm NL}$ and $g_{\rm NL}$ in the commonly used language) contribute to $\zeta_N$ with opposite signs (and are comparable in absolute value); this suggests that truncating the expansion at the quadratic order without including also the cubic term may lead to incorrect results.
Finally, the last remark is that, as anticipated in the introduction, the condition $\sigma_0 \ll 1$ will crucially depend on the assumption about the power spectrum of the Gaussian variable $\zeta_{\rm G}$. 
We will come back shortly on this point. We anticipate that in the case of a broad power spectrum the condition $\sigma_0 \ll 1$ will be badly violated. The previous discussion, therefore, suggests that in such a case the perturbative analysis based on $\zeta_{\rm G}$ will not  be applicable. 
The right-most panel of  fig.\,\ref{fig:CnCoeff} shows that if one takes $\sigma_0 = 0.5$ then any hierarchy between the various $c_n(r_{\rm dec})$ coefficients gets already lost.

The statistics of $\zeta$ is easy to get. 
We can obtain the PDF of the NG variable $\zeta$ starting from the gaussian PDF 
of $\zeta_{\rm G}$ (which we shall denote in the following as 
$\textrm{P}_{\rm G}$). Conservation of probability gives\,\cite{Sasaki:2006kq}
\begin{align}
\textrm{P}(\zeta,r_{\rm dec}) = 
\textrm{P}_{\rm G}\big[
\zeta_{\rm G}^{(+)}(\zeta,r_{\rm dec})
\big]\left|
\frac{d\zeta_{\rm G}^{(+)}}{d\zeta}
\right| + 
\textrm{P}_{\rm G}\big[
\zeta_{\rm G}^{(-)}(\zeta,r_{\rm dec})
\big]\left|
\frac{d\zeta_{\rm G}^{(-)}}{d\zeta}
\right|\,,\label{eq:ExactPDF}
\end{align}
with 
\begin{align}\label{eq:ZetaRoots}
\zeta_{\rm G}^{(\pm)}(\zeta,r_{\rm dec}) & = 
\frac{2r_{\rm dec}}{3}\bigg[
-1 \pm \sqrt{
\bigg(\frac{3+r_{\rm dec}}{4r_{\rm dec}}\bigg)e^{3\zeta} +
\bigg(\frac{3r_{\rm dec}-3}{4r_{\rm dec}}\bigg)e^{-\zeta} 
}
\bigg]\,.
\end{align}
As a consistency check, integrating the PDF we find the total probability 
\begin{align}
\int_{\zeta_{\rm min}(r_{\rm dec})}^{\infty}\textrm{P}(\zeta,r_{\rm dec})d\zeta = 1,\,~~~~~~~~~~~
{\rm with}~~~ \zeta_{\rm min}(r_{\rm dec}) \equiv 
\frac{1}{4}\log\bigg(
\frac{3-3r_{\rm dec}}{3+r_{\rm dec}}
\bigg)\,.
\end{align}
Mathematically, the condition $\zeta \geqslant \zeta_{\rm min}(r_{\rm dec})$ follows from requiring eq.\,(\ref{eq:ZetaRoots}) to be real.

\subsection{The distribution of curvature perturbations}\label{sec:IntroStat}
The statistics of the NG variable $\zeta$ is, therefore,  
completely determined by the statistics of $\zeta_{\rm G}$. Being the latter a gaussian random field, 
we only need to specify its variance $\sigma_0$ (assuming vanishing mean value) according to
\begin{align}\label{eq:GaussianPDF}
\textrm{P}_{\rm G}(\zeta_{\rm G}) = \frac{1}{\sqrt{2\pi}\sigma_0}\exp\bigg[
-\frac{\zeta_{\rm G}^2}{2\sigma_0^2}
\bigg]\,.
\end{align}  
The problem we face is that we are not interested in the statistics of $\zeta$ but rather in the statistics of 
the so-called density contrast field\,\cite{Harada:2015yda}
\begin{align}\label{eq:NonLinearDelta}
\delta(\vec{x},t) =  
-
\frac{2}{3}\Phi
\left(
\frac{1}{aH}
\right)^2 
e^{-2 \zeta(\vec{x})}
\bigg[
\nabla^2\zeta(\vec{x}) + \frac{1}{2} \partial_{i}\zeta(\vec{x})
 \partial_{i}\zeta(\vec{x})
\bigg]\,,~~~~~~~~~~{\rm with}~~~~\zeta(\vec{x}) = \log\big\{X[r_{\rm dec},\zeta_{\rm G}(\vec{x})]\big\}\,.
\end{align}
where we highlighted the spatial dependence of the random field $\zeta(\vec{x})$ and $\Phi$ captures the 
dependence of the relation between curvature and density contrast on the equation of state.
Written in terms of a constant
equation of state parameter 
$\omega = p/\rho$ (with $\omega =1/3$ for a radiation-dominated Universe), one finds
\begin{align}
 \Phi \equiv \frac{3(1+\omega)}{5+3\omega}\,,
 \label{eq:DefPhi}
\end{align}
and  $\Phi = 2/3$ in a radiation fluid.

The time dependence in eq.\,(\ref{eq:NonLinearDelta}) comes from the scale factor $a=a(t)$ and the Hubble rate $H=H(t)$. 
The density contrast 
can be defined more precisely as $\delta(\vec{x},t) \equiv \delta\rho(\vec{x},t)/\rho_b(t)$ where 
$\rho_b(t)$ is the average background radiation energy density
$\rho_b(t) = 3H(t)^2/8\pi$
and $\delta\rho(\vec{x},t) \equiv \rho(\vec{x},t)- \rho_b(t)$ its perturbation. 
We are interested in the evaluation of eq.\,(\ref{eq:NonLinearDelta}) 
at $t = t_H$, that is the time when
curvature perturbations re-enter the horizon. 

Eq.\,(\ref{eq:NonLinearDelta}) tells us two important things.  
First, information about first and second spatial derivatives of $\zeta(\vec{x})$ -- which, in turn, are also random variables -- is relevant to determine the 
spatial distribution of $\delta(\vec{x},t)$.
Second, additional NGs are present because of the 
non-linear relation between $\zeta$ and $\delta$\,\cite{DeLuca:2019qsy,Young:2019yug}.  

At the gaussian level, including information about the spatial derivatives of a random field is easy (see e.g.\,\cite{Bardeen:1985tr}). 
This is because the joint ten-dimensional probability density function (PDF) for the gaussian variable $\zeta_{\rm G}$ and its spatial derivatives 
takes the multi-normal form (under the assumption of spatial homogeneity and isotropy)
 \begin{align}\label{eq:GaussianPDF}
 {\scalebox{0.95}{$
\textrm{P}_{\rm G}(\zeta_{\rm G},\zeta_{{\rm G},i},\zeta_{{\rm G},ij})
 d\zeta_{\rm G}d\zeta_{{\rm G},i}d\zeta_{{\rm G},ij} = 
\bigg[\prod_{i=x,y,z}^{}\textrm{P}_{\rm G}(\zeta_{{\rm G},i})\bigg]
\bigg[\prod_{ij=xy,yz,xz}\textrm{P}_{\rm G}(\zeta_{{\rm G},ij})\bigg]
\textrm{P}_{\rm G}(\zeta_{\rm G},\zeta_{{\rm G},xx},\zeta_{{\rm G},yy},\zeta_{{\rm G},zz})
d\zeta_{\rm G}d\zeta_{{\rm G},i}d\zeta_{{\rm G},ij}$
}}\,,
 \end{align}
where we are using the short-hand notation $\zeta_{{\rm G},i} = \partial_{i}\zeta_{{\rm G}}$, 
$\zeta_{{\rm G},ij} = \partial_{ij}\zeta_{{\rm G}}$,
 and where
 \begin{align}
 \textrm{P}_{\rm G}(\zeta_{{\rm G},i}) = 
 \frac{1}{\sqrt{\pi \sigma_1^2}}\exp\left(-\frac{\zeta_{{\rm G},i}^2}{\sigma_1^2}\right)\,,~~~{i=x,y,z}\,,~~~~~~~~~
 \textrm{P}_{\rm G}(\zeta_{{\rm G},ij}) = 
  \frac{2}{\sqrt{\pi \sigma_2^2}}\exp\left(-\frac{4\zeta_{{\rm G},ij}^2}{\sigma_2^2}\right)\,,~~~{ij=xy,yz,xz}\,,
 \end{align}
 with
   \begin{align}
  \textrm{P}_{\rm G}(\zeta_{\rm G},\zeta_{{\rm G},xx},\zeta_{{\rm G},yy},\zeta_{{\rm G},zz}) = 
  \frac{\exp\left(
 -\frac{1}{2}\tilde{\zeta}_{\rm G}^{\rm T} \tilde{C}^{-1}\tilde{\zeta}_{\rm G}
  \right)}{(2\pi)^{3/2}\sqrt{{\rm det}\tilde{C}}}\,,\hspace{0.3cm}
\tilde{C} \equiv   
\bordermatrix{     
  & {\scriptstyle \zeta_{\rm G}}         & {\scriptstyle \zeta_{{\rm G},xx}}     & {\scriptstyle \zeta_{{\rm G},yy}} & {\scriptstyle \zeta_{{\rm G},zz}} & \cr
    {\scriptstyle \zeta_{\rm G}}         & \sigma_0^2 & -\sigma_1^2/2 & -\sigma_1^2/2  &  -\sigma_1^2/2 \cr
    {\scriptstyle  \zeta_{{\rm G},xx} }  & -\sigma_1^2/2 & 3\sigma_2^2/8 & \sigma_2^2/8 & \sigma_2^2/8 \cr
    {\scriptstyle  \zeta_{{\rm G},yy}  } &  -\sigma_1^2/2 & \sigma_2^2/8 & 3\sigma_2^2/8 & \sigma_2^2/8 \cr
    {\scriptstyle  \zeta_{{\rm G},zz}  } &  -\sigma_1^2/2 & \sigma_2^2/8 & \sigma_2^2/8 & 3\sigma_2^2/8
}\,,
 \end{align}
with $\tilde{\zeta}_{\rm G} \equiv (\zeta_{{\rm G}},\zeta_{{\rm G},xx},\zeta_{{\rm G},yy},\zeta_{{\rm G},zz})^{\rm T}$.

The variances $\sigma_i^{2}$ can be computed integrating the power spectrum of curvature perturbations, $P_{\zeta}(k)$. 
The latter refers to the gaussian field $\zeta_{\rm G}$.
Furthermore, we include  the Fourier transform of the top-hat window
function in real space $W(k,R)$ (used to smooth the field over a finite volume of size set by the length scale $R$) and the linear transfer function $T(k,\tau)$ (which describes, being $\tau$ the conformal time, the linear evolution  of sub-horizon scales). In full generality, the variances are
\begin{align}
\sigma_j^{2} = \int 
\frac{dk}{k}
W^2(k,R) 
T^2(k,\tau) 
P_{\zeta}(k) 
k^{2j}\,,~~~~~~~~{\rm with}~~~~~~~~
\left\{
\begin{array}{ccc}
W(k,R)  & =  & 3\big[
\frac{\sin(kR) - kR\cos(kR)}{(kR)^3}\big]  \\
 &   &   \\
T(k,\tau)  & = & 3\big[
\frac{\sin(k\tau/\sqrt{3}) - (k\tau/\sqrt{3})\cos(k\tau/\sqrt{3})}{(k\tau/\sqrt{3})^3}\big]  
\end{array}.
\right.\label{eq:Variances}
\end{align} 
Notice that the expression of $T(k,\tau)$ in eq.\,(\ref{eq:Variances}) is strictly valid during radiation domination
with constant degrees of freedom.
We note that $\lim_{kR\gg 1}W(k,R) = 0$ and $\lim_{kR \ll 1}W(k,R) = 1$ meaning that, for fixed $R$, modes with comoving wavenumber $k\gg 1/R$ are smoothed away by $W(k,R)$; similarly, $\lim_{k\tau \gg 1}T(k,\tau) = 0$ and $\lim_{k\tau \ll 1}T(k,\tau) = 1$ meaning that $T(k,\tau)$ has the effect of smoothing out sub-horizon modes,
playing the role of the pressure gradients and dissipative effects.

 For definiteness, as far as the power spectrum of the gaussian curvature perturbation 
 field is concerned, 
 in this paper we shall focus on the two representative (and widely used) cases
 \begin{align}
{\rm log \mbox{-} normal\,power\,spectrum:}~~~~~~~& P_{\zeta}(k) = 
\frac{A}{\sqrt{2\pi}\sigma}\exp\left[
-\frac{\log^2(k/k_{\star})}{2\sigma^2}
\right]\,,\label{eq:LogNo}\\
{\rm broad\,power\,spectrum:}~~~~~~~& P_{\zeta}(k) = A\,\Theta(k - k_{\rm min})\,\Theta(k_{\rm max} - k)\,,
\label{eq:Bro}
 \end{align}
where $\Theta$ is the Heaviside step function. 
The log-normal functional form is a proxy for a power spectrum with a narrow peak, that is a 
peak characterized by a definite wavenumber $k_{\star}$ (with amplitude $A$ and width 
controlled by the parameter $\sigma$). In the broad case, on the contrary, 
the power spectrum does not possess a definite peak wavenumber but it extends 
over the large window in between $k_{\rm min}$ and $k_{\rm max}\gg k_{\rm min}$.
\begin{figure}[b]
\begin{center}
$$\includegraphics[width=.495\textwidth]{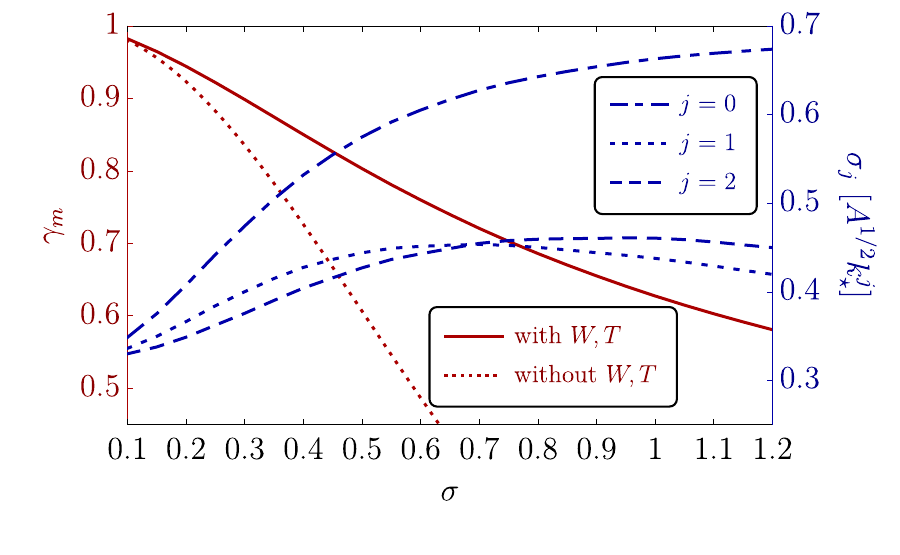}~
\includegraphics[width=.495\textwidth]{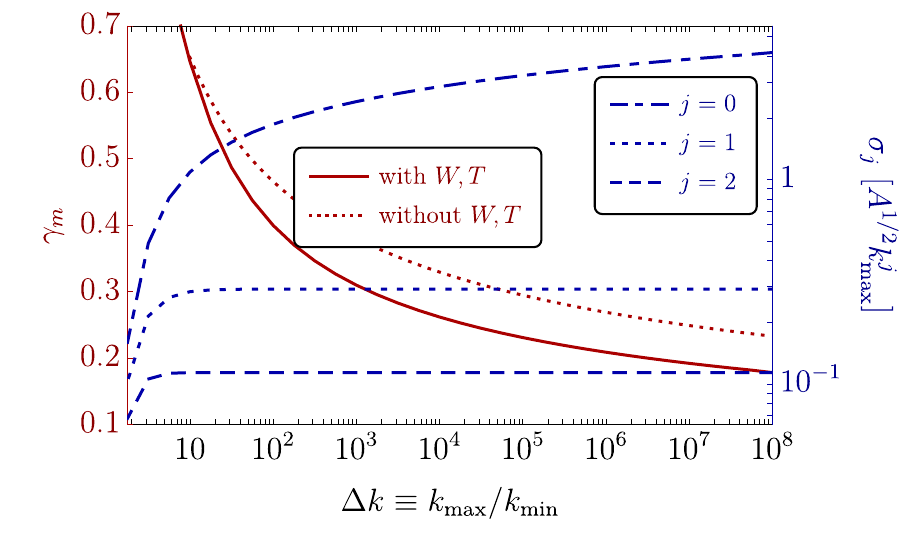}$$\vspace{-0.5cm}
\caption{\em  
Variances (right-side axis) and $\gamma_m$ (left-side axis) computed according to, respectively, 
eq.\,(\ref{eq:Variances}) and eq.\,(\ref{eq:gammam}) in the case of the log-normal power spectrum (\textbf{\textit{left panel }}, as function of $\sigma$) and 
the broad power spectrum (\textbf{\textit{right panel }}, as function of $k_{\rm max}/k_{\rm min}$). 
As far as the variances are concerned, we show the cases with $j=0,1,2$. 
As far as $\gamma_m$ is concerned, the dotted red lines represent the analytic results given 
in eqs.\,(\ref{eq:anal1},\,\ref{eq:anal2}) in the absence of both the window and transfer function. 
We remark that in the computation of the variances window and transfer functions are included.
 }\label{fig:SpectralFeatures}  
\end{center}
\end{figure}
We compute numerically the variances in eq.\,(\ref{eq:Variances}) for the two power spectra in 
eq.\,(\ref{eq:LogNo}) and eq.\,(\ref{eq:Bro}). Furthermore, for future reference, we also compute the dimensionless parameter 
\begin{align}\label{eq:gammam}
\gamma_m \coloneqq \frac{\sigma_1^2}{\sigma_0\sigma_2}\,.
\end{align}
We show our results in fig.\,\ref{fig:SpectralFeatures}. 
The parameter $\gamma_m$ is controlled by the width of the power spectrum. 
This point can be made more transparent if we temporarily neglect the presence of the window and transfer functions; in such a case, 
we get the following analytic 
expressions for the variances
\begin{align}
{\rm log\,normal\,power\,spectrum:}~~~~~~~&\sigma_j^2 = A k_{\star}^{2j} e^{2j^2\sigma^2}~~~~~~~~~~~~~\,\,
\Longrightarrow~~~ \gamma_m = e^{-2\sigma^2}\,,\label{eq:anal1}\\
{\rm broad\,power\,spectrum:}~~~~~~~& \sigma_j^2 = \frac{A k_{\rm max}^{2j}(1-\Delta k^{-2j})}{2j}
~~~
\Longrightarrow~~~\gamma_m = \frac{\Delta k^2 - 1}{\sqrt{(\Delta k^4 - 1)\log(\Delta k)}}\,,\label{eq:anal2}
\end{align}
where $\Delta k \coloneqq k_{\rm max}/k_{\rm min} \gg 1$. 
In both cases, when the power spectrum has a large width (that is in the formal limits $\sigma,\Delta k \to \infty$) we have that 
$\gamma_m \to 0$. 
Furthermore, in both cases we have that $0<\gamma_m < 1$. In the log-normal case, $\gamma_m \to 1$ if $\sigma\to 0$; 
in the broad case, the relevant limit is $\lim_{\Delta k\to 1^{+}}\gamma_m = 1$.

 Qualitatively, these behaviors are respected in the presence of the window and transfer functions, that we include in  
 fig.\,\ref{fig:SpectralFeatures} 
 following ref.\,\cite{Musco:2020jjb}. 
 More precisely, with the window function $W(k,R)$ we smooth the curvature field over a region of size $R = r_m$ and we set the time-scale that enters in the transfer function $T(k,\tau)$ to be $\tau=r_m$. 
The scale $r_m$ measures the characteristic scale of
the density perturbation, and can be defined as the location where the
compaction function is maximized.
In fig.\,\ref{fig:SpectralFeatures} we use the numerical values for $k_{\star}r_m$ 
(as far as the log-normal power spectrum is concerned) and $k_{\rm max}r_m$ (in the broad case) derived as in 
ref.\,\cite{Musco:2020jjb}.

If we now combine fig.\,\ref{fig:SpectralFeatures} and fig.\,\ref{fig:CnCoeff}, we gain a very important intuition. If we consider a very narrow power spectrum (like the log-normal power spectrum in eq.\,(\ref{eq:LogNo}) 
with small $\sigma$ or the 
double-step power spectrum in eq.\,(\ref{eq:Bro}) with $1 <\Delta k \ll 10$) then we expect the expansion in eq.\,(\ref{eq:ZetaSeries}) to be a valid approximation already for small $N$. 
This is because we have in these cases $\sigma_0 \ll 1$. 
On the contrary, for  broader power spectra that violate the condition  $\sigma_0 \ll 1$ we expect a non perturbative analysis based on the full NG curvature perturbation variable $\zeta$ to be needed. 
Remarkably, this intuition---based on very simple consideration mostly involving the curvature perturbation instead of the density contrast---will turn out to be true. We shall elaborate more quantitatively the case of broad power spectra in section\,\ref{sec:TheBroad}.

Inferring the ten-dimensional joint PDF of $\zeta$ starting from the multi-normal distribution in 
eq.\,(\ref{eq:GaussianPDF}) is a very complicated task.\footnote{See ref.\,\cite{Riccardi:2021rlf} for a general strategy in this direction 
which is, however, realistically affordable if one limits the analysis to the quadratic approximation.} 
As anticipated, to make matters worse one is actually interested in computing the PBHs mass fraction (see e.g.\,\cite{Sasaki:2018dmp})
\begin{align}\label{eq:SimpleBeta}
\beta = \int_{\delta_c}^{\infty}
\mathcal{K}(\delta -\delta_c)^{\gamma}\,
\textrm{P}_{\delta}(\delta)d\delta\,,
\end{align}
where $\textrm{P}_{\delta}(\delta)$ is the PDF of the density contrast field that
 we integrate above a certain threshold above which over-densities are expected to collapse and form black holes. 
 The scaling-law factor for critical collapse $\mathcal{K}(\delta -\delta_c)^{\gamma}$ 
 accounts for the mass of the primordial black hole at formation written 
 in units of the horizon mass at the time of horizon re-entry\,\cite{Choptuik:1992jv,Evans:1994pj,Musco:2012au}.
  During radiation domination, the constants $\mathcal{K}$ and $\gamma$
have been numerically found to be given by $\mathcal{K} = O(1\div 10)$ (see e.g.\,\cite{Young:2019yug}) and $\gamma \simeq 0.36$.
 For definiteness, we take $\mathcal{K} \simeq 3.3$ for a log-normal power spectrum and $\mathcal{K} \simeq 4.36$ for a broad power spectrum\,\cite{Muscoinprep} .
 
 Eq.\,(\ref{eq:SimpleBeta}) is valid within the Press–Schechter formalism.  
 Computing $\beta$ in eq.\,(\ref{eq:SimpleBeta}) implies that one should {\it i)} extract the joint PDF of $\zeta$ from that of $\zeta_{\rm G}$ and 
 {\it ii)} find a way to use it to compute the statistics of collapsing region 
 characterised by extreme values
 of density contrast.

\subsection{Comparison between peaked and broad power spectrum: preliminary considerations}\label{sec:TheBroad}

Before entering in the main part of our analysis, it is instructive to pause for a moment and compare, 
in a simplified working setup, the main differences that are expected when comparing the 
two cases of peaked and broad power spectrum. We already built some intuition in the previous section, which we shall now further corroborate in a slightly more complicated setup. 
For simplicity's sake, we will keep focusing the discussion on the curvature perturbation field 
$\zeta$ 
with a ``logarithmic functional form''
as the one in eq.\,(\ref{eq:MasterX}). However, compared to what we have done in the preceding section, the point now is to translate our intuition about the key role of $\sigma_0$ from the mere computation of the coefficients 
$c_n(r_{\rm dec})$ to a more realistic (even though not yet fully rigorous) computation of the PBHs mass fraction.

We consider as formal analogue of eq.\,(\ref{eq:SimpleBeta}) the following integral
\begin{align}
\beta_{\rm NG} = \int_{\zeta_c}^{\infty}
\textrm{P}(\zeta,r_{\rm dec})d\zeta = 
\int_{-\infty}^{+\infty}
\textrm{P}_{\rm G}(\zeta_{\rm G})
\Theta\{\log\big[X(r_{\rm dec},\zeta_{\rm G})\big] - \zeta_c\}d\zeta_{\rm G}\,,\label{eq:ProxyBeta}
\end{align}
where our main focus is on the statistics of the $\zeta$ field (so that we did not include any scaling factor). 
Notice that, because of eq.\,(\ref{eq:ExactPDF}), this integral can be computed exactly once we fix the value of $r_{\rm dec}$ and the 
characteristic width $\sigma_0$ that controls the gaussian distribution of $\zeta_{\rm G}$. 
In the second equality in eq.\,(\ref{eq:ProxyBeta}) we recast the integration over the 
NG field $\zeta$ into  
 an integration over the gaussian component $\zeta_{\rm G}$ subject to the constraint imposed by the 
  the Heaviside step function $\Theta$. 
  In light of the power-series expansion proposed  in eq.\,(\ref{eq:ZetaSeries}), we introduce the approximated quantity 
\begin{align}
\beta_{\rm NG}^{(N)} = 
\int_{-\infty}^{+\infty}
\textrm{P}_{\rm G}(\zeta_{\rm G})
\,\Theta\left[
\sum_{n = 1}^{N} c_n(r_{\rm dec}) \zeta_{\rm G}^{n} - \zeta_c \right]d\zeta_{\rm G}\,.\label{eq:ProxyBetaApprrox}
\end{align}  
Our goal is the comparison between eq.\,(\ref{eq:ProxyBeta}) and 
eq.\,(\ref{eq:ProxyBetaApprrox}).
In the end, we aim to answer two reasonable questions: 
\begin{itemize}
    \item [{\it i)}]
at which order the truncated power series 
gives a $\beta_{\rm NG}^{(N)}$ that converges to the exact value given by eq.\,(\ref{eq:ProxyBeta})? 
\item [{\it ii)}]
does this conclusion depend on the form of the power spectrum?
\end{itemize}
To answer these questions the key point to bear in mind is that a power-series expansion of the form
\begin{align}
\sum_{n = 1}^{\infty} c_n(r_{\rm dec}) \zeta_{\rm G}^{n} = \log\big[X(r_{\rm dec},\zeta_{\rm G})\big]\,,\label{eq:ConvergenceTest}
\end{align} 
must be always accompanied, to make mathematical sense, by the information about its radius of convergence $R$, that is 
the region of values $-R<\zeta_{\rm G} < + R$ in which the above equality is strictly valid. 
This is a crucial information given that in eq.\,(\ref{eq:ProxyBetaApprrox}) we integrate, in principle, 
over all real values of $\zeta_{\rm G}$.
We address the computation of the radius of convergence $R$ for the above expansion in appendix\,\ref{app:Radius}, and 
we shall focus here on the implications.  
\begin{figure}[!h!]
\begin{center}
$$\includegraphics[width=.52\textwidth]{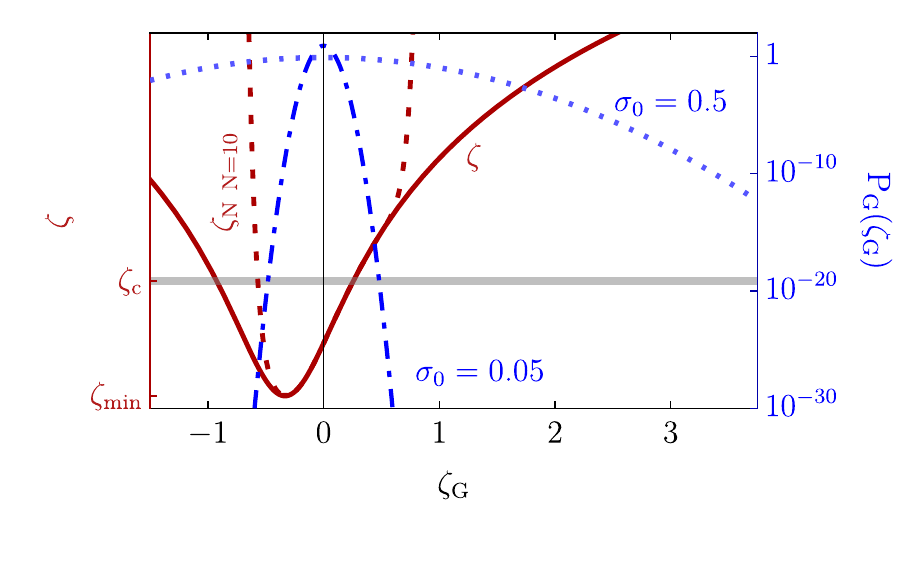}
\quad\quad\includegraphics[width=.52\textwidth]{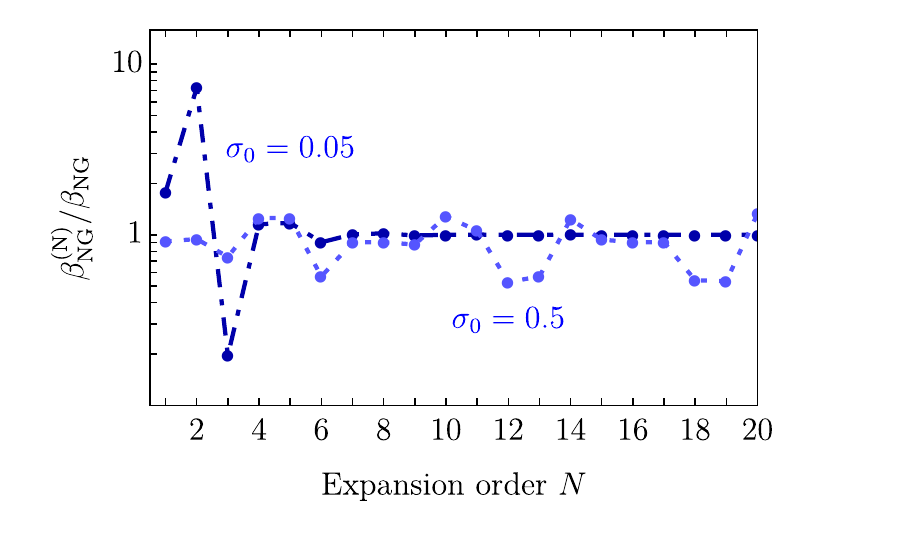}$$
\caption{\em  
We fix $r_{\rm dec}=0.5$.
	\textbf{\textit{	Left panel. }} On the left-side $y$-axis we plot the function 
$\zeta = \log\big[X(r_{\rm dec},\zeta_{\rm G})\big]$ as function of the Gaussian field $\zeta_{\rm G}$ (solid red line) and its power-series expansion $\zeta_{10}$ (dashed red line). The two lines 
are superimposed and indistinguishable only for a small range of values of $\zeta_{\rm G}$ centered around the origin. 
On the right-side $y$-axis we plot the Gaussian PDF (with zero mean) for two representative values of its variance $\sigma_0$ (blue dotted line, $\sigma_0 = 0.5$, and blue dot-dashed line, $\sigma_0 = 0.05$). 
	\textbf{\textit{	Right panel. }} We plot the ratio 
$\beta_{\rm NG}/\beta_{\rm NG}^{(N)}$ (with $\beta_{\rm NG}$ and $\beta_{\rm NG}^{(N)}$ defined, respectively, as in eq.\,(\ref{eq:ProxyBeta}) and eq.\,(\ref{eq:ProxyBetaApprrox})) as function of the expansion order $N$. We plot both cases with 
$\sigma_0 = 0.5$ and $\sigma_0 = 0.05$.
 }\label{fig:PeakedBroad}  
\end{center}
\end{figure}
Consider the situation displayed in the left panel of fig.\,\ref{fig:PeakedBroad}. 
In this plot we show the graph of $\zeta = \log\big[X(r_{\rm dec},\zeta_{\rm G})\big]$ as function of the gaussian 
variable $\zeta_{\rm G}$ (solid red line). For definiteness, we take $r_{\rm dec}=0.5$. 
In addition, we show the graph of $\zeta_N$ in eq.\,(\ref{eq:ZetaSeries}) for some value of $N$ (dashed red line; we take $N= 10$ but 
the exact 
value of $N$ is not of crucial importance for the following argument).

From the comparison between $\zeta_N$ and $\zeta$, it is evident that the graph of $\zeta_N$ perfectly agrees with that of $\zeta$ 
for values of $\zeta_{\rm G}$ that lie into an interval centred in the origin and not too wide. 
This is because this interval lies within the radius of convergence of 
eq.\,(\ref{eq:ConvergenceTest}). However, as soon as we get too far from the origin 
the agreement between $\zeta_N$ and $\zeta$ gets unavoidably lost: in the language of eq.\,(\ref{eq:ConvergenceTest}) 
this happens because we go outside the interval of convergence of the power-series expansion. 
We refer to appendix\,\ref{app:Radius} for a careful discussion about the computation of the radius of convergence of 
eq.\,(\ref{eq:ConvergenceTest}) (see fig.\,\ref{fig:Radius} for a more detailed version of 
the left panel of fig.\,\ref{fig:PeakedBroad}).

The key point is the following. In eq.\,(\ref{eq:ProxyBetaApprrox}), the 
values of $\zeta_{\rm G}$ over which we integrate are weighted by the 
gaussian distribution $\textrm{P}_{\rm G}(\zeta_{\rm G})$ with width $\sigma_0$. 
We remind that, as illustrated in fig.\,\ref{fig:SpectralFeatures}, the broader the power spectrum the bigger the value 
of $\sigma_0$ compared to higher momenta $\sigma_i$. 
In the left panel of fig.\,\ref{fig:PeakedBroad} we superimpose in blue the profile of the gaussian distribution 
 $\textrm{P}_{\rm G}(\zeta_{\rm G})$ for two different benchmark values of $\sigma_0$, namely 
 $\sigma_0 = 0.05$ and $\sigma_0 = 0.5$. We argue the following points: 
\begin{itemize}
\item[$\circ$] If the power spectrum is very narrow -- thus $\sigma_0$ very small, like in the case $\sigma_0 = 0.05$ 
in our example -- 
the dominant  
contribution to the integral in eq.\,(\ref{eq:ProxyBetaApprrox}) comes from a region of $\zeta_{\rm G}$ that lies 
within the radius of convergence of 
eq.\,(\ref{eq:ConvergenceTest}). 
Consequently, we expect an excellent agreement between eq.\,(\ref{eq:ProxyBeta}) and 
eq.\,(\ref{eq:ProxyBetaApprrox}) already for relatively small values of $N$. 
\item[$\circ$] If the power spectrum gets broader -- thus for increasing 
$\sigma_0$, like in the case $\sigma_0 = 0.5$ 
in our example -- in the computation of the integral in eq.\,(\ref{eq:ProxyBetaApprrox}) 
the gaussian distribution will give more weight to values of $\zeta_{\rm G}$ that are outside the 
region of convergence of eq.\,(\ref{eq:ConvergenceTest}). 
In this case, the agreement between eq.\,(\ref{eq:ProxyBeta}) and 
eq.\,(\ref{eq:ProxyBetaApprrox}) will get fatefully lost. 
Notice that, in this situation, going to arbitrary increasing values of $N$ 
will not fix the problem simply because outside the region of convergence talking about 
the equality in eq.\,(\ref{eq:ConvergenceTest}) does not make any sense at all.
\end{itemize}
To further corroborate these considerations, in the right panel of fig.\,\ref{fig:PeakedBroad} we show the value of $\beta_{\rm NG}^{(N)}/\beta_{\rm NG}$ 
for increasing $N$ and for the two values of $\sigma_0$ discussed before. 
In the case    $\sigma_0 = 0.05$ we see that we have an excellent agreement between 
the full and the approximated computation already for $N\gtrsim 4$.  
On the contrary, if we consider $\sigma_0 = 0.5$ we see that the agreement between 
$\beta_{\rm NG}^{(N)}$ and $\beta_{\rm NG}$ starts deteriorating precisely because 
we are now including with some sizable probability the integration over a non-convergent region.

From this simplified exercise, we draw the following conclusions.
If we consider a very narrow power spectrum (like the log-normal power spectrum in eq.\,(\ref{eq:LogNo}) 
with small $\sigma$ or the 
double-step power spectrum in eq.\,(\ref{eq:Bro}) with $1 <\Delta k \ll 10$) we expect that the power-series expansion in eq.\,(\ref{eq:ZetaSeries}) will give 
reliable results. 
However, in the case of a broad power spectrum (like 
the log-normal power spectrum in eq.\,(\ref{eq:LogNo}) with large $\sigma$ or the 
double-step power spectrum in eq.\,(\ref{eq:Bro}) with $\Delta k \gg 1$) 
we expect that the power-series expansion in eq.\,(\ref{eq:ZetaSeries}) will give 
	unreliable results. 
	
	In the case of a broad power spectrum, therefore,
	understanding how to compute  eq.\,(\ref{eq:SimpleBeta}) without relying on the power-series expansion in 	
eq.\,(\ref{eq:ZetaSeries}) is of pivotal importance.

\section{The NG PBH mass fraction at all orders}\label{sec:NonGauBeta}

After the propaedeutic discussion of section\,\ref{sec:NGIntro}, we are now ready to tackle the main part of this work, that is the computation of the PBH mass fraction $\beta$ including both the non-linear relation between $\zeta$ and $\delta$ and the intrinsic NG of $\zeta$ in the form given by eq.\,(\ref{eq:MainF}). 
The key aspect is encoded in the equality\,\cite{Musco:2018rwt}
\begin{align}\label{eq:Dicotomy}
\underbrace{\frac{1}{V_b(r_m,t_H)}
\int_{S^2_{R_m}}d\vec{x}\,
\delta\rho(\vec{x},t_H)
= \frac{\delta M(r_m,t_H)}{M_b(r_m,t_H)} }
_{
\rm volume\,\,averaged\,\,density\,\,contrast
}
=\underbrace{
\frac{2\left[M(r,t) - M_b(r,t)\right]}{R(r,t)} 
= \mathcal{C}(r_m)}_{\rm 
compaction\,\,function}\,,
\end{align}
To understand the meaning of this equation, we first clarify few definitions. The areal radius is defined by
$R(r,t) \equiv a(t)re^{\zeta(r)}$ where $r$ is the radial coordinate 
that appears in the expression of the metric in the
uniform density slicing $ds^2 = -dt^2 + a(t)^2 e^{2\zeta(r)}(dr^2 + r^2d\Omega^2)$; 
$\zeta(r)$ is the conserved comoving curvature perturbation defined on
super-Hubble scales, and written assuming spherical symmetry. 
Said differently, the line element $ds^2$ describes  a locally perturbed region that would eventually collapse to a PBH; 
such a region will be very rare, and it may be approximated
by a spherically symmetric region of positive curvature. 
$M(r,t)$ is the Misner-Sharp mass within a sphere of radius $R(r,t)$
and $M_b(r,t)$ is the background mass within the same areal radius calculated
with respect to the background energy density. We have 
\begin{equation}
 M_b(r,t) = \rho_b(t)V_b(r,t)
 ~~~~~~~~~\text{with} ~~~~~~~~~
 V_b(r,t) = (4\pi/3)R(r,t)^3.   
\end{equation}
The quantity $\delta M(r,t) = M(r,t) - M_b(r,t)$ is the mass excess within a spherical region of
areal radius $R$.
Eq.\,(\ref{eq:Dicotomy}) is evaluated at $t = t_H$, that is the time when
curvature perturbations re-enter the horizon.
We have the relation 
\begin{align}\label{eq:MangioExp}
a(t_H)H(t_H)r_m e^{\zeta(r_m)} = 1\,,
\end{align}
that defines implicitly the length scale $r_m$.
The latter can be computed as the scale at which the compaction function is maximized. 
In this paper, as far as the computation of $r_m$ is concerned, 
we shall use the results of ref.\,\cite{Musco:2020jjb}. 
Concretely, we consider
 \begin{align}
{\rm log\,normal\,power\,spectrum:}~~~~~~~&
r_m k = \kappa(\sigma)\,,\label{eq:Logrk}
\\
{\rm broad\,power\,spectrum:}~~~~~~~&
r_m k = \kappa = 4.49\,,\label{eq:Broadrk}
\end{align}
where in the first equation $\kappa$ is a function of the log-normal width (that we take from ref.\,\cite{Musco:2020jjb}) while in the second one $\kappa$ takes the constant value valid for the broad power spectrum in eq.\,(\ref{eq:Bro}), cf. ref.\,\cite{Musco:2020jjb}.
Furthermore, it is possible to relate the scale $r_m$ to the horizon mass $M_H$ 
via its relation with $k$  
\begin{align}
M_H \simeq 17 M_{\odot}
\left(
\frac{g_{\star}}{10.75}
\right)^{-1/6}\left(\frac{k/\kappa}{10^6 {\rm Mpc}^{-1}}\right)^{-2}
~~~~~
\Longrightarrow
~~~~~
r_m = \frac{\kappa}{k} \simeq 2.4\times 10^{-7} {\rm Mpc}
\left(
\frac{g_{\star}}{10.75}
\right)^{1/12}
\left(\frac{M_H}{M_{\odot}}\right)^{1/2},
\label{eq:HorizonMass}
\end{align}
that can be used to trade $r_m$ for $M_H$. 
In eq.\,(\ref{eq:HorizonMass})
$g_{\star}$ is the number of degrees of freedom of relativistic
particles with 
$g_{\star} = 106.75$ deep in the radiation epoch. 
We will take into account the temperature-dependence of $g_{\star}$ in section\,\ref{sec:PBHGW} when discussing the collapse of PBHs taking place across the quark-hadron QCD phase transition.

Let's come back to eq.\,(\ref{eq:Dicotomy}).
On the left side, we have the density contrast field averaged over a spherical region of areal radius $R_m\equiv a(t_H)r_me^{\zeta(r_m)}$.
On the right side, we have the compaction function $\mathcal{C}(r)$, defined as twice the local mass excess over the areal radius, evaluated at the scale $r_m$ at which it is maximized (which is a crucial condition behind the validity of eq.\,(\ref{eq:Dicotomy})). 
This last point is crucial. 
As discussed in ref.\,\cite{Musco:2018rwt},
the gravitational collapse that triggers the formation of a PBH takes place
when the maximum of the compaction function $\mathcal{C}(r_m)$ is larger than a certain threshold
value. 
Therefore, 
eq.\,(\ref{eq:Dicotomy}) shows that PBH formation 
can be investigated by studying directly statistics of the compaction function, and compute the probability of it overcoming the threshold $\mathcal{C}_{th}$, derived 
in numerical simulations of collapses of the volume-averaged density contrast $\delta_{c}$. 
We have
 \begin{align}
{\rm log\,normal\,power\,spectrum:}~~~~~~~&
\mathcal{C}_{th}=\delta_c = \delta_c(\sigma)\,,\label{eq:deltacLog}
\\
{\rm broad\,power\,spectrum:}~~~~~~~&
\mathcal{C}_{th}=\delta_c = 0.56\,.\label{eq:deltacBroad}
\end{align}
As discussed in\,\cite{Germani:2018jgr,Musco:2018rwt,Musco:2020jjb}, the threshold $\mathcal{C}_{th}$ (or $\delta_c$) depends on the shape of the curvature power spectrum. 
In eqs.\,(\ref{eq:deltacLog},\,\ref{eq:deltacBroad}) we use the results of ref.\,\cite{Musco:2020jjb} that include the non-linear effects between curvature perturbations and density contrast.  
Furthermore, it should be noted that eqs.\,(\ref{eq:deltacLog},\,\ref{eq:deltacBroad}) are strictly valid during the radiation epoch. 
In section\,\ref{sec:PBHGW} we will include, in the case of the broad power spectrum, deviations with respect to the value $\delta_c = 0.56$ due to the quark-hadron QCD phase transition.

One final remark is in order. 
As discussed, we compute $\delta_c$ using the results of ref.\,\cite{Musco:2020jjb} that include the effect of the non-linear relation between curvature perturbations and density contrast. However, 
ref.\,\cite{Musco:2020jjb} assumes that curvature perturbations follow a Gaussian statistics. 
We expect that the presence of primordial NG will impact the value of $\delta_c$.\footnote{See Refs.\,\cite{Kehagias:2019eil,Escriva:2022pnz} for some efforts in this direction, showing the variation of threshold is around few percent for $|f_\text{\tiny NL}| <{\cal O}(5)$.}
Including (consistently) primordial NG in the computation of $\delta_c$ is a task that goes well beyond the purpose of this work, and will be addressed in a separate publication. We will come back to the relevance of this point in section\,\ref{sec:Final}.

\subsection{Threshold statistics on the compaction function: a prescription}\label{sec:CompFunc}

In this section we elaborate and extend the approach studied in ref.\,\cite{Young:2022phe} to account for the 
presence of primordial NGs.

This approach puts in the foreground the role of the compaction function in the process of PBH formation.
However, as we shall explain, 
the approach of ref.\,\cite{Young:2022phe} is strictly valid only for a simplified monochromatic power spectrum of curvature perturbations, and  
accounts for primordial NGs considering separately the quadratic and cubic approximations
(separately means that in the cubic approximation $f_{\rm NL}$ is set to zero while $g_{\rm NL}\neq 0$). 
As discussed in section\,\ref{sec:NGIntro}, in realistic curvaton models we expect 
the full series in eq.\,(\ref{eq:ZetaSeries}) to be relevant.
Motivated by this point, in this section we extend the analysis of ref.\,\cite{Young:2022phe}.

The compaction function is generically defined as twice the local mass excess over the areal radius 
\begin{align}\label{eq:DefinitionCompaction}
\mathcal{C}(r,t) = \frac{2\left[M(r,t) - M_b(r,t)\right]}{R(r,t)} =
\frac{2}{R(r,t)}\int_{S^2_R} d^{3}\vec{x}
\left[\rho(\vec{x},t) - \rho_b(t)\right] = 
\frac{2}{R(r,t)}
\underbrace{\int_{S^2_R} d^{3}\vec{x}\,\rho_b(t)\,\delta(\vec{x},t)}_{\coloneqq\,\delta M(r,t)}\,.
\end{align}
On super-horizon scales, and adopting the 
 gradient expansion approximation,
 the density contrast is given by eq.\,(\ref{eq:NonLinearDelta}) that we now rewrite 
assuming spherical symmetry 
\begin{align}\label{eq:SphericalDelta}
\delta(r,t) = 
-\frac{2}{3}
\Phi
\left(\frac{1}{aH}\right)^2 
e^{-2\zeta(r)}\left[
\zeta^{\prime\prime}(r) + \frac{2}{r}\zeta^{\prime}(r) + \frac{1}{2}\zeta^{\prime}(r)^2
\right]\,.
\end{align}
Using the above expression and integrating over the radial coordinate, 
eq.\,(\ref{eq:DefinitionCompaction}) takes the form
\begin{align}\label{eq:CompactionFull}
\mathcal{C}(r) = 
-2\Phi\,r\,\zeta^{\prime}(r)\left[
1 + \frac{r}{2}\zeta^{\prime}(r)
\right] = 
\mathcal{C}_1(r) - \frac{1}{4\Phi}\mathcal{C}_1(r)^2\,,~~~~~~~~~~~~~
\mathcal{C}_1(r) \coloneqq -2\Phi\,r\,\zeta^{\prime}(r)\,,
\end{align} 
where $\mathcal{C}_1(r)$ defines the so-called linear component of the compaction function. 
Notice that the compaction function becomes time-independent and
eq.\,(\ref{eq:CompactionFull}) automatically includes the full non-linear relation between $\delta$ and $\zeta$. 
The length scale $r_m$ is defined as the scale at which the compaction function is maximized, and, therefore, 
it verifies the condition 
\begin{equation}
\mathcal{C}^{\prime}(r_m) = 0
~~~~~~~~~~\text{that is }~~~~~~~~~~
\zeta^{\prime}(r_m) + r_m\zeta^{\prime\prime}(r_m) = 0    
\end{equation}
in terms of 
the comoving curvature perturbation. 
If we define $\mathcal{C}_{\rm max} = \mathcal{C}(r_m)$ as the value of the compaction at the position of the maximum, 
PBHs form if the maximum of the compaction function is above some threshold value, 
$\mathcal{C}_{\rm max} > \mathcal{C}_{\rm th}$.
If we consider the averaged mass excess within a spherical region of areal radius $R$, that is the ratio
$\delta M(r,t)/M_b(r,t)$, a direct computation shows that\,\cite{Musco:2018rwt} 
\begin{align}
\frac{\delta M(r,t)}{M_b(r,t)} = \frac{1}{V_b(r,t)}
\int_{S^2_R} d^{3}\vec{x}\,\delta\rho(\vec{x},t)
~~~~~~~
\Longrightarrow~~~~~~~
\delta_m \coloneqq \frac{\delta M(r_m,t_H)}{M_b(r_m,t_H)} = \mathcal{C}(r_m) = 3\delta(r_m,t_H)\,,
\label{eq:MainCompa}
\end{align}
where the last equality follows from eq.\,(\ref{eq:SphericalDelta})  evaluated at horizon crossing
together with the condition $\zeta^{\prime}(r_m) + r_m\zeta^{\prime\prime}(r_m) = 0$ that defines $r_m$. 
Eq.\,(\ref{eq:MainCompa}) shows that the peak value of the compaction function $\mathcal{C}(r_m)$ equals 
$\delta_m$, that is 
the density contrast volume-averaged over a spherical region of areal radius set by the length scale $r_m$.  

We now elaborate on the presence of primordial NGs that are encoded in the relation 
$\zeta = F(\zeta_{\rm G})$ (see eq.\,\eqref{eq:MainF}).
The linear component of the compaction function takes the form
\begin{align}\label{eq:C1expl}
\mathcal{C}_1(r) = -2\Phi\,r\,\zeta_{\rm G}^{\prime}(r)\,
\frac{dF}{d\zeta_{\rm G}} = 
\mathcal{C}_{\rm G}(r)\,
\frac{dF}{d\zeta_{\rm G}}\,,~~~~~~~~{\rm with}~~~
\mathcal{C}_{\rm G}(r) \coloneqq
-2\Phi\,r\,\zeta_{\rm G}^{\prime}(r)\,.
\end{align}
 Consequently, the compaction function reads
 \begin{align}\label{eq:CCgau}
\mathcal{C}(r) = 
\mathcal{C}_{\rm G}(r)\,
\frac{dF}{d\zeta_{\rm G}} 
 - \frac{1}{4\Phi}
 \mathcal{C}^2_{\rm G}(r)
 \left(\frac{dF}{d\zeta_{\rm G}}
 \right)^2\,.
 \end{align}
The compaction function depends on both the gaussian linear component $\mathcal{C}_{\rm G}$ and the gaussian  
 curvature perturbation $\zeta_{\rm G}$. 
 Both these random variables are gaussian; $\zeta_{\rm G}$ is gaussian by definition while  
 $\mathcal{C}_{\rm G}$ is defined by means of the derivative of the gaussian variable $\zeta_{\rm G}$.

 We start from the two-dimensional joint PDF of $\zeta_{\rm G}$ and $\mathcal{C}_{\rm G}$, which can be written as
 \begin{align}\label{eq:PDFCompa}
 \textrm{P}_{\rm G}(\mathcal{C}_{\rm G},\zeta_{\rm G}) 
 = \frac{1}{(2\pi)\sqrt{\det\Sigma}}
 \exp\left(
 -\frac{1}{2}Y^{\rm T}\Sigma^{-1}Y
 \right)\,,~~~~~~~{\rm with}~~Y =\left(
\begin{array}{c}
 \mathcal{C}_{\rm G}  \\
  \zeta_{\rm G}    
\end{array}
\right)\,,~~~~~{\rm and}~~~
\Sigma =
\left(
\begin{array}{cc}
 \langle\mathcal{C}_{\rm G}\mathcal{C}_{\rm G}\rangle & \langle\mathcal{C}_{\rm G}\zeta_{\rm G}\rangle   \\
 \langle\mathcal{C}_{\rm G}\zeta_{\rm G}\rangle & \langle\zeta_{\rm G}\zeta_{\rm G}\rangle     
\end{array}
\right)\,,
 \end{align}
 where $\Sigma$ is the covariance matrix. The entries of $\Sigma$ are\,\cite{Young:2022phe}
 \begin{align}
 \langle\mathcal{C}_{\rm G}\mathcal{C}_{\rm G}\rangle & = \sigma_c^2 =
  \frac{4\Phi^2}{9}\int_0^{\infty}\frac{dk}{k}
  (kr_m)^4 W^2(k,r_m) T^2(k,r_m) P_{\zeta}(k)\,,\label{eq:Var1}
   \\
 \langle\mathcal{C}_{\rm G}\zeta_{\rm G}\rangle & = \sigma_{cr}^2 = 
 \frac{2\Phi}{3}\int_0^{\infty}\frac{dk}{k}(kr_m)^2
 W(k,r_m)
 W_s(k,r_m) T^2(k,r_m) P_{\zeta}(k)\,,
  \\
  \langle\zeta_{\rm G}\zeta_{\rm G}\rangle & = \sigma_r^2 =   \int_0^{\infty}\frac{dk}{k}
  W_s^2(k,r_m) T^2(k,r_m) P_{\zeta}(k)\,,\label{eq:Var3}
 \end{align}
where $W_s(k,r) = \sin(kr)/kr$ while $W(k,R)$ and $T(k,\tau)$ are given in eq.\,(\ref{eq:Variances}).  
As explained in the above preamble, the key point to build the variance of the compaction function, is to consider the density contrast volume-averaged over a spherical region of size set by $r_m$. This is done by means of the top-hat smoothing $W(k,R)$ in eq.\,(\ref{eq:Variances}). Concretely, we have set $R=r_m$ and all variances in this section are evaluated according to this choice.
Furthermore, we also set $\tau = r_m$ in the computation of the linear transfer function as required when computing the variance at the time of horizon crossing of the scale $r_m$\,\cite{Musco:2018rwt}.

After computing the inverse of $\Sigma$ and its determinant, and completing the square in the argument of the exponential function,
 eq.\,(\ref{eq:PDFCompa}) can be recast in the form
\begin{align}\label{eq:PDFCompa2}
 \textrm{P}_{\rm G}(\mathcal{C}_{\rm G},\zeta_{\rm G}) =
 \frac{1}{(2\pi)\sigma_c\sigma_{r}\sqrt{1-\gamma_{cr}^2}}
 \exp\left(
 -\frac{\zeta_{\rm G}^2}{2\sigma_r^2}
 \right)
 \exp\left[
 -\frac{1}{2(1-\gamma_{cr}^2)}\left(
 \frac{\mathcal{C}_{\rm G}}{\sigma_c} - \frac{\gamma_{cr}\zeta_{\rm G}}{\sigma_r}
 \right)^2
 \right]\,,
 \end{align}
where 
\begin{align}
\gamma_{cr} \coloneqq \frac{\sigma_{cr}^2}{\sigma_c \sigma_r}\,.\label{eq:GammacrDef}
\end{align} 
If we take the limit $\gamma_{cr}\to 1$ in eq.\,(\ref{eq:PDFCompa2}), we find the distribution
\begin{align}
\lim_{\gamma_{cr}\to 1}\textrm{P}_{\rm G}(\mathcal{C}_{\rm G},\zeta_{\rm G}) = 
\frac{1}{\sqrt{2\pi}\sigma_c\sigma_{r}} \exp\left(
 -\frac{\zeta_{\rm G}^2}{2\sigma_r^2}
 \right)\delta\left(\mathcal{C}_{\rm G}/\sigma_c -  \zeta_{\rm G}/\sigma_r\right)\,,
\end{align} 
where the last delta function forces the relation 
$\zeta_{\rm G} = \sigma_r\mathcal{C}_{\rm G}/\sigma_c$ once the integral over $\zeta_{\rm G}$ is performed. 
This is the essence of the so-called high-peak limit of ref.\,\cite{Young:2022phe}. 
This limit is strictly valid if one takes $\gamma_{cr}\to 1$ that is the limit in which one takes a 
monochromatic power spectrum of curvature perturbations (that gives precisely $\gamma_{cr} = 1$). 

The main goal of this section is to go beyond the high-peak limit. This is crucial as we are interested in the case of a broad power spectra for which the high-peak limit is not applicable. 
We take on this problem in a very simple way.
The prescription to compute the NG abundance (without taking the limit $\gamma_{cr}\to 1$)
is summarised in the following:
\begin{mynamedbox2}{
NG PBH mass fraction
adopting 
threshold statistics on the compaction function}
\begin{align}
\beta_{\rm NG} & = \int_{\mathcal{D}}\mathcal{K}(\mathcal{C} - \mathcal{C}_{\rm th})^{\gamma}
\textrm{P}_{\rm G}(\mathcal{C}_{\rm G},\zeta_{\rm G})d\mathcal{C}_{\rm G} d\zeta_{\rm G}\,,\label{eq:CompactionIntegral}
\\
 \textrm{P}_{\rm G}(\mathcal{C}_{\rm G},\zeta_{\rm G}) & =
 \frac{1}{(2\pi)\sigma_c\sigma_{r}\sqrt{1-\gamma_{cr}^2}}
 \exp\left(
 -\frac{\zeta_{\rm G}^2}{2\sigma_r^2}
 \right)
 \exp\left[
 -\frac{1}{2(1-\gamma_{cr}^2)}\left(
 \frac{\mathcal{C}_{\rm G}}{\sigma_c} - \frac{\gamma_{cr}\zeta_{\rm G}}{\sigma_r}
 \right)^2
 \right]\,,
 \\
 \mathcal{D} & = 
\left\{
\mathcal{C}_{\rm G},\,\zeta_{\rm G} \in \mathbb{R}~:~~
\mathcal{C}(\mathcal{C}_{\rm G},\zeta_{\rm G}) > \mathcal{C}_{\rm th}  
~\land~\mathcal{C}_1(\mathcal{C}_{\rm G},\zeta_{\rm G}) < 2\Phi
\right\}\,,\label{eq:RegionD}
\end{align} 
\end{mynamedbox2}
\noindent
where $\mathcal{C}$ is written implicitly in terms of $\mathcal{C}_{\rm G}$ and $\zeta_{\rm G}$ by means of 
eq.\,(\ref{eq:CCgau});
 the two conditions in the integration domain $\mathcal{D}$ (with again $\mathcal{C}_1$ written in terms of $\mathcal{C}_{\rm G}$ and $\zeta_{\rm G}$ according to eq.\,(\ref{eq:C1expl})) select the so-called type-I perturbations.
We implement in eq.\,(\ref{eq:CompactionIntegral}) the two conditions that define $\mathcal{D}$ by means 
of Heaviside step functions.
Eq.\,(\ref{eq:CompactionIntegral}) is the main result of this paper: It describes the abundance of collapsing peaks 
in the context of threshold statistics and gives an exact expression in the presence of primordial NGs.

Instead of working with the full resummed expression $\zeta = \log\big[X(\zeta_{\rm G})\big]$, it is also possible to 
consider its power-series expansion in eq.\,(\ref{eq:ZetaSeries}) truncated at some given order $N$. 
The same integral in eq.\,(\ref{eq:CompactionIntegral}) will give an expression for $\beta_{\rm NG}^{(N)}$ at order $N$. This will be used to test the convergence of the series exapnsion in relation to the width of the spectrum.

In practice, at the prize of a slightly more complicated numerical integration (but at the end the computation of 
$\beta$ would have been numerical anyway) we can go beyond the high-peak limit and consider generic 
functional dependences for local type NGs.
In the following, we are going to present 
worked examples by specializing our analysis 
to the expression $\zeta = \log\big[X(\zeta_{\rm G})\big]$.

\subsection{The log-normal power spectrum}

\begin{figure}[h]
\begin{center}
\boxed{{\rm  
Non~linearities~+~primordial~NG:~~\textbf{{\color{venetianred}Threshold~statistics~on~the~compaction~function}}~(log\text{-}normal}~P_{\zeta})}
\vspace{0.25cm}
\includegraphics[width=1\textwidth]{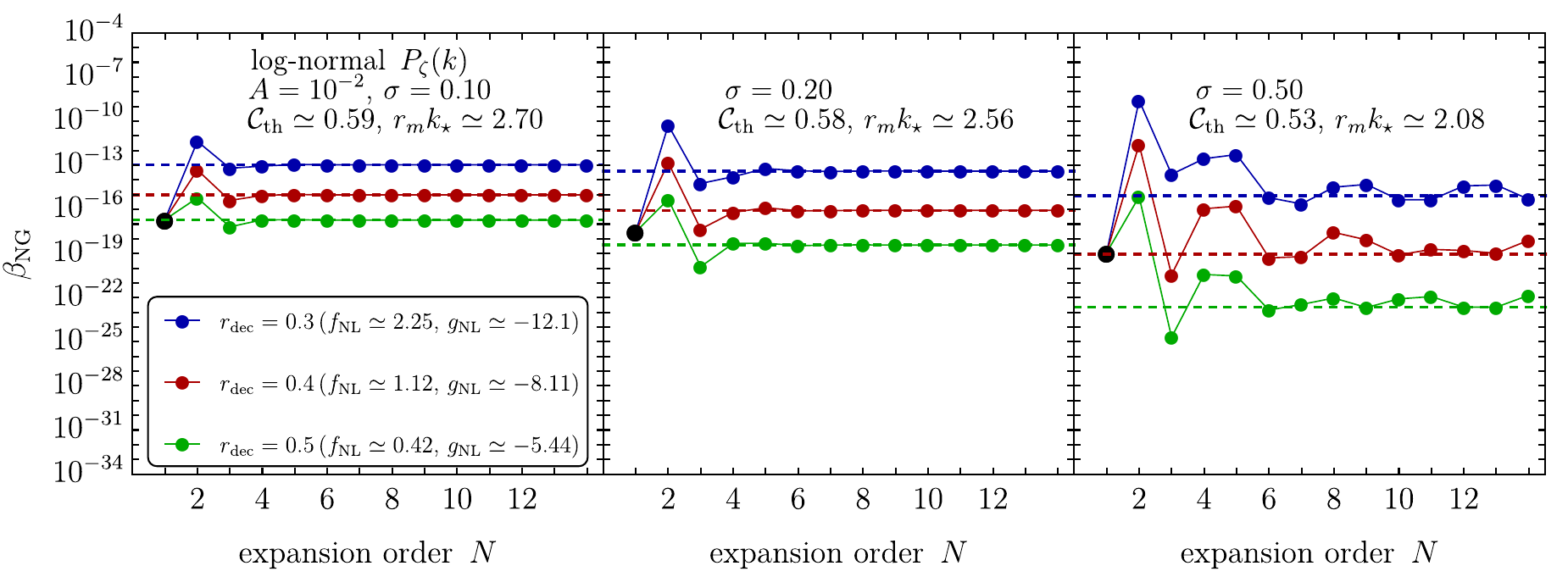}
\caption{\em 
Computation of the primordial mass fraction $\beta$ of the Universe stored into PBHs at the formation time done using the prescription in eq.\,(\ref{eq:CompactionIntegral}).
We consider the log-normal power spectrum in eq.\,(\ref{eq:LogNo}) with fixed amplitude $A$ and 
three different values of $\sigma$; from left to right, we take $\sigma = 0.1$, $\sigma = 0.2$ and $\sigma = 0.5$. 
For each case, we investigate the impact of primordial NGs for three benchmark values of $r_{\rm dec}$.
This result includes both non-linearities and 
NGs of primordial origin motivated by the curvaton model, i.e. eq.\,(\ref{eq:MasterX}). 
We account for primordial NGs both perturbatively (by means of 
the power-series expansion in eq.\,(\ref{eq:ZetaSeries}) truncated at some finite 
order $N$ - filled dots)
and non-perturbatively (with the full resummed result in eq.\,(\ref{eq:MasterX}) - horizontal 
dashed lines). 
  }\label{fig:LogNoCompa}  
\end{center}
\end{figure}

First, we consider the log-normal power spectrum in eq.\,(\ref{eq:LogNo}).
In fig.\,\ref{fig:LogNoCompa} we plot the value of $\beta_{\rm NG}$, computed by means of eq.\,(\ref{eq:CompactionIntegral}), for different values of $\sigma$ and $r_{\rm dec}$ (see caption for details). 
We compare $\beta_{\rm NG}$ (dashed lines) with its perturbative expansion $\beta_{\rm NG}^{(N)}$ 
for different values of the expansion order $N$ (indicated in the $x$-axis in fig.\,\ref{fig:LogNoCompa}).

\begin{figure}[h]
\begin{center}
$$\includegraphics[width=.495\textwidth]{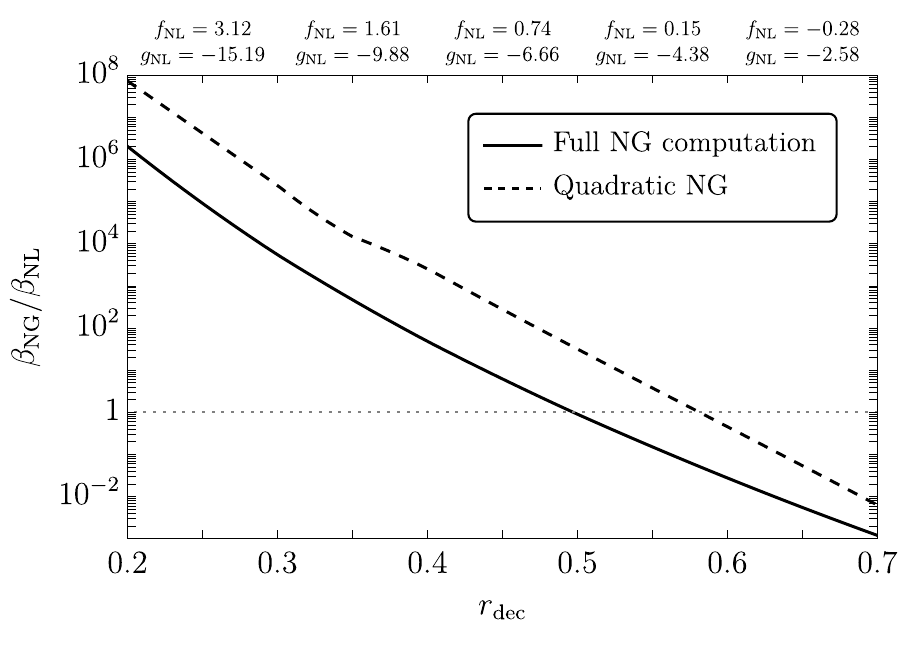}~
\includegraphics[width=.495\textwidth]{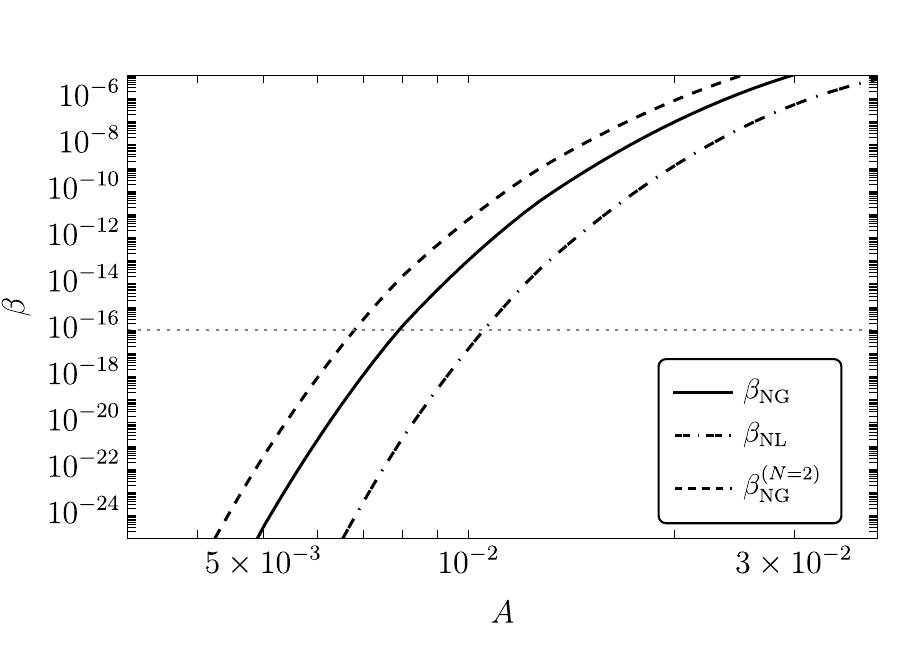}$$\vspace{-0.5cm}
\caption{\em 
	\textbf{\textit{	Left panel. }} 
Ratio $\beta_{\rm NL}/\beta_{\rm NG}$ as function of the parameter $r_{\rm dec}$.
We evaluate $\beta_{\rm NG}$ at the quadratic order in the primordial NGs 
(see eq.\,(\ref{eq:ZetaSeries}) with $N=2$) and considering the full resummed expression for $\zeta$ 
(see eq.\,(\ref{eq:MasterX})). 
We focus on the log-normal power spectrum in eq.\,(\ref{eq:LogNo}) with fixed values of $\sigma$ and $A$. 
	\textbf{\textit{	Right panel. }} We take $r_{\rm dec} = 0.3$ and compute $\beta$ as function of the amplitude 
$A$ of the log-normal power spectrum (while $\sigma$ is kept fixed at the same reference value $\sigma = 0.1$ used in the left-side plot). 
We compare $\beta_{\rm NL}$ with 
$\beta_{\rm NG}$, computing the latter both at the quadratic order and with fully resummed primordial NGs. 
 }\label{fig:LogNoTest}  
\end{center}
\end{figure}

We find that quadratic NGs with positive $f_{\rm NL}$ enhance
the value of $\beta$, as expected.  
However, already the inclusion of the cubic term reverse 
this result. This is because the cubic correction derived from eq.\,(\ref{eq:ZetaC}) is characterised by a large and negative 
coefficient $g_{\rm NL}$. 
 This trend nicely fits with what discussed in section\,\ref{sec:NGIntro}. 
 Given the analyzed values of $\sigma$ in fig.\,\ref{fig:LogNoCompa}, we are in the situation in which $\sigma_0 \ll 1$ and we indeed expected a good convergence of the NG expansion.
Therefore, the main conclusion we draw from this figure is the following: if we limit the analysis to narrow spectra (e.g. $\sigma = 0.10$, $\sigma = 0.20$) we find that the perturbative approach quickly converges to the exact result.

The net effect of primordial NGs displayed in fig.\,\ref{fig:LogNoCompa} depends on the specific value of $r_{\rm dec}$. 
To better illustrate this point, we show 
in the left panel of fig.\,\ref{fig:LogNoTest} 
the ratio between the results which only includes non-linearities
$\beta_{\rm NL}$ (and Gaussian primordial curvature perturbation)
and the NG result $\beta_{\rm NG}$ as function of $r_{\rm dec}$. 
 It is interesting to remark that, for $r_{\rm dec} \lesssim 0.6$, 
 the quadratic approximation for $\beta_{\rm NG}$ 
 systematically gives a result that is larger than $\beta_{\rm NL}$ (thus 
 $\beta_{\rm NG}/\beta_{\rm NL} > 1$). 
 This is because, as we can see from eq.\,\eqref{eq:ZetaQ}, the quadratic coefficient $f_{\rm NL}$ at $r_{\rm dec} = (\sqrt{10} - 2)/2 \approx 0.58$ changes from positive to negative values (for completeness, in the left panel of fig.\,\ref{fig:LogNoTest} we indicate on the top $x$-axis the values of the first two coefficients $f_{\rm NL}$ and $g_{\rm NL}$). From this perspective, the previous result is consistent with the expectation that local quadratic primordial NG with $f_{\rm NL} < 0$ ($f_{\rm NL} > 0$) tend to suppress (enhance) the PBH abundance compared to the case in which they are absent.
However, we find that including primordial NG at all-orders 
 modifies this conclusion: we find that
 $\beta_{\rm NG}/\beta_{\rm NL} > 1$ if 
 $r_{\rm dec} \lesssim 0.5$ 
 (instead of $r_{\rm dec} \lesssim 0.6$).

 In the right panel of fig.\,\ref{fig:LogNoTest} we set $r_{\rm dec} = 0.3$ and consider the computation 
 of $\beta$ as function of the  amplitude $A$. 
 As expected, we find that orders of magnitude differences in the computation of the mass fraction $\beta$ imply relatively small changes of $A$ because of the leading order scaling $\beta \sim e^{-1/P_{\zeta}}$. 
 For instance, we find that a relative change $\Delta A \simeq -25\%$ in the amplitude of the power spectrum (compared to the non-linear case) is needed in order to fit the reference value $\beta = 10^{-16}$ (we note that including only quadratic NG gives $\Delta A \simeq -40\%$). 

Let us summarize our findings in the case of the log-normal power spectrum.
\begin{itemize}
    \item[{\it i)}] Truncating the computation of the PBHs mass fraction $\beta_{\rm NG}$ at the quadratic order (that is, only including $f_{\rm NL}$) does not give reliable results, cf. fig.\,\ref{fig:LogNoCompa} and the left panel of fig.\,\ref{fig:LogNoTest}.
    \item[{\it ii)}] We find that the power-series expansion $\beta_{\rm NG}^{(N)}$ in eq.\,(\ref{eq:ZetaSeries}) does converge on the full result for appropriate $N$. 
    The optimal-truncation order $N$ depends on the width $\sigma$ of the log-normal, cf. fig.\,\ref{fig:LogNoCompa}. 
    For instance, if $\sigma = 0.1$ we find that $N=4$ gives the correct result while if 
    $\sigma = 0.5$ one should push the perturbative expansion up to $N\gtrsim 10$.
    \item[{\it iii)}] When translated in terms of 
    the amplitude $A$ of the power spectrum, 
    the various orders of magnitude changes in the PBH mass fraction are reabsorbed by re-tuning of $A$, which is typically less than a factor of 2 in the range of  $r_{\rm dec}$ that we explored.
\end{itemize}

\subsection{The broad power spectrum}

We now consider the broad power spectrum of curvature perturbations in eq.\,(\ref{eq:Bro}). 
This kind of power spectrum, widely used as a benchmark model in theoretical studies, 
recently attracted some 
attention also on the phenomenological side. As shown in ref.\,\cite{DeLuca:2020agl}, 
PBHs obtained assuming a broad power spectrum of the type in eq.\,(\ref{eq:Bro})  may comprise 
the totality of dark matter observed in the present-day Universe and, at the same time, generate, as a second-order effect, a detectable signal 
of gravitational waves that matches, both in frequency and amplitude, the tentative signal recently reported by the NANOGrav Collaboration. 
We also refer to ref.\,\cite{Franciolini:2022pav} for an explicit implementation of the idea in the context of USR inflationary models.
 The computation in refs.\,\cite{DeLuca:2020agl,Franciolini:2022pav} 
does assume the absence of NGs of primordial origin (while it fully does include NGs from non-linearities). 
However, primordial NGs may be expected in concrete models that feature a broad power spectrum.

\begin{figure}[t]
\begin{center}\vspace{-2cm}
\includegraphics[width=.98\textwidth]{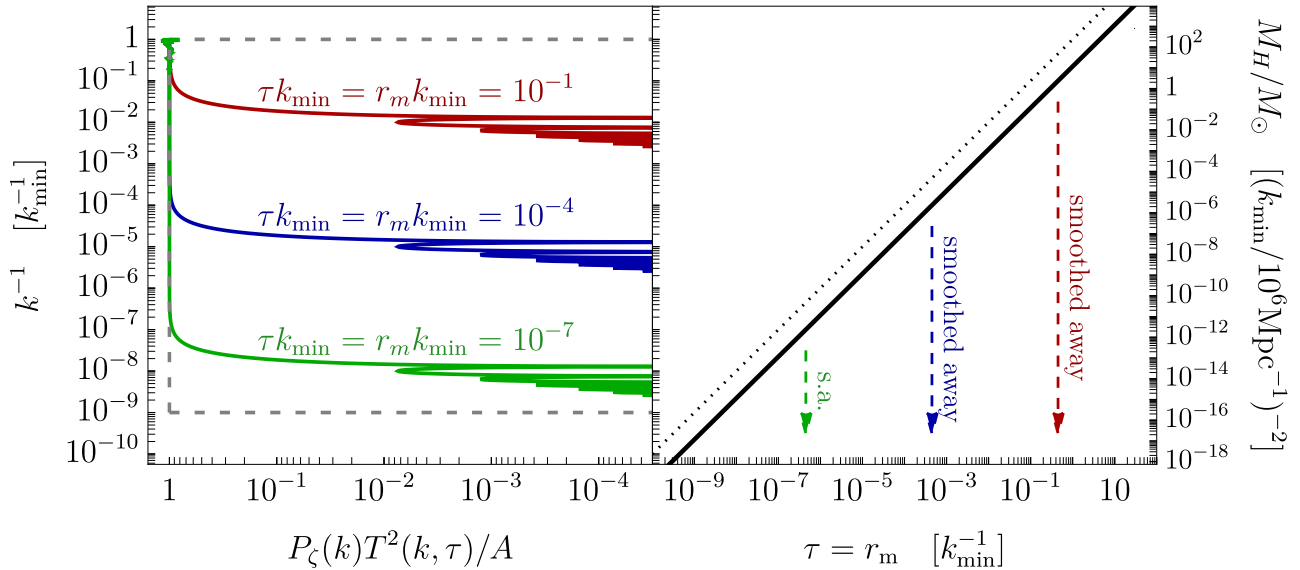}
\caption{\em 
We plot (left-side $y$ axis, right panel) the relation $k^{-1} = r_m/\kappa$ in eq.\,(\ref{eq:Broadrk}) as function of $r_m$. Both length scales $k^{-1}$ and $r_m$ are written in units of $k_{\rm min}^{-1}$;
from eq.\,(\ref{eq:HorizonMass}), we read the corresponding horizon mass $M_H$ (rigth-side $y$ axis). 
We set $\tau = r_m$ and plot (left panel, rotated by 90$^{\circ}$ so to share the same $y$ axis with the right panel) the broad power spectrum in eq.\,(\ref{eq:Bro}) normalized to $1$, convoluted with  the 
the transfer function, and shown as function of
$k^{-1}$. 
The gray dashed line represents  the broad power spectrum without the transfer function.
We take $\Delta k = 10^9$, and consider three different choices of the comoving time-scale $\tau$.
As a whole, the figure shows that as time passes by one moves along the black line from left to right in the right panel and modes with 
decreasing $k$ (i.e. larger wavelengths and horizon mass $M_H$) enter the expanding cosmological horizon and are smoothed away by the transfer function. 
Consequently, at late times associated to the formation of heavier PBHs, the broad power spectrum effectively behaves as a peaked one (cf., for instance, the red curve in the left panel with $\tau = 10^{-1}k_{\rm min}^{-1}$) since the majority of modes, deep in the sub-horizon regime, have been smoothed away. 
 }\label{fig:SmoothedPS}  
\end{center}
\end{figure}

\begin{figure}[t]
\begin{center}
\boxed{{\rm Non~linearities
~+~primordial~NG:~~\textbf{{\color{venetianred}Threshold~statistics~on~the~compaction~function}}~(broad}~P_{\zeta})}\vspace{0.25cm}
\includegraphics[width=1\textwidth]{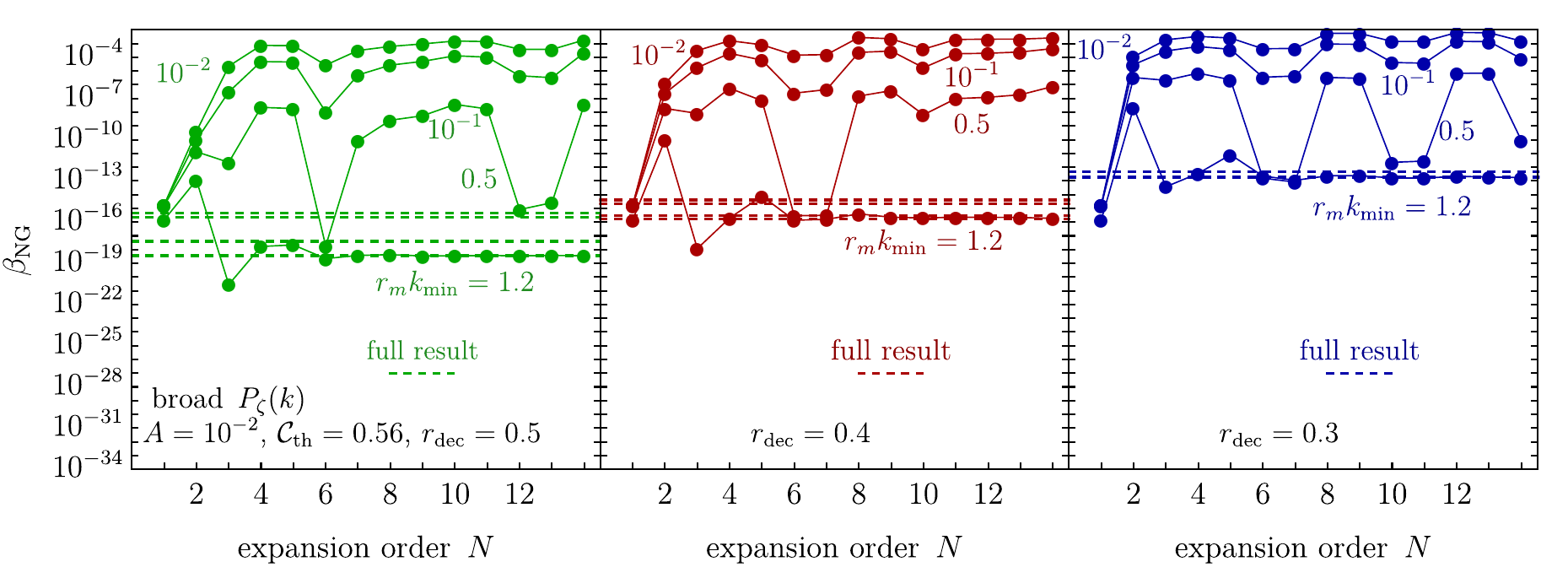}
\caption{\em 
Left ($r_{\rm dec}=0.5$), central ($r_{\rm dec}=0.4$) and right ($r_{\rm dec}=0.3$) panels. 
We plot the NG mass fraction $\beta_{\rm NG}$ computed according to eq.\,(\ref{eq:CompactionIntegral}) and its  expansion $\beta_{\rm NG}^{(N)}$ in as function of the expansion order $N$. 
We consider different choices for $r_m k_{\rm min}$, with $\tau = R \equiv r_m$ in the linear transfer function and window function.
 }\label{fig:BroadCompa}  
\end{center}
\end{figure}

This is the case, for instance, of the curvaton model studied in ref.\,\cite{Inomata:2020xad,Ferranteinprep} in which the curvature perturbation 
has the functional form in eq.\,(\ref{eq:MasterX}).
These considerations motivate the analysis proposed in this paper.  
Motivated by the toy model for the power spectrum proposed in ref.\,\cite{DeLuca:2020agl}, we allow for a hierarchy 
 $k_{\rm max}\gg k_{\rm min}$ up to values $k_{\rm max}/k_{\rm min} = O(10^{9})$.
Let us illustrate our results, summarized in fig.\,\ref{fig:SmoothedPS} and in the three panels of fig.\,\ref{fig:BroadCompa}.

In the right panel of fig.\,\ref{fig:SmoothedPS}, 
we show the relation $k^{-1} = r_m/\kappa$ in eq.\,(\ref{eq:Broadrk}) as function of $r_m$ (see caption). On the right-side $y$ axis, we show the horizon mass $M_H$ that corresponds to a given value of $k^{-1}$, cf. eq.\,(\ref{eq:HorizonMass}). We assume $g_{\star} = 106.75$ (radiation-dominated universe).
We also show (left side of the plot, rotated by 90$^\circ$ to match the same $y$ axis with the right panel) the convolution $P_{\zeta}(k)T^2(k,\tau)$ for the broad power spectrum $P_{\zeta}(k)$ in eq.\,(\ref{eq:Bro}) as function of 
    $k_{\rm min}/k$ (and normalized to unit in amplitude). We plot the convoluted power spectrum for different values of 
    $\tau k_{\rm min}$. 
    This plot shows the effect of the linear transfer function on the broad power spectrum: sub-horizon modes with $\tau k \gg 1$ are suppressed. 
    Intuitively, what happens is that as time passes by (that is, for increasing $\tau$, going from the bottom green line with $\tau k_{\rm min} = 10^{-7}$ to the top red one with $\tau k_{\rm min} = 10^{-1}$) more and more modes with decreasing $k$ (thus larger comoving wavelengths) re-enter the horizon 
    and are subject to pressure gradients which effectively deplete their corresponding power. 
    A very similar effect is provided by the window function $P_{\zeta}(k)W^2(k,R)$: 
    modes with $kR\gg 1$ are smoothed away by the volume average. 
    
The above considerations are very interesting since they imply that, if we set $\tau = R \equiv r_m$ and consider the case $r_m k_{\rm min} = O(1)$, almost all modes of the power spectrum are suppressed, and 
    the latter effectively looks like a very peaked one. 
    This point, in light of our discussion in section\,\ref{sec:NGIntro}, is crucial since we argued that the broadness of the power spectrum controls the convergence of the power-series expansion. 
We are now in the position to test the intuition developed in section\,\ref{sec:NGIntro}, at that time only based on considerations regarding the NG coefficients $c_n(r_{\rm dec})$ and their role with respect to the curvature perturbation field.

In the three panels of fig.\,\ref{fig:BroadCompa} we compare the NG PBHs mass fraction $\beta_{\rm NG}$ in eq.\,(\ref{eq:CompactionIntegral})
with its expansion $\beta_{\rm NG}^{(N)}$ as function of $N$. 
We consider different choices of $r_m k_{\rm min}$. 
    If we take $r_mk_{\rm min} = O(1)$, we find that $\beta_{\rm NG}^{(N)}$ converges to the full result given by 
    $\beta_{\rm NG}$ for $N$ large enough. 
    This is expected because, as explained before, for $r_m k_{\rm min} = O(1)$ the broad power spectrum is effectively peaked since the majority of its modes have $r_m k \gg 1$ and are smoothed away. 
    However, if we consider smaller and smaller values of $r_m k_{\rm min}$ then an increasing number of modes contributes to the variances, and the convergence of $\beta_{\rm NG}^{(N)}$ gets quickly lost. 
    As explained in section\,\ref{sec:NGIntro}, in this situation the computation of $\beta_{\rm NG}^{(N)}$ based on the power-series expansion is simply wrong (the series does not converge for whatever $N$) and one is forced to use the full result $\beta_{\rm NG}$.

We arrive at a very important conclusion. 
We are eventually interested in the computation of the PBHs mass fraction at different scales (that  is for different values of the PBH mass). 
In other words, if we combine eq.\,(\ref{eq:Broadrk}) and eq.\,(\ref{eq:HorizonMass}) we find
\begin{align}
    r_m k_{\rm min} \approx \left(\frac{M_H}{M_{\odot}}\right)^{1/2}
    \left(\frac{k_{\rm min}}{10^6\,{\rm Mpc}^{-1}}\right)\,,
\end{align}
and the computation of the abundance entails scanning over all the relevant horizon masses $M_H$. 
Only for values of $M_H$ such that above relation gives $r_m k_{\rm min} = O(1)$ we expect the perturbative computation based on $\beta_{\rm NG}^{(N)}$ to give reliable results; 
however, as soon as one takes smaller and smaller values for the horizon mass $M_H$ such that 
$r_m k_{\rm min} \ll 1$, we expect that the computation of $\beta_{\rm NG}^{(N)}$ based on the power-series expansion $\zeta_N$ is not trustable, for whatever expansion order $N$.  
The reason is precisely the one discussed in section\,\ref{sec:TheBroad}:
if the power spectrum is very broad (that is the case with $r_m k_{\rm min} \ll 1$), we go outside the radius of convergence of the series and beyond the range of validity of the power-series expansion $\zeta_N$.
For this reason, we now abandon the computation based on $\beta_{\rm NG}^{(N)}$ and focus on the full result described by $\beta_{\rm NG}$.

\begin{figure}[t]
\begin{center}
\boxed{{\rm NG~from~non~linearities
~+~primordial~NG:~~\textbf{{\color{venetianred}Threshold~statistics~on~the~compaction~function}}~(broad}~P_{\zeta})}\vspace{-0.25cm}
\includegraphics[width=0.495\textwidth]{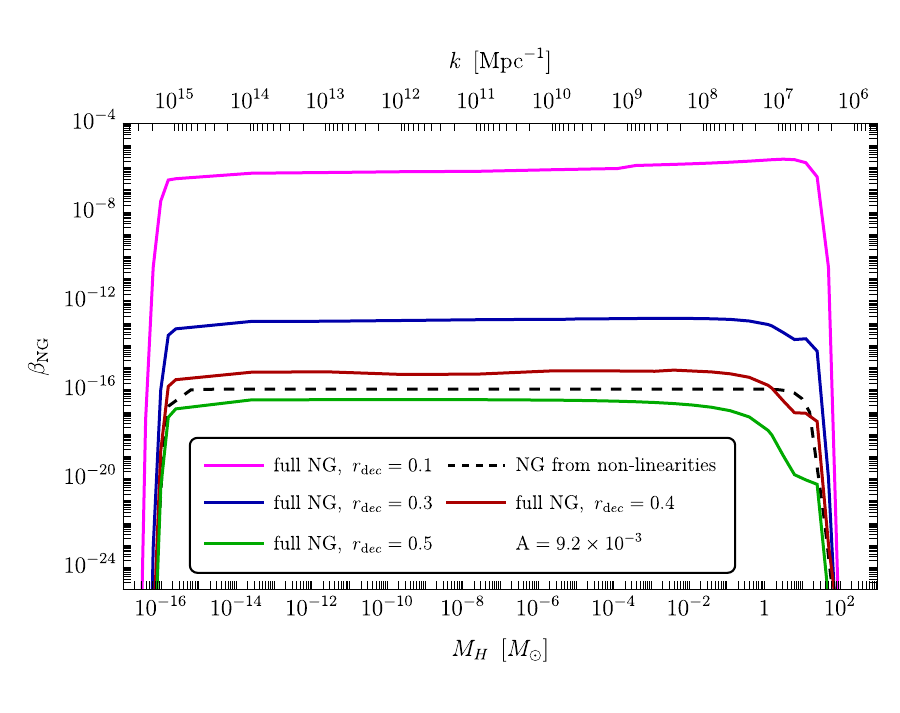}
\includegraphics[width=0.495\textwidth]{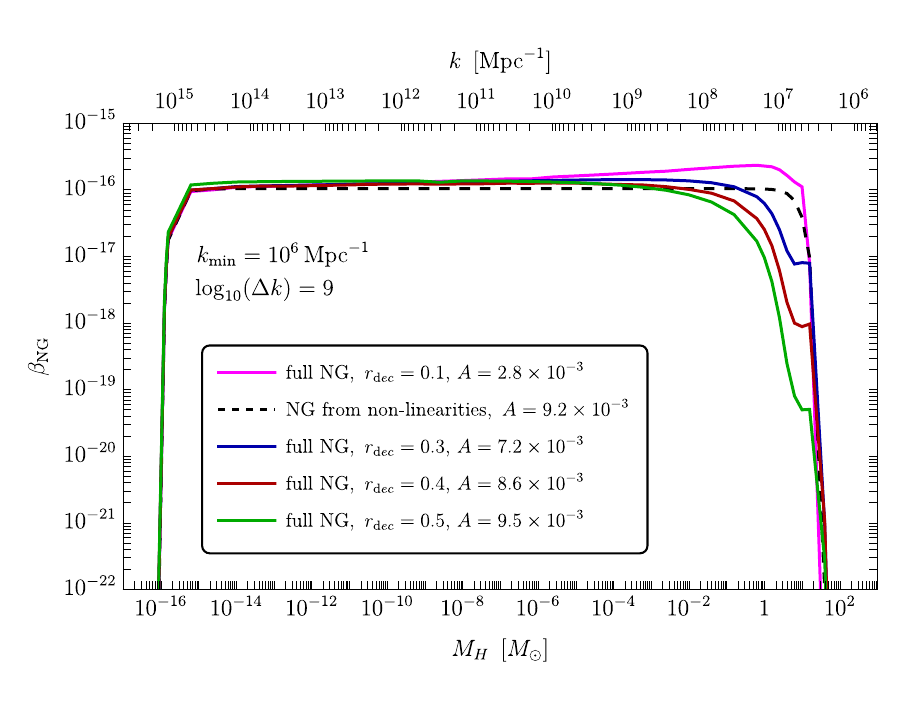}
\caption{\em  
Computation of $\beta_{\rm NG}$ as a function of the horizon mass $M_H$ for different values of $r_{\rm dec}$. 
We fix $k_{\rm min} =10^6$ ${\rm Mpc}^{-1}$ and $\Delta k = 10^{9}$. 
The dashed black line corresponds to eq.\,(\ref{eq:CompactionIntegral}) in which we only include the effect non-linearities (i.e. assuming the absence of primordial NGs).
\textbf{\textit{Left panel.}}
The amplitude $A = 9.2 \times 10^{-3}$ is fixed in order to get $\beta_{\rm NG} \simeq 10^{-16}$ at $M_H = 10^{-15}\,\,M_{\odot}$ in the presence of only non-linearities.
Keeping the amplitude fixed, we show how several values of $r_{\rm dec}$ influence $\beta_{\rm NG}$. 
\textbf{\textit{Right panel.}}
For each $r_{\rm dec}$ we tune the amplitude $A$ in such a way that
$\beta_{\rm NG} \simeq 10^{-16}$ at $M_H \simeq 10^{-15}\,\,M_{\odot}$.
 }\label{fig:BroCompa2}  
\end{center}
\end{figure}

After having gained intuition on the convergence as a function of the smoothing scale when broad spectra are considered, let us analyse the impact of NGs from the curvaton model on the computation of $\beta_{\rm NG}$ as a function of the horizon mass $M_H$ in eq.\,(\ref{eq:HorizonMass}). 
This is shown in fig.\,\ref{fig:BroCompa2}. 
We explore the benchmark values $r_{\rm dec}=0.1,0.3,0.4,0.5$. 
As far as the power spectrum is concerned, 
we fix $k_{\rm min} = 10^6$ ${\rm Mpc^{-1}}$ and $\Delta k = 10^{9}$ while the amplitude $A$ is tuned, for each $r_{\rm dec}$
in such a way that we get $\beta_{\rm NG} \simeq 10^{-16}$ at $M_H = 10^{-15}\,\,M_{\odot}$. 
This choice is motivated by the fact that the above value of $\beta_{\rm NG}$ is what is typically needed in order to get a sizable abundance of asteroid-mass PBHs. On the top $x$-axis, we convert $M_H$ into $k$ according to eq.\,(\ref{eq:HorizonMass}) to highlight the relevant scales in $[{\rm Mpc^{-1}}]$. 

We emphasize two important points:
\begin{itemize}
\item[$\circ$] 
First of all, 
we find that increasing the level of NG (that is decreasing the value of $r_{\rm dec}$ from $0.5$ to $0.1$) has the net effect of enhancing the overall mass fraction $\beta_{\rm NG}$. 
More explicitly, in fig.\,\ref{fig:BroCompa2}, in order to fit the same reference value $\beta_{\rm NG} \simeq 10^{-16}$, the amplitude $A$ decreases from 
$A\simeq 9.5\times 10^{-3}$ (for $r_{\rm dec} = 0.5$) to $A\simeq 2.8\times 10^{-3}$ (for $r_{\rm dec} = 0.1$). This trend is analogous to the one observed with a positive $f_{\rm{NL}}$ in a large PBH abundance.
\item[$\circ$] 
The second effect is more subtle, and regards the behavior of $\beta_{\rm NG}$ near the cut-off at large horizon mass. Looking at fig.\,\ref{fig:BroCompa2}, it is evident that the right-side of the distribution is altered by NG if compared to the part of $\beta_{\rm NG}$ at smaller $M_H$. 
In other words, the primordial NG breaks the $M_H$-independence of $\beta_{\rm NG}$ that would be expected on the basis of the scale invariance of eq.\,(\ref{eq:Bro}) (and that is verified, for instance, by the computation that only includes NG from non-linearities, cf. the black dashed line in fig.\,\ref{fig:BroCompa2} that remains approximately constant as function of $M_H$).
This is an interesting effect that is worth investigating.

\end{itemize}

\subsubsection{On the breaking of scale invariance}

\begin{figure}[t]
\begin{center}
\includegraphics[width=0.495\textwidth]{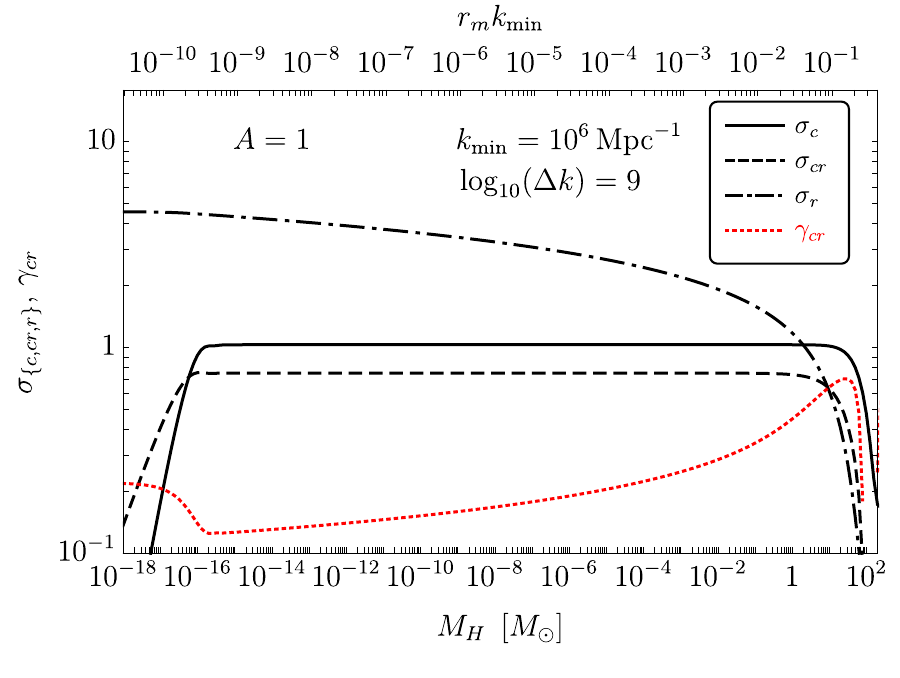}
\includegraphics[width=0.495\textwidth]{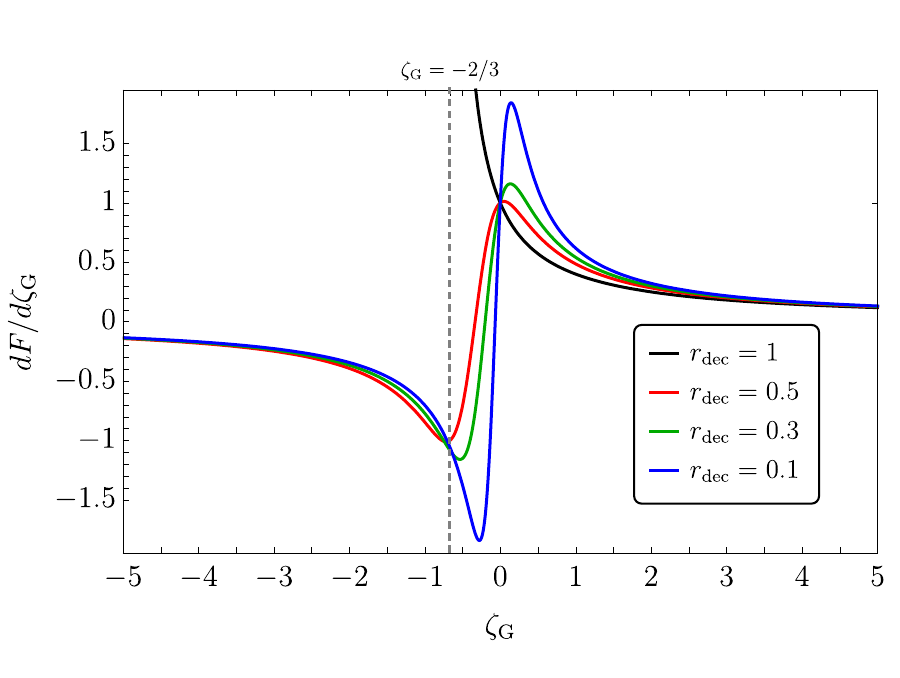}
\caption{\em 
\textbf{\textit{Left panel.}} 
Variances $\sigma_{c,cr,r}$ in eqs.\,(\ref{eq:Var1}-\ref{eq:Var3}) as function of the horizon mass $M_H$. We take the parameters 
$k_{\rm min}= 10^{6}$ ${\rm Mpc^{-1}}$  and $\Delta k=10^{9}$, and normalize to 1 the amplitude of the power spectrum (we remind that the variances $\sigma_{c,cr,r}$ simply scale as $\sqrt{A}$). We also plot the correlation parameter $\gamma_{cr}$ (which, on the contrary, does not depend on $A$).
\textbf{\textit{Right panel.}} 
Functional dependence of $dF/d\zeta_{\rm G}$ on the Gaussian variable $\zeta_{\rm G}$ for different values of $r_{\rm dec}$. We consider the explicit case of the curvaton field.
 }\label{fig:discussion_broadinv_intro}  
\end{center}
\end{figure}

\begin{figure}[t]
\begin{center}
\includegraphics[width=1\textwidth]{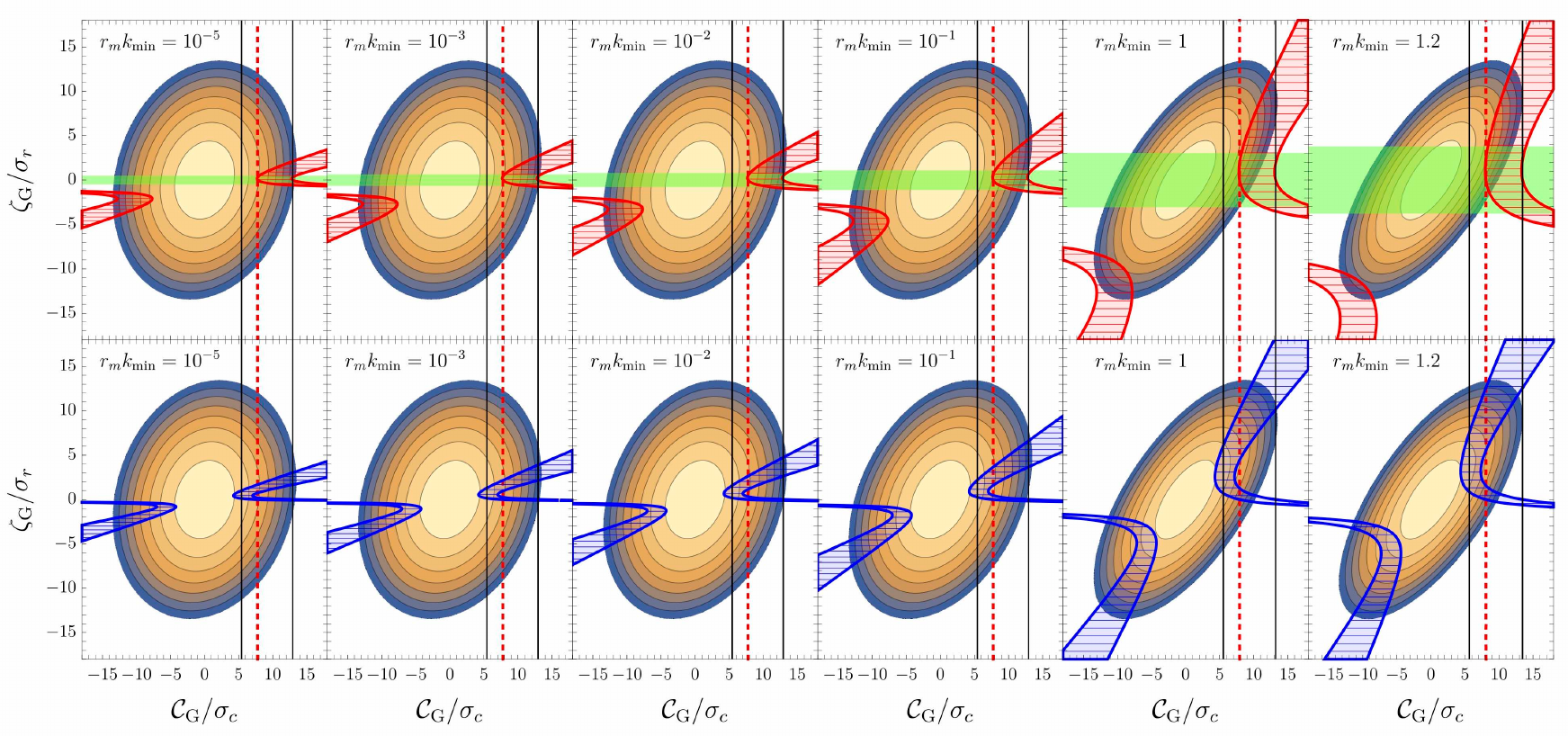}
\caption{\em 
\textbf{\textit{Top row.}} 
Contour lines indicate values of the probability density $P_\text{G} ({\cal C}_\text{G}, \zeta_\text{G})$. From the inner circle to the last purple one we locate values equally spaced in log scale in between $10^{-5}$ and $10^{-40}$.
The region between the red solid lines is the parameter space defined as $\mathcal{D}$ in eq.\,(\ref{eq:RegionD}). This is the region over which the PDF $P_\text{G} ({\cal C}_\text{G}, \zeta_\text{G})$ is integrated in the computation of the PBH mass fraction as it corresponds to perturbations above the threshold. 
Inside the horizontal region marked in green the power-series expansion in eq.\,(\ref{eq:ZetaSeries}) does converge to the full resummed result in eq.\,(\ref{eq:MasterX}).
The region limited by the two vertical black lines  is defined by the condition $\mathcal{C}_{\rm th} < \mathcal{C}_{\rm G} < 4/3$ (pure Gaussian case) while the region on the right side of the vertical dashed red line is limited by the condition $\mathcal{C}_{\rm th} < \mathcal{C}_{\rm G} - 
(3/8)\mathcal{C}_{\rm G}^2< 2/3$ (thus including NG from non-linearities). Notice that the right-edge of the above boundary falls outside the range of the $x$-axis. 
We consider the benchmark value $r_{\rm dec} = 0.5$ while $k_{\rm min} = 10^{6}$ $ {\rm Mpc}^{-1}$ and $\Delta k = 10^9$.  
The amplitude of the curvature power spectrum is fixed to the value $A= 10^{-2}$.
\textbf{\textit{Bottom row.}} 
We consider $r_{\rm dec} = 0.1$.
}\label{fig:discussion_broadinv}  
\end{center}
\end{figure}

In the computation of the PBH mass fraction, the $M_H$-dependence enters implicitly in eq.\,(\ref{eq:CompactionIntegral}) via the variances $\sigma_{c,cr,r}^2$ in eqs.\,(\ref{eq:Var1}-\ref{eq:Var3}). 
In the left panel of fig.\,\ref{fig:discussion_broadinv_intro}, we show the three variances $\sigma_{c,cr,r}^2$, together with their combination $\gamma_{cr}$ in 
eq.\,(\ref{eq:GammacrDef}), as function of the horizon mass $M_H$.  
The figure shows that 
while $\sigma_{c}$ and $\sigma_{cr}$ remain constant (apart from threshold effects) for a scale invariant spectrum, the value of $\sigma_{r}$ changes drastically within the first decade in $r_m k_{\textrm{min}}$ (looking at the plot starting from the right-side at $r_m k_{\textrm{min}} = O(1)$ values, see labels on the top $x$-axis) while only logarithmically in the remaining decades.
The $M_H$-dependence of $\sigma_{c}$, therefore, reflects the 
$M_H$-dependence observed in the right panel of fig.\,\ref{fig:BroCompa2} with the value of $\beta_{\rm NG}$ as function of $M_H$ becoming nearly $M_H$-invariant going towards smaller horizon mass.

We remind that  $\sigma_{r}$ is the variance of the Gaussian variable $\zeta_{\rm G}$, cf. eq.\,(\ref{eq:Var3}). 
In light of the above result, it is, therefore, crucial to better understand how $\sigma_r$ enters in the computation of the PBHs mass fraction.
Consider again  eq.\,(\ref{eq:CompactionIntegral}). The change of variable 
$\zeta_{\rm G}\to \nu_{\rm G}\equiv \zeta_{\rm G}/\sigma_{r}$ seems to completely reabsorb the explicit dependence on $\sigma_{r}$. However, and crucially, 
in the presence of primordial NGs 
$\sigma_{r}$ appears back through the term $dF/d\zeta_{\rm G}$ in the non-linear relation in eq.\,(\ref{eq:CCgau}). Furthermore, 
 the coefficient $\gamma_{cr}$, which measures the correlation between $\zeta_{\rm G}$ and $\mathcal{C}_{\rm G}$, will also retain some $M_H$ dependence. 
 These simple observations suggest that a full understanding of the 
 effect of NGs requires to consider both the functional dependence of 
 $dF/d\zeta_{\rm G}$ on the Gaussian variable $\zeta_{\rm G}$ 
 and the two-dimensional probability distribution $\textrm{P}_{\rm G}(\mathcal{C}_{\rm G},\zeta_{\rm G})$.

First, consider $dF/d\zeta_{\rm G}$; we focus on the case of the curvaton, 
eq.\,(\ref{eq:MasterX}). 
In the right panel of fig.\,\ref{fig:discussion_broadinv_intro}, we show the behavior of $dF/d\zeta_{\rm G}$ as function of $\zeta_{\rm G}$ for different values of $r_{\rm dec}$. 
For $r_{\rm dec}<1$,
$dF/d\zeta_{\rm G}$ transits from small negative to small positive values through an heartbeat transition located at around $\zeta_{\rm G} \simeq 0$. 
The case $r_{\rm dec} = 1$ is somewhat special; in this case, 
$dF/d\zeta_{\rm G}$ is positive and 
grows for decreasing $\zeta_{\rm G}$ with the latter that is subject to the condition  $\zeta_{\rm G} > - 2/3$. 
This restriction is a consequence of 
the presence of a branch-point singularity that, for $r_{\rm dec} = 1$, is located on the real axis (while for 
$r_{\rm dec}<1$ it acquires an imaginary part and moves away from the real axis, cf. appendix\,\ref{app:Radius}).

In fig.\,\ref{fig:discussion_broadinv} we show iso-contours of the two-dimensional probability distribution $\textrm{P}_{\rm G}(\mathcal{C}_{\rm G},\zeta_{\rm G})$. 
From left to right, different panels correspond to increasing  values of $r_m k_{\rm min}$. This is to capture the effect of the $M_H$-dependence on $\gamma_{cr}$. 
For $r_m k_{\rm min} = 1$ (right-most panel, large horizon mass) the power spectrum is very peaked and the correlation among 
$\mathcal{C}_{\rm G}$ and $\zeta_{\rm G}$ maximal ($\gamma_{cr} \to 1$, cf. the left panel of fig.\,\ref{fig:discussion_broadinv_intro}). For smaller $r_m k_{\rm min}$ (broader power spectrum and small horizon mass), the level of correlation decreases and the ellipses in fig.\,\ref{fig:discussion_broadinv} rotate towards a diagonal configuration ($\gamma_{cr} \to 0$).
The top (bottom) row refers to $r_{\rm dec} = 0.5$ ($r_{\rm dec} = 0.1$).
The regions shaded in red (top row) and blue (bottom row) in  fig.\,\ref{fig:discussion_broadinv} are selected by the condition 
$ \mathcal{D} = 
\left\{
\mathcal{C}_{\rm G},\,\zeta_{\rm G} \in \mathbb{R}~:~~
\mathcal{C}(\mathcal{C}_{\rm G},\zeta_{\rm G}) > \mathcal{C}_{\rm th}  
~\land~\mathcal{C}_1(\mathcal{C}_{\rm G},\zeta_{\rm G}) < 2\Phi
\right\}$ (cf. eq.\,(\ref{eq:RegionD})) in the presence of both NGs and non-linearities. 
For comparison we also show:
the region defined by the condition
$\mathcal{C}_{\rm th} < \mathcal{C}_{\rm G} < 4/3$ and limited by the two vertical black lines (pure Gaussian case) and 
the region defined by the condition $\mathcal{C}_{\rm th} < \mathcal{C}_{\rm G} - 
(3/8)\mathcal{C}_{\rm G}^2< 2/3$ 
that lies on the right-side of the dashed red line is (including NG from non-linearities; notice that the right-side of this constraint lies outside the range of the $x$-axis).

We are in the position to combine the two effects. For simplicity, we start from the benchmark case with  $r_{\rm dec} = 0.5$. 
Many important lessons can be drawn.
\begin{itemize}
    \item[{\it i)}] First of all we notice that, 
    going from the pure Gaussian case to the case in which only NGs from non-linearities 
    are included, the region over which the PDF 
    is integrated shifts towards larger values of  $\mathcal{C}_{\rm G}$ (where the PDF is smaller). Consequently,  
    including non-linearities decreases the PBH abundance in agreement with what found in ref.\,\cite{Young:2019yug,DeLuca:2019qsy}.
    \item[{\it ii)}] Second, we recover, from a different perspective, what we have already learned about the validity of the perturbative approach. 
  As discussed before, the perturbative approach is applicable in a very small range of values for $\zeta_{\rm G}$ centered around $\zeta_{\rm G} = 0$ and  within the radius of convergence of the power series expansion ($-0.173< \zeta_{\rm G} < +0.173$ for $r_{\rm dec} = 0.5$, cf. appendix\,\ref{app:Radius}). 
  If we consider 
  $r_m k_{\rm min} = O(1)$ (right-most panel in fig.\,\ref{fig:discussion_broadinv}) we see that part of the region that is selected by the condition $(\mathcal{C}_{\rm G},\,\zeta_{\rm G})\in \mathcal{D}$ (and over which we integrate the PDF in eq.\,(\ref{eq:CompactionIntegral})) lies away from the convergence of the power series expansion but mostly  overlaps with negligible probability; on the contrary, the small part of the overlapping region in which the 
  probability is sizable (that is, 
  within the ellipses colored in yellow) lies within the convergence condition. 
  Consequently, as expected, 
  we find that for $r_m k_{\rm min} = O(1)$ the region inside the radius of convergence of the power-series expansion captures the relevant contribution to the PBH formation probability.
 
  On the contrary, if we move in fig.\,\ref{fig:discussion_broadinv} towards smaller 
  $r_m k_{\rm min}$ we see that already for $r_m k_{\rm min} = 10^{-1}$ 
  the region selected by the condition $\mathcal{C}(\mathcal{C}_{\rm G},\zeta_{\rm G}) > \mathcal{C}_{\rm th}$ 
  has large probability 
  $\textrm{P}_{\rm G}(\mathcal{C}_{\rm G},\zeta_{\rm G})$ outside the radius of convergence thus invalidating, as expected, the applicability of the perturbative approach.
  \item[{\it iii)}] 
  In the presence of primordial NGs, we expect a smaller abundance of PBHs compared to the case in which only non-linearities are included. 
  This is indeed confirmed by the computation in the left panel of fig.\,\ref{fig:BroCompa2} (the line corresponding to the mass fraction with $r_{\rm dec} = 0.5$ lies below the dashed black line).   
  This is due to the effect of NGs in the non-linear relation,  eq.\,(\ref{eq:CCgau}). 
  There are two key points.
  
  The first one is that
  the constraint $\mathcal{D}$ in 
  eq.\,(\ref{eq:RegionD}) admits the solution (with $\Phi =2/3$)
\begin{align}
\frac{4(1-\sqrt{1-3\mathcal{C}_{\rm th}/2})}{3} 
< \mathcal{C}_{\rm G}\frac{dF}{d\zeta_{\rm G}} <
\frac{4}{3}\,,\label{eq:NGConstraint}
\end{align} 
that is a trivial generalization of the one we get in the case in which only non-linearities are included, 
$4(1-\sqrt{1-3\mathcal{C}_{\rm th}}/2)/3 
< \mathcal{C}_{\rm G} <
4/3$.  
However, eq.\,(\ref{eq:NGConstraint}) now has two branches depending on the sign of $dF/d\zeta_{\rm G}$
 since, with the only exception of the special case with $r_{\rm dec} = 1$, $dF/d\zeta_{\rm G}$ takes both positive and negative values (cf. fig.\,\ref{fig:discussion_broadinv_intro}, right panel). 
 As shown in fig.\,\ref{fig:discussion_broadinv} (the two regions in red and blue) both these possibilities are realized with negative 
 $dF/d\zeta_{\rm G}$ (thus negative $\zeta_{\rm G}$, cf. fig.\,\ref{fig:discussion_broadinv_intro}, right panel) that requires negative $\mathcal{C}_{\rm G}$ and viceversa.
 
 The second key point is the following.
 Looking at the right panel of fig.\,\ref{fig:discussion_broadinv_intro}, we notice that 
 $|dF/d\zeta_{\rm G}| \leqslant 1$ for $r_{\rm dec} = 0.5$. 
 More in detail, we find that 
 $dF/d\zeta_{\rm G} \simeq 1$ for $\zeta_{\rm G} \simeq 0.05$, 
 $dF/d\zeta_{\rm G} \simeq -1$ for $\zeta_{\rm G} \simeq -0.7$
 with 
 $|dF/d\zeta_{\rm G}|$ that becomes smaller as $|\zeta_{\rm G}|$ increases. 
 Consequently, as we move towards larger $|\zeta_{\rm G}|$, the variable $\mathcal{C}_{\rm G}$ is forced -- in order to compensate in eq.\,(\ref{eq:NGConstraint})  the change of $dF/d\zeta_{\rm G}$  -- to take larger values compared with the non-linear case. As a direct consequence, the 
 part of the parameter space that is selected by eq.\,(\ref{eq:NGConstraint}) falls in a region with smaller PDF thus smaller PBH mass fraction.
 
 \item[{\it iv)}] The discussion at point {\it iii)} immediately explains also why we get a larger PBH mass fraction for some other values of $r_{\rm dec}$. 
 Consider, for instance, the case $r_{\rm dec} = 0.1$. 
 From the left panel of fig.\,\ref{fig:BroCompa2} we know that in this case the mass fraction of PBHs is much larger compared to the pure non-linear case. 
 This is because, as evident from the right panel of fig.\,\ref{fig:discussion_broadinv_intro}, in this case it is possible to have $|dF/d\zeta_{\rm G}| > 1$ in a small range of values of $\zeta_{\rm G}$ corresponding to the heartbeat transition. Consequently, when $|dF/d\zeta_{\rm G}| > 1$ the variable $\mathcal{C}_{\rm G}$ takes smaller values thus falling in a region in which the PDF is larger. This is evident by inspecting in fig.\,\ref{fig:discussion_broadinv} the interplay between the ellipses of constant PDF with the region selected by the integration constraint $\mathcal{D}$.
\end{itemize}

To sum up, there are two important take-home messages we learn from the above discussion. 
{\it a)} In the presence of primordial NG, $\beta_{\rm NG}$ depends on $M_H$ (that is, the
formation of PBHs of various masses does not happen with equal probability as na\"{\i}vely expected on the basis of a scale-invariant power spectrum). 
This is a model-independent statement.
{\it b)} The size and impact of the effect, on the contrary, is model-dependent, and controlled by $dF/d\zeta_{\rm G}$. We explored the physics-case of the curvaton field but we stress that our analysis can be easily applied, step-by-step, to other models schematically indicated in eq.\,\eqref{eq:Schema}.

\subsubsection{Summary}

 It should be noted that we assumed a radiation-dominated universe ($\omega = 1/3$, $g_* = 106.75$) in all computations carried out in this section. This comment is particularly relevant as far as the results presented in fig.\,\ref{fig:BroCompa2} are concerned. The reason is that in this figure we considered horizon mass as large as $M_H \sim O(1)$ $M_{\odot}$ values where the effects of the quark-hadron QCD phase transition on the equation of state of the universe become relevant. However, in this section we decided to focus only on the impact of NG on the computation of the PBH mass fraction in order to better highlight their role. In the phenomenological discussion of section\,\ref{sec:PBHGW}, we will include the effects of the quark-hadron QCD phase transition.

 Let us summarize our findings in the case of the broad power spectrum.
\begin{itemize}

\item[{\it i)}] The perturbative approach based on eq.\,(\ref{eq:ZetaSeries}) is not applicable (at any order $N$). 
    We claim that only the non-perturbative computation based on eq.\,(\ref{eq:CompactionIntegral}) captures the correct result.

\item[{\it ii)}] Primordial NG leave a peculiar imprint 
on both the overall abundance (for fixed spectral amplitude) as well as on the PBHs mass fraction as a function of the horizon mass $M_H$, cf. fig.\,\ref{fig:BroCompa2}. 
Because of the scale invariance of the broad power spectrum in eq.\,(\ref{eq:Bro}), one would na\"{\i}vely expect a constant (that is, $M_H$-independent) mass fraction, as in the case in which primordial NG are not included. 
    However, we find that the inclusion of primordial NG changes this expectation by introducing a distinctive  $M_H$-dependence. 
This can help distinguishing a NG signature even if 
only the PBH mass function is considered, 
as it is not fully degenerate with a rescaling of the amplitude $A$ of perturbations. However, a residual degeneracy with the power spectrum tilt remains, that can only be broken with the observation of both PBH mass distribution and GWs. 
We will discuss these consequences in more depth in the following section. 
\end{itemize}

It is worth noting that there exists an interesting exception to the result discussed at point {\it i)}. This is the case of the curvaton model with $r_{\rm dec} = 1$. 
In this case, we find that the region of integration $\mathcal{D}$ always falls within the radius of convergence of the power series expansion (even in the case of very small $r_m k_{\rm min}$).  
Consequently, we verified numerically that the perturbative approach based on eq.\,(\ref{eq:ZetaSeries}) is always applicable for sufficiently large $N$. In the context of concrete curvaton models that generate a sizable abundance of PBHs, the case $r_{\rm dec} = 1$ seems difficult to be realized (cf. ref.\,\cite{Ferranteinprep}). However, it is interesting to note that the functional form of 
$\zeta=\zeta(\zeta_{\rm G})$ for $r_{\rm dec} = 1$ (that is, eq.\,(\ref{eq:MasterX2}))  looks remarkably similar to the one found in the case of USR models (cf. our schematic in eq.\,(\ref{eq:Schema})). 
It is, therefore, possible that the above argument could be applied to USR models that feature a broad power spectrum.

\section{Impact on PBH phenomenology}\label{sec:PBHGW}

We now proceed to apply the machinery worked out in the above sections to describe the implications for the phenomenological observables associated with PBHs. This brief discussion is intended to highlight the impact of NGs on the relation between the PBH abundance and the amplitude of the GW spectrum. 
It is well known in literature that a population of PBHs with masses in the range $10^{-16}M_\odot \lesssim M_{\rm PBH}\lesssim10^{-10}M_\odot$ could account for the totality of dark matter observed in our Universe\,\cite{Montero-Camacho:2019jte,Carr:2020gox}. We will focus on two classes of models (with a narrow and broad mass distributions) where PBHs explain the majority of the dark matter in our universe in the form of asteroidal mass PBHs.

\subsection{Computation of the PBH mass distribution}
We start by computing the mass distribution of PBHs at the end 
of the formation era, 
which is directly derived from the mass fraction $\beta_{\rm NG}$ as
\begin{align}
f_{\rm PBH}(M_{\rm PBH}) 
\equiv
\frac{1}{\Omega_{\rm  DM}} 
\frac{d\Omega_{\rm PBH}}
{d\log M_{\rm PBH}}\,,
~~~~~~~~
{\rm with}
~~~~~~~~
\Omega_{\rm PBH} = 
\int
d \log M_{H} \left(\frac{M_{\rm eq}}{M_{H}} \right)^{1/2}\beta_{\rm NG}(M_{H})\,,
\label{eq:diffmassfraction}
\end{align}
where $M_{\rm eq} \approx 2.8\times 10^{17}\,\,M_{\odot}$ is the horizon mass at the time of matter-radiation equality and 
$\Omega_{\rm  DM}$ is the cold dark matter density of the universe ($\Omega_{\rm  DM} \simeq 0.12\,h^{-2}$ with $h = 0.674$ for the Hubble parameter). 
Our task is the generalization of the computation carried out in ref.\,\cite{Young:2019yug,DeLuca:2019qsy,Franciolini:2022tfm},
that only includes non-linearities and assumes gaussian primordial curvature perturbations, to the case in which local primordial NG are also present 
with the generic functional form $\zeta = F(\zeta_{\rm G})$. 
This computation is part of the original results of this work, and, for this reason, we give a few details about its derivation.

The differential fraction of PBHs with mass $M_{\rm PBH}$ over the totality of dark matter at present day can be computed by rearranging the integration \eqref{eq:CompactionIntegral}.
Requiring over-threshold perturbations ${\cal C} \geq {\cal C}_{\rm th}$, and using eq.\,\eqref{eq:CCgau}, one finds that 
 the critical values of the linear component ${\cal C}_\text{G}$ is
\begin{align}
    {\cal C}_{\text{G}, {\rm th}, \pm} = 
    2\Phi \left(\frac{dF}{d\zeta_{\rm G}} \right)^{-1}
    \left(1 \pm \sqrt{1-\frac{{\cal C}_\text{th}}{\Phi}}\right)\,.
\end{align}
so we limit our integration to values in the range
\begin{align}
     {\cal C}_{\text{G}, {\rm th}, - } \leq  {\cal C}_{\text{G}} \leq 2\Phi \left(\frac{dF}{d\zeta_{\rm G}} \right)^{-1},
\end{align}
where the minus sign was chosen as we focus only on the type-I branch~\cite{Musco:2020jjb} of solutions.
Also, by inverting the definition of mass during the critical collapse,
we get the explicit relation between ${\cal C} $ and the horizon mass $M_H$ as
\begin{align}
M_{\rm PBH} = \mathcal{K}M_H\left[
    \left({\cal C} - \frac{1}{4\Phi}{\cal C}^2\right)
    - {\cal C}_{\rm th}
    \right]^{\gamma}
~~~~~~\text{or}~~~~~~
{\cal C}_\text{G}  = 
2\Phi 
\left(\frac{dF}{d\zeta_{\rm G}} \right)^{-1} 
\left[
1 - \sqrt{1 - \frac{{\cal C}_{\rm th}}{\Phi} - 
\frac{1}{\Phi}\left(
 \frac{M_{\rm PBH}}{\mathcal{K}M_H}
\right)^{1/\gamma}}
    \right]\,.
\end{align}
Requiring the argument of the square-root to be real coincide with the requirement of having threshold perturbations.
Also, the condition ${\cal C} \leq 2 \Phi$ indirectly selects a maximum mass $M_\text{PBH}$ that can be formed at each epoch, where the epoch of formation is parametrized by $M_H$.
With this change of variable, we can express the integration over $ d{\cal C}_{\rm G} d \zeta_{\rm G}$ in eq.\,\eqref{eq:CompactionIntegral} in terms of $d{M}_{\rm PBH}d \zeta_{\rm G}$, and then consider the differential mass fraction \eqref{eq:diffmassfraction}. 
Finally, the abundance of a given PBH mass $M_{\rm PBH}$ comes out of the integration across the possible epochs of formation, parametrized by $M_H$. Therefore, we obtain
\begin{mynamedbox2}{
NG mass distribution of PBHs adopting threshold statistics on the compaction function
}
\begin{align}
 f_{\rm PBH}(M_{\rm PBH})=
 \frac{1}{\Omega_{\rm DM}}
 \int_{M_{\rm H}^{\rm min}(M_\text{PBH})}
 d \log M_{\rm H} \left(\frac{M_{\rm eq}}{M_{\rm H}}\right)^{1/2}
 \left[
 1 - \frac{{\cal C}_{\rm th}}{\Phi} - 
\frac{1}{\Phi}\left(
 \frac{M_{\rm PBH}}{\mathcal{K}M_H}
\right)^{1/\gamma} 
\right]^{-1/2}
\nn \\
 \times 
  \frac{{\cal K}}{\gamma}
 \left(\frac{M_{\rm PBH}}{\mathcal{K} M_{\rm H}}\right)^\frac{1+\gamma}{\gamma}
 \int d\zeta_\text{G}
 P_{\rm G}({\cal C}_{\rm G}(M_{\rm PBH},\zeta_\text{G}), \zeta_\text{G}|M_{\rm H})
\left(\frac{dF}{d\zeta_{\rm G}} \right)^{-1}  ,
\end{align}
\end{mynamedbox2}
\noindent
where the integrand also includes the determinant of the Jacobian and the horizon mass dependence of 
the PDF $P_{\rm G}(M_{\rm H})$ 
is inherited by the smoothing scale $r_m(M_H)$ controlling the variances 
$(\sigma_{c},\sigma_{cr}, \sigma_{r})$, fixing the horizon crossing epoch.
As shown in\,\cite{Franciolini:2022tfm,Muscoinprep}, $\gamma(M_H)$, $\mathcal{K}(M_H)$, $\delta_c(M_H)$ and $\Phi(M_H)$ are functions of the mass of the horizon for epochs close to the QCD phase transitions (around the formation of solar mass PBHs). 
At that epoch, we also account for the variation of $g_{\star}(M_H)$
relating spectral modes to the horizon mass $M_H$ in eq.\,\eqref{eq:HorizonMass} \cite{PhysRevD.81.104019}.
The total fraction of PBHs is given by the integral 
\begin{equation}
f_{\rm PBH}=\int f_{\rm PBH}(M_{\rm PBH})d\log M_{\rm PBH}.
\end{equation}

In fig.\,\ref{fig:GWfPBH} we show the following constraints (see ref.\,\cite{Green:2020jor} for a review and\,\href{github.com/bradkav/PBHbounds}{\faGithub/bradkav/PBHbounds}).
Envelope of evaporation constraints (see also \cite{Saha:2021pqf,Laha:2019ssq,Ray:2021mxu}): EDGES\,\cite{Mittal:2021egv}, 
CMB\,\cite{Clark:2016nst}, INTEGRAL\,\cite{Laha:2020ivk,Berteaud:2022tws}, 511 keV\,\cite{DeRocco:2019fjq}, Voyager\,\cite{Boudaud:2018hqb}, 
EGRB\,\cite{Carr:2009jm};
microlensing constraints from the Hyper-Supreme Cam (HSC), ref.\,\cite{Niikura:2017zjd}; 
microlensing constraints from EROS, ref.\,\cite{EROS-2:2006ryy}; 
microlensing constraints from OGLE, ref.\,\cite{Niikura:2019kqi}; 
Icarus microlensing event, ref.\,\cite{Oguri:2017ock}; 
constraints from modification of the CMB spectrum due to accreting PBHs, ref.\,\cite{Serpico:2020ehh} (see also \cite{Piga:2022ysp});
direct constraints on PBH-PBH mergers with LIGO, as recently derived in ref.\,\cite{Franciolini:2022tfm} (see also \cite{Kavanagh:2018ggo,Hall:2020daa,Wong:2020yig,Hutsi:2020sol,DeLuca:2021wjr,Franciolini:2021tla}).

\subsection{NG impact on the amplitude of the induced stochastic gravitational wave background}

As already mentioned, PBHs production is related to a signal of induced GWs. Such GWs are produced as a second order effect by scalar perturbations which re-enter the horizon and collapse to form a PBH\,\cite{Tomita:1975kj,Matarrese:1993zf,Acquaviva:2002ud,Mollerach:2003nq,Ananda:2006af,Baumann:2007zm,Domenech:2021ztg}.
The spectral density of induced GWs associated to PBHs production can be computed as\,\cite{Espinosa:2018eve}
\begin{equation}\label{eq:OmegaGW}
\Omega_{{\rm GW}}=\frac{c_{g} \Omega_{r}}{36} \int_{0}^{\frac{1}{\sqrt{3}}} d t \int_{\frac{1}{\sqrt{3}}}^{\infty} d s\left[\frac{\left(t^{2}-1 / 3\right)\left(s^{2}-1 / 3\right)}{t^{2}-s^{2}}\right]^{2}\left[\mathcal{I}_{c}(t, s)^{2}+\mathcal{I}_{s}(t, s)^{2}\right] P_{\zeta}\left[\frac{k \sqrt{3}}{2}(s+t)\right] P_{\zeta}\left[\frac{k \sqrt{3}}{2}(s-t)\right],
\end{equation}
with 
\begin{equation}
c_g	\equiv \frac{g_*(M_H)}{g_{*}^0}
\left( \frac{g_{*S}^0}{g_{*S} (M_H)}\right) ^{4/3}
\approx 0.4
\end{equation} 
 ($g_{*S}$ and $g_{*}$ being the effective degrees of freedom of thermal radiation and the superscript $^0$ stands for present day values),  $\Omega_r$ current radiation density and $\mathcal{I}_{c}(t, s)$, $\mathcal{I}_{s}(t, s)$ are transfer functions computed analytically assuming a perfect radiation cosmological background (see, for example, ref.\,\cite{Espinosa:2018eve}). 
 Notice therefore eq.\,(\ref{eq:OmegaGW}) is strictly valid only during radiation domination. In principle, we should also consider the evolution of the relativistic degrees of freedom as the QCD phase transition takes place. We neglect this effects and just assume the Universe to be radiation dominated at all the scales relevant for GWs production
 (see Ref.~\cite{Abe:2020sqb} for and estimation of the magnitude of this modulation).
 We also neglect possible primordial non-Gaussian corrections in the computation of the scalar-induced GW signal, which we leave for future work (we refer to refs.\,\cite{Cai:2018dig,Yuan:2020iwf,Adshead:2021hnm,Abe:2022xur,Chang:2022nzu} for a discussion about these effects).

 As shown in ref.\,\cite{DeLuca:2020agl}, a broad power spectrum, like the one in eq.\,(\ref{eq:Bro}), produces a population of PBHs which can explain the totality of dark matter and, at the same time, are associated with a signal of second order GWs compatible with the event detected by the NANOGrav collaboration. It is crucial to stress the fact that hereafter $\Omega_{\rm GW}$ is computed in gaussian approximation, as done in ref.\,\cite{DeLuca:2020agl}.
 Let us stress that, on the contrary to the computation of the abundance which is sensitive to the tail of the distribution, 
 the GW emission is mostly controlled by the characteristic amplitude of perturbations, and thus better captured by the leading order computation we report here. 
 Therefore, the effect of NG corrections is indirect and it comes from the re-normalized amplitude of the power spectrum induced by NGs. The improvement with respect to ref.\,\cite{DeLuca:2020agl} is that in this work the new value for the amplitude of the power spectrum also takes into account primordial NGs, by means of the exact relation in eq.\,(\ref{eq:MasterX}).

\begin{figure}[t]
	\begin{center}
\includegraphics[width=.49\textwidth]{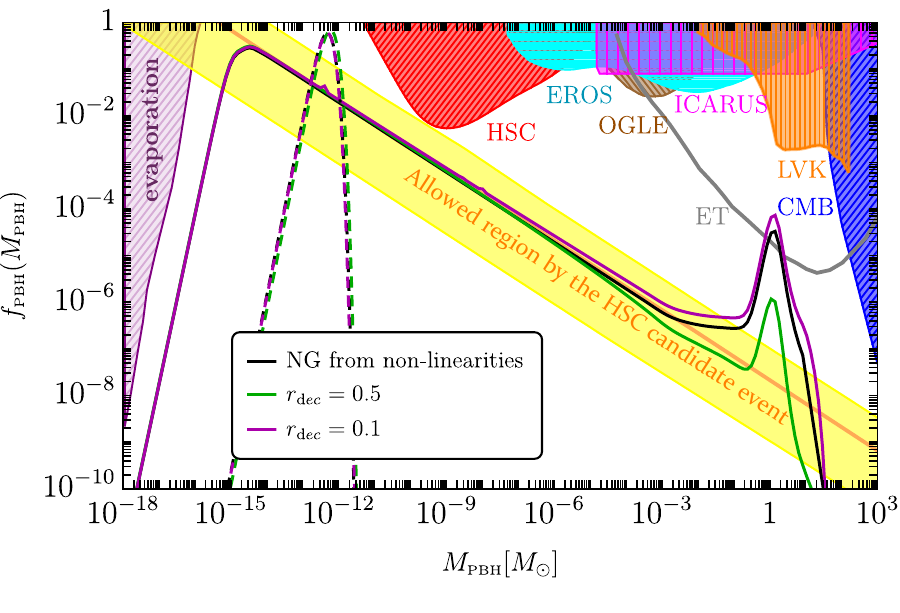}
\includegraphics[width=.49\textwidth]{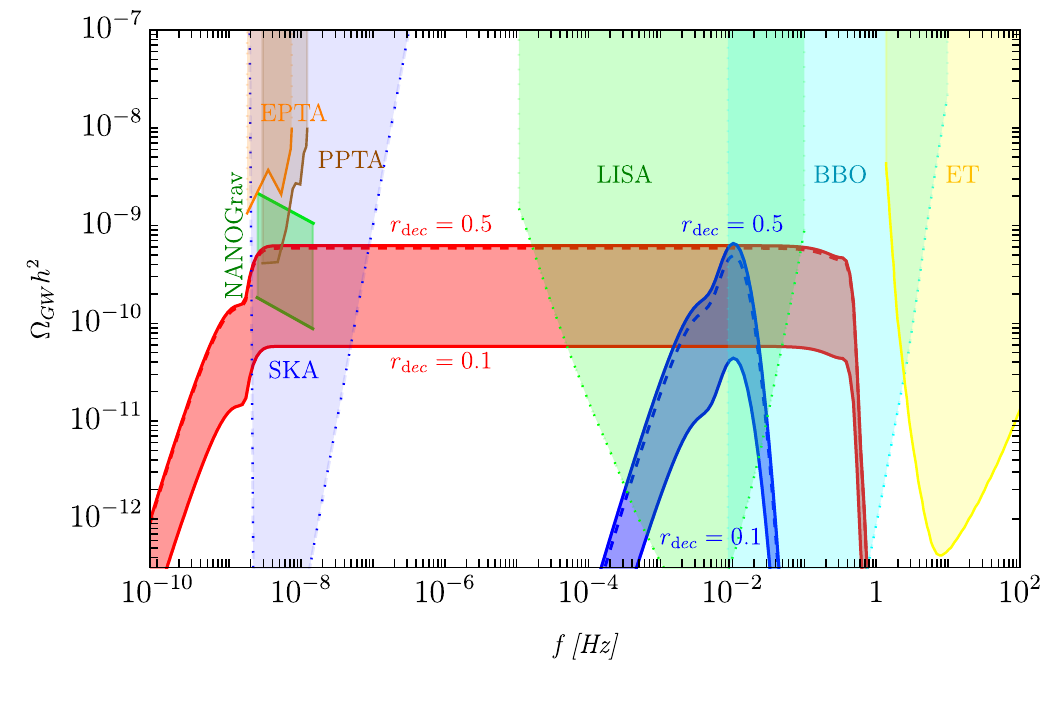}
		\caption{\em
\textbf{\textit{Left panel.}}
Mass function resulting from a broad power spectrum 
(solid lines, with $k_{\rm min}=10^{6}$ $ {\rm Mpc}^{-1}$ and $ \Delta k =10^{8.5}$ $ {\rm Mpc}^{-1}$) and from a
Log-Normal power spectrum 
(dashed lines, with $\sigma=0.5$ and $k_{\rm peak}=6 \times 10^{12}$ $ {\rm Mpc}^{-1}$) with different amount of NGs, i.e. values of $r_{\rm dec}$. 
The amplitude of the power spectrum has been re-scaled for each 
value of $r_{\rm dec}$ such that PBHs comprise the totality of dark matter. 
The most stringent experimental constraints are shown 
(for a recent comprehensive review on constraints see\,\cite{Carr:2020gox,Green:2020jor} and description in the text). Minimum testable abundance from PBH mergers
with the Einstein Telescope as derived in \cite{DeLuca:2021hde} (see also \cite{Ng:2022agi,Martinelli:2022elq}). 
\textbf{\textit{Right panel.}} Spectral density of gravitational waves computed with $r_{\rm dec}=0.1$ and $r_{\rm dec}=0.5$ both with the broad spectrum (red lines) and with the log-normal spectrum (blue lines). 
The coloured dashed lines indicate instead the results when only contributions coming from non-linearities are present. 
The plot also shows the constraints coming from the NANOGrav 12.5 yrs experiment\,\cite{NANOGrav:2020bcs}, EPTA\,\cite{Lentati:2015qwp}, PPTA\,\cite{Shannon:2015ect}, and future sensitivity for planned experiments like SKA\,\cite{Zhao:2013bba}, LISA\,\cite{LISA:2022kgy}, BBO/DECIGO\,\cite{Yagi:2011wg} and ET
(power law integrated sensitivity curves as derived in ref.~\cite{Bavera:2021wmw}). 
}\label{fig:GWfPBH} 
	\end{center}
\end{figure}

On the left side in fig.\,\ref{fig:GWfPBH}, we show the PBHs abundance 
computed as described in the preceding sections. 
We have taken $r_{{\rm dec}}=0.5$ and $r_{{\rm dec}}=0.1$ as benchmark values for primordial NGs, 
since they are typical values in expected in realistic curvaton models, 
see e.g. ref\,\cite{Ferranteinprep}. 
For each value of $r_{{\rm dec}}$, 
we re-scale the amplitude of the power spectrum in order to keep $f_{\rm PBH} \simeq 1$ fixed.  
Notice that the integral is always dominated by the asteroidal mass range, where there are no constraints on the abundance\,\cite{Carr:2020gox}. 
As shown on the right side in fig.\,\ref{fig:GWfPBH}, such a re-scaling reflects quadratically on the amplitude of $\Omega_{\rm GW}\approx A^2$ according to eq.\,(\ref{eq:OmegaGW}). 
As we
can see from fig.\,\ref{fig:GWfPBH}, in the case of a broad power spectrum we are able to get simultaneously the totality of dark matter and a gravitational wave signal compatible with the NANOGrav 12.5 years experiment \,\cite{NANOGrav:2020bcs} for most of
the possible values for $r_{{\rm dec}}$.
We also repeat the same computation for a narrow power spectrum, i.e. a log-normal function with $\sigma=0.50$. Results are shown in fig.\,\ref{fig:GWfPBH}. We choose the amplitude of the spectrum in order to obtain $f_{\rm PBH} \simeq 1$. 
Consequently, the corresponding signal of second-order GWs falls within the range of sensitivity of the future experiments LISA and BBO.
Overall, the effect of NGs can modify the predicted SGWB associated to PBH dark matter by one order if magnitude in curvaton models.

\subsubsection{Implications for PBHs in the stellar mass range}

\begin{figure}[t]
	\begin{center}
		$$\includegraphics[width=.46\textwidth]{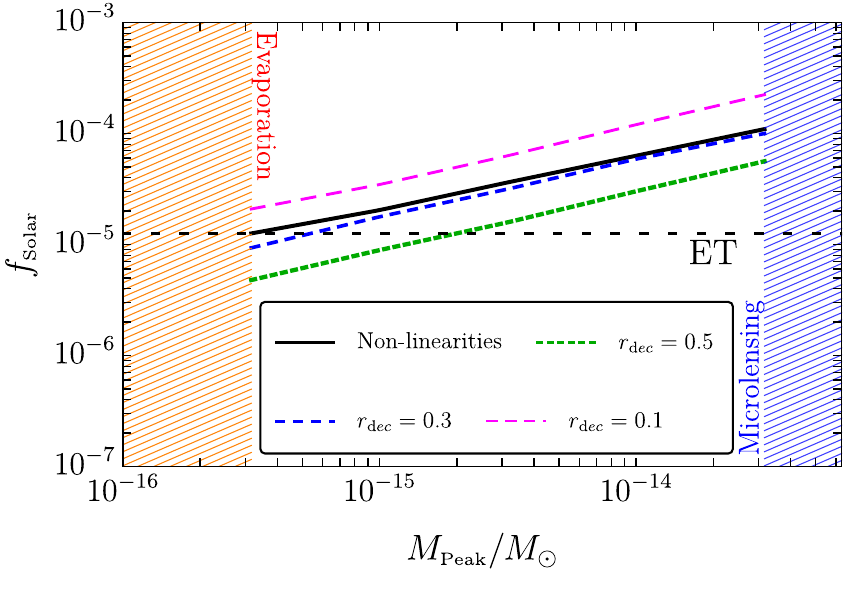}
		\qquad\qquad\includegraphics[width=.46\textwidth]{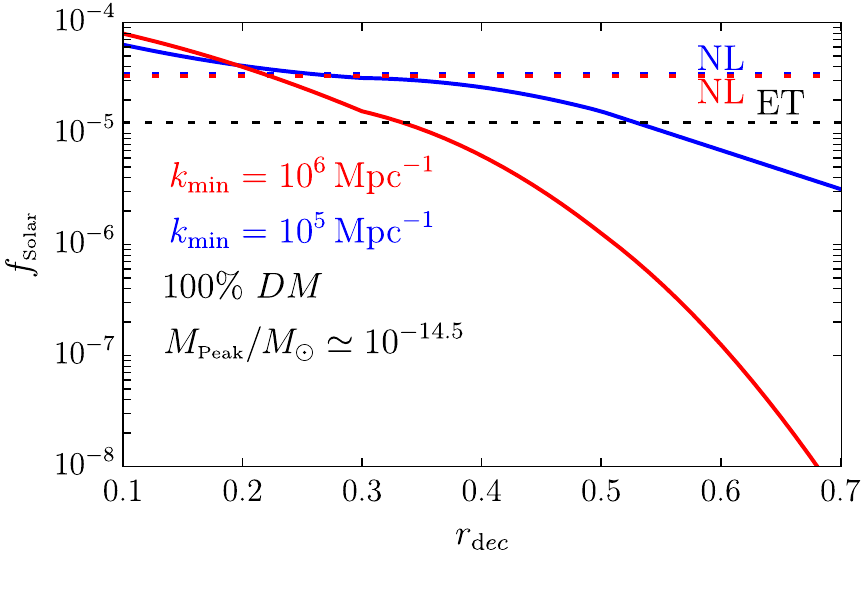}$$
		\caption{\em
	\textbf{\textit{	Left panel. }} 
Height of the peak at the QCD Mass scale $f_\text{Solar}$,
assuming a broad power spectrum with $k_{\rm min}=10^{5}$ $ {\rm Mpc}^{-1}$ as a function of the location of the asteroidal mass peak dominating the contribution to the dark matter. 
We compare the result obtianed considering only the effect of non-linearities with three different values of $r_{\rm dec}$. 
The dashed regions at the sides represent positions of the main peak which are ruled out by the constraints given by evaporation or microlensing (see fig.\,\ref{fig:GWfPBH}). 
\textbf{\textit{	Right panel. }} 
Height of the peak at the QCD scale for different values of $r_{\rm dec}$. 
We choose as benchmark values 
$k_{{ \rm min}}=10^{6}$ ${\rm  Mpc}^{-1}$ (red) 
and $k_{{ \rm min}}=10^{5}$ ${\rm  Mpc}^{-1}$ (blue) respectively.
The plot shows a decreasing trend of $f_\text{Solar}$ when $r_{\rm dec}$ increases, resulting from the effect of NG reducing the abundance of high masses. 
The dashed lines represent the height of the peak at the QCD scale computed by taking into account only non-linearities while the dashed black one shows the lower bound of the sensitivity of the Einstein Telescope (ET) experiment to PBH mergers as derived in \cite{DeLuca:2021hde}
(see also \cite{Punturo:2010zz,Chen:2019irf,Ng:2022agi,Martinelli:2022elq}).
}
\label{fig:parScan}  
\end{center}
\end{figure}

The left-hand plot in fig.\,\ref{fig:GWfPBH} shows the presence, around solar masses, of a second peak in $f_\text{PBH}(M_\text{PBH})$ which, as widely known in literature, is typically caused by the QCD phase transition \cite{Jedamzik:1996mr,Byrnes:2018clq,Muscoinprep}.  
With the inclusion of NGs in a scenario characterised by a broad, and nearly scale invariant, power spectrum, the PBH abundance is further modulated and can be enhanced/suppressed compared to the case of gaussian primordial curvature perturbations. 
We highlight few main trends:
\begin{itemize}
    \item[$\circ$] 
Moving the position of the main peak of asteroidal mass PBHs in the allowed window between evaporation and microlensing constraints while keeping $f_\text{PBH} =  1$, 
one finds that the height of the QCD peak, dubbed $f_\text{Solar} = f_\text{PBH}(1.25 M_\odot)$, 
changes approximately with a power law 
$f_\text{Solar} \propto M_{\rm PBH}^{1/2}$.
This is shown in the left panel of fig.\,\ref{fig:parScan}
and it is caused by the leading order behaviour of the mass distribution $f_{\rm PBH} \propto M_{\rm PBH}^{1/2}$ obtained for scale invariant mass fraction $\beta(M_H)$.
Indeed, choosing $k_{{\rm min}} = 10^5$ ${\rm Mpc}^{-1}$, 
$\beta_{\rm NG}$ still remains approximately scale invariant
close to the QCD epoch. 
Hence, imposing PBHs account for the totality of dark matter, 
a shift to the right of the mass characterising 
the main peak $M_{\rm Peak}$ 
cause an increase of $f_\text{Solar}$. 
\item[$\circ$]
When the large scale cut-off $k_\text{min}$ is moved closer to the QCD mass scale, 
one obtains larger violations of the scale invariance around $M_{\rm PBH} \approx M_\odot$ (see fig.\,\ref{fig:BroCompa2}). 
Therefore, larger values of $k_{\rm min}$ leads to deviations from the overall trend highlighted in the previous point 
and $f_\text{Solar}$ inherits a stronger dependence to $r_\text{dec}$. This is highlighted in fig.\,\ref{fig:parScan} 
where we show results for 
$k_\text{min} =10^{5}$  ${\rm Mpc}^{-1}$ in blue while 
$k_\text{min} =10^{6}$  ${\rm Mpc}^{-1}$ in red.
\item[$\circ$]
As we can see from the right plot in fig.\,\ref{fig:BroCompa2}, the value of $r_{{\rm dec}}$ controls the deviation from a 
scale independent $\beta_{\rm NG}$.
This translates into modified secondary peak amplitude $f_\text{Solar}$. 
The right-hand plot in fig.\,\ref{fig:parScan} shows how $f_\text{Solar}$ varies for different values of $r_{\rm dec}$. Interestingly enough, the height of the peak grows as $r_{\rm dec}$ decreases (that is, for larger primordial NG).
This is nothing but the effect already discussed in figs.\,\ref{fig:BroadCompa},\,\ref{fig:BroCompa2} (at the level of the mass fraction $\beta_{\rm NG}$) but now reflected on the computation of $f_{\rm PBH}$.
\end{itemize}
\noindent
It is important to note that
an enhanced effect of NGs on the secondary peak would 
allow for a larger merger rate of stellar mass mergers that may be potentially constrained at current and future ground base GW detectors \,\cite{Pujolas:2021yaw,DeLuca:2021hde,Franciolini:2022tfm}.

\section{Discussion and conclusions}\label{sec:Final}

In this work we have developed an exact formalism for the computation of PBHs abundance in the presence of local NG in the curvature perturbation field by including both
\begin{itemize}
    \item[$\circ$] NG arising from the non-linear relation between the density contrast and the curvature perturbation field;
    \item[$\circ$] NG of primordial origin by means of an exact functional form $\zeta = F(\zeta_{\rm G})$, replacing the customary perturbative expansion.
\end{itemize}
Our approach is very generic and can be applied to a plethora of models, modulo being able to find an exact expression for primordial NG, i.e. the explicit form of $F(\zeta_{\rm G})$. 
We have included such NG corrections in the computation of the abundance with an approach based on threshold statistics on the compaction function, with a prescription described in  eq.\,(\ref{eq:CompactionIntegral}).
This is the main result of our paper.

To give an example of the application of our prescription,  we focused our analysis on curvaton models in which primordial NG is given by the logarithmic expression in eq.\,(\ref{eq:MasterX})\,\cite{Sasaki:2006kq}. Being able to compute abundance in an exact way, we  compared our result $\beta_{\rm NG}$ with the usual quadratic approximation, in which $\zeta$ is expanded into a power series of $\zeta_{\rm G}$, $\beta_{\rm NG}^{(N)}$, and only the first non-trivial term is kept. We found that stopping at second order in the expansion leads to a flawed result. Indeed, independently from the shape of the power spectrum and from the adopted approach, a positive $f_{\rm NL}$ enhances the abundance of PBHs by many orders of magnitude with respect to the primordiarly Gaussian result (that is, $\beta$ computed by only including NG from non-linearities). 
In general, this result is turned completely upside down by higher-order corrections and by the exact result itself.

In our numerical analysis we considered a log-normal and a step-function power spectrum. For narrow power spectra we observe that the perturbative result $\beta_{\rm NG}^{(N)}$ nicely converges to the exact one $\beta_{\rm NG}$ and that the narrower the spectrum the quicker the convergence. On the other hand, as the spectrum under examination gets broader, convergence is immediately lost. 
The reason is that, as explained in section\,\ref{sec:TheBroad}, for wide spectra contributions outside the radius of convergence of the expansion in eq.\,(\ref{eq:ZetaSeries}) are no longer suppressed and the perturbative result looses any mathematical ground. 
This analysis persuaded us that non-perturbative techniques are the only sensible way to compute PBHs abundance in the case of power spectra which are not extremely peaked.

To show the implications of our results, we  studied the phenomenological observables associated with PBHs. 
Requiring that $\beta_{\rm NG}$ accounts for the totality of dark matter fixes the amplitude of the spectrum and, consequently, 
that of the signal of second-order GWs, making the inclusion of NGs an observable effect. 
For a fixed value of the abundance but a different amount of NGs, 
we find that in the worked example inspired by the curvaton model, the amplitude of the SGWB induced at second order can vary up to one order of magnitude. 
In the case of a broad spectrum, with the inclusion of NGs 
we are able to produce a population of PBHs which can explain the totality of dark matter and, at the same time, generate a signal of second order GWs compatible with the candidate signal reported by the NANOGrav collaboration. 
For a narrow spectrum, we draw similar conclusions 
but with  the signal of stochastic GWs which is now peaked 
in the region of sensitivity of LISA and BBO.

We studied the impact of NGs on the secondary peak of the 
mass distribution around the the QCD epoch (i.e. $M_{\rm PBH} \approx M_\odot$) for several values of the parameter $r_{\rm dec}$, which controls the size of NGs. 
For a broad power spectrum with $k_{{\rm min}}=10^{5}$ ${\rm Mpc}^{-1}$, moving the characteristic mass of the main peak in $f_{\rm PBH}$ in the range $3 \times 10^{-15}M_{\odot}\lesssim M_{\rm PBH}\lesssim 3 \times 10^{-14}M_{\odot}$ 
to avoid observational bounds on $f_{{\rm PBH}}$ and requiring each time $f_{\rm PBH}\simeq 1$, 
we observed how such secondary peak varies and
potentially producing a detectable population of PBH mergers 
detectable at  future GW experiments such as ET \cite{Punturo:2010zz,Chen:2019irf,DeLuca:2021hde,Ng:2022agi}.
Differently from a scenario without non-gaussianities \,\cite{DeLuca:2020ioi}, 
we find that sizeable modifications to the amplitude of the QCD peak induced by a braking of the scale invariance of $\beta(M_H)$,
which is controlled by the value of 
$r_{{\rm dec}}$ within the curvaton scenario.

Throughout this work we referred to ref.\,\cite{Musco:2020jjb,Franciolini:2022tfm} for the evaluation of the threshold for collapse, where ${\cal C}_\text{th}$
is computed only including the effect of NG from non-linearities. 
In principle, also NG of primordial origin affect the threshold value. 
The impact of second-order primordial NG has been discussed by refs.\cite{Kehagias:2019eil,Escriva:2022pnz} in the case of a monochromatic spectrum, finding sub-percent corrections to its value. 
In light of the results presented in our work about the validity of the perturbative approach, we are fully aware that a computation of the threshold beyond the quadratic approximation is needed to consistently take into account NGs at all stages of the computation of PBHs abundance. 
We leave this as future work.

We conclude insisting again on the take-home message of our work: NGs have a significant impact on observables and have to be taken into account in a correct way if one wants to make contact with phenomenology. 
The technique developed in the main body of this paper is a step towards the direction of refining the computation of the PBHs abundance, both for narrow and broad power spectra when primordial NGs are included in the model.

\begin{acknowledgments}
We thank I. Musco and S. Young for interesting discussions. 
The research of A.U. was supported in part by the MIUR under contract 2017\,FMJFMW (``{New Avenues in Strong Dynamics},'' PRIN\,2017).
G.F. acknowledges financial support provided under the European
Union's H2020 ERC, Starting Grant agreement no.~DarkGRA--757480 and under the MIUR PRIN programme, and support from the Amaldi Research Center funded by the MIUR program ``Dipartimento di Eccellenza" (CUP:~B81I18001170001).
This work was supported by the EU Horizon 2020 Research and Innovation Programme under the Marie Sklodowska-Curie Grant Agreement No. 101007855.
\end{acknowledgments}

\appendix

\section{On the convergence of the power-series expansion}\label{app:Radius}

Consider the power series expansion of eq.\,(\ref{eq:MasterX})
\begin{align}
\sum_{n = 1}^{\infty} c_n(r_{\rm dec}) \zeta_{\rm G}^{n} = \log\big[X(r_{\rm dec},\zeta_{\rm G})\big]\,.
\label{eq:PoweSeriesExpansion}
\end{align}
The above equality is valid only within the radius of convergence of the series expansion.
First, consider for simplicity the case with $r_{\rm dec}=1$. The coefficients $c_n(1)$ take the form
\begin{align}
c_n(1) = \frac{(-1)^{n+1}}{n(2/3)^{n-1}}\,,
\end{align}
and the radius of convergence is given by
\begin{align}
R = \lim_{n\to \infty}\left|\frac{c_n(1)}{c_{n+1}(1)}\right| = \frac{2}{3}\,,
\end{align}
so that we have
\begin{align}
\sum_{n = 1}^{\infty} c_n(1) \zeta_{\rm G}^{n} = 
 \log\big[X(1,\zeta_{\rm G})\big] = \frac{2}{3}\log\bigg(
1+ \frac{3}{2}\zeta_{\rm G}
\bigg)\,,~~~~~{\rm for}\,\,\,-\frac{2}{3} < \zeta_{\rm G} < +\frac{2}{3}\,.
\end{align}
 This result makes perfectly sense since the function 
 $\log(
1+ 3\zeta_{\rm G}/2)$
has a singularity when $\zeta_{\rm G} = -2/3$ and the series expansion centered in $\zeta_{\rm G} = 0$ is limited  
by its presence. 

Consider now the case $r_{\rm dec}\neq 1$. 
For definiteness, we consider $r_{\rm dec} = 0.5$. 
In this case, we find that the radius of convergence is given by $R \simeq 0.173$. 
Unfortunately, 
we do not have analytic expressions for the coefficients 
 $c_n(r_{\rm dec})$ with $r_{\rm dec} \neq 1$ so the aforementioned result must be derived and understood without relying on standard methods.
 
First of all, let us visualize the situation with the help of a graph.
\begin{figure}[!h!]
\begin{center}
$$\includegraphics[width=.45\textwidth]{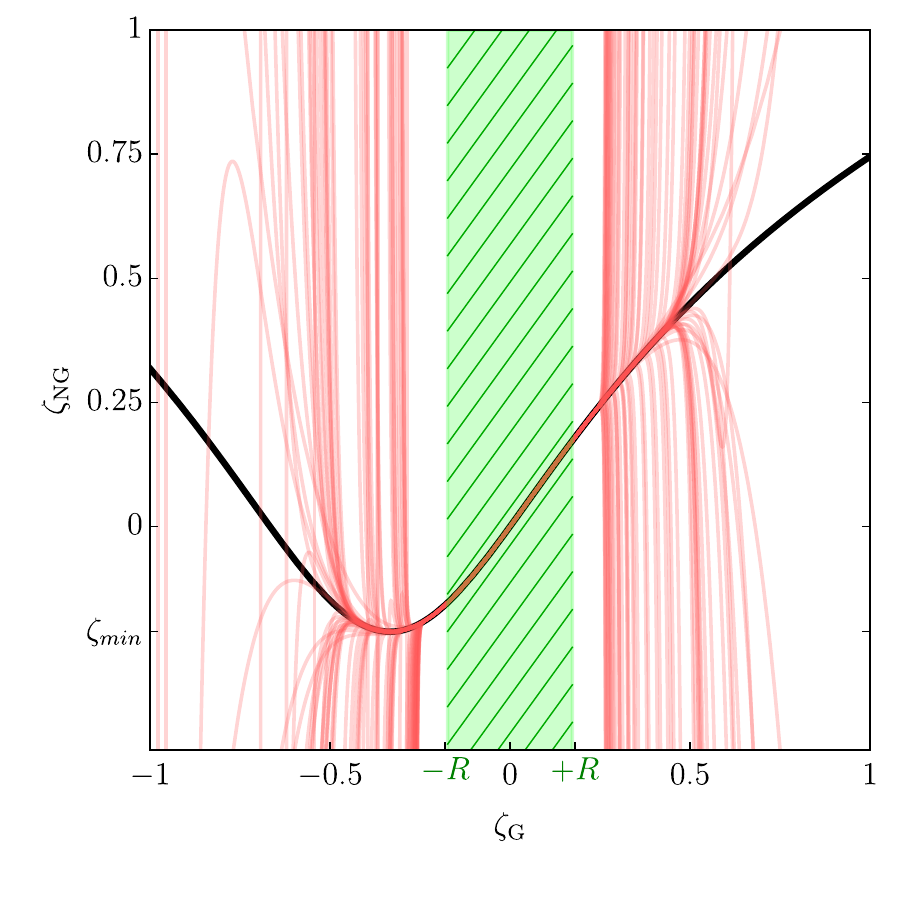}
\qquad\qquad\includegraphics[width=.45\textwidth]{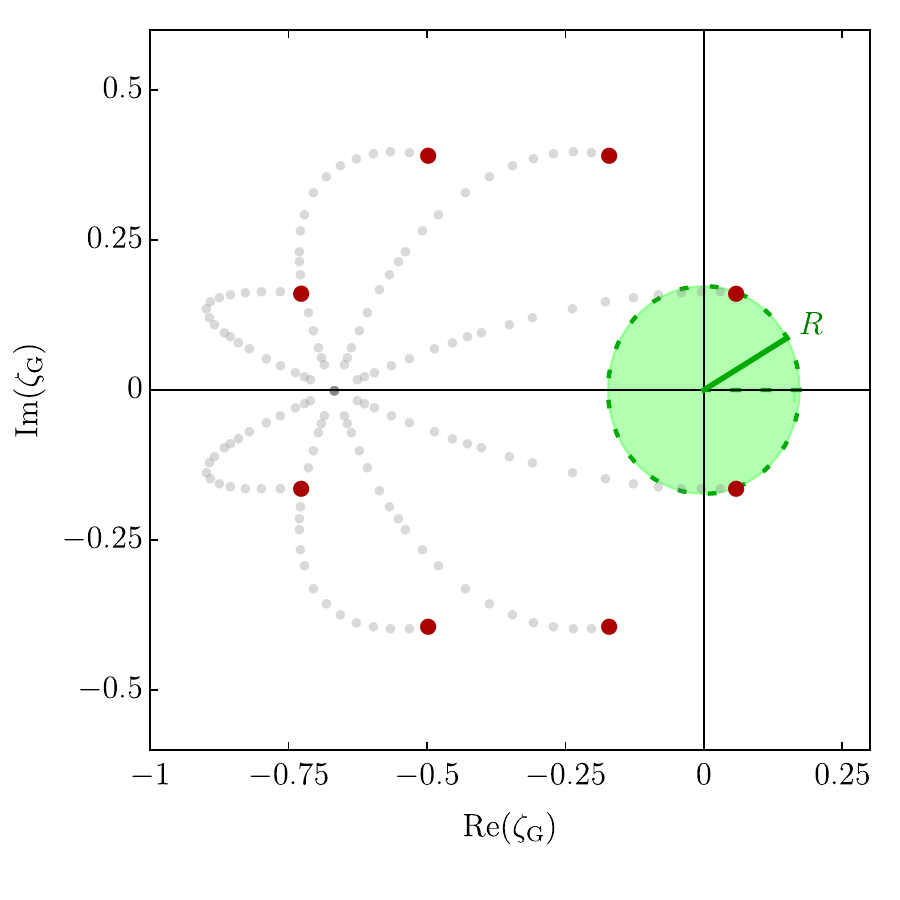}$$
\caption{\em  
	\textbf{\textit{	Left panel. }}Comparison between the function $\zeta =  \log\big[X(r_{\rm dec},\zeta_{\rm G})\big]$ with $r_{\rm dec}=0.5$ (solid black line) 
and its power series expansion $\sum_{n = 1}^{N} c_n(r_{\rm dec}) \zeta_{\rm G}^{n}$ for increasing 
values of $N$ (light red lines). The region 
shaded in green is limited by the condition $-R< \zeta_{\rm G} < +R$ with radius of convergence 
$R \simeq 0.173$. 
	\textbf{\textit{	Right panel. }}The red dots mark the position of the branch points of the function 
$P(r_{\rm dec},\zeta_{\rm G})$ in eq.\,(\ref{eq:EqRip}) with $r_{\rm dec}=0.5$. To visualize these points, 
the independent variable $\zeta_{\rm G}$ is promoted to be complex. 
The green region is the disk of convergence with radius $R \simeq 0.173$; the latter measures the 
distance from the center of the expansion (that is the origin of the complex plane in this case) of the closest branch point. 
The trajectories marked by light gray dots indicate how the branch points move as we increase $r_{\rm dec}$ from $r_{\rm dec}=0.5$ to $r_{\rm dec}=1$.
 }\label{fig:Radius}  
\end{center}
\end{figure} 
 In the left panel of fig.\,\ref{fig:Radius} we compare, as function of $\zeta_{\rm G}$, the full 
 expression $\log\big[X(r_{\rm dec},\zeta_{\rm G})\big]$ (solid black line) with its power-series expansion 
 $\sum_{n = 1}^{N} c_n(r_{\rm dec}) \zeta_{\rm G}^{n}$ truncated at increasing values of $N$ (light red lines); 
 the region shaded in green is limited by $-R<\zeta_{\rm G}<+R$ with $R \simeq 0.173$. 
 As clear from this plot, within the radius of convergence the equality in 
 eq.\,(\ref{eq:PoweSeriesExpansion}) holds true but as soon as we consider values of 
 $\zeta_{\rm G}$ outside the green band the power-series badly diverges from the exact result. 
 This plot gives a numerical confirmation that indeed $R \simeq 0.173$. 
 
 However, it is natural to ask what is the origin of this number and how we computed it. 
 At first sight, this is indeed puzzling. 
 In the case $r_{\rm dec} = 1$, as previously discussed, the meaning of the radius of convergence was clear given the 
 singularity of the function $\log(
1+ 3\zeta_{\rm G}/2)$: the radius of convergence was just the distance from the origin of the singularity 
 (including which analyticity would be lost).
 
However, if we take $r_{\rm dec} = 0.5$ the function 
$\log\big[X(r_{\rm dec},\zeta_{\rm G})\big]$ does not present any problem. This is indeed evident 
from the very same plot in the left panel of fig.\,\ref{fig:Radius} that we just discussed: 
the black curve is a perfectly 
smooth curve. What is the obstruction that sets $R \simeq 0.173$? 
To answer this question, let us discuss a simple example. 
Consider the geometric series
\begin{align}
\sum_{k=0}^{\infty}(-x)^{2k} = \frac{1}{1+x^2}\,,~~~~~{\rm for}\,\,\,-1<x<1\,.
\end{align}
The radius of convergence is $R=1$. However, if we plot the function $f(x) = 1/(1+x^2)$ we notice 
that this is a perfectly smooth curve that peaks at the origin and dies off at $\pm \infty$ without any singularity 
on the real axis.
Notice that this situation is qualitatively very similar to the one we discussed before.
 What limits $R=1$? The (well-known) answer is that, because of analytical continuation, 
 the function $f(x) = 1/(1+x^2)$ actually knows about its singularities in the complex plane 
 located at $x = \pm i$. The radius of convergence, therefore, is still measuring the distance from 
 the origin of the closest singularity with the difference that now the singularity in question 
 is displaced from the real axis.  

Something analogue happens in our case. 
The radius of convergence 
$R \simeq 0.173$
measures the distance from the center of the closest branch-point singularity of the square root 
that enters in the function
 \begin{align}
P(r_{\rm dec},\zeta_{\rm G}) \equiv 
\frac{(2r_{\rm dec} + 3\zeta_{\rm G})^{4}}{16 r_{\rm dec}^2} + \sqrt{
(1-r_{\rm dec})^3(3+r_{\rm dec}) + \frac{(2r_{\rm dec} + 3\zeta_{\rm G})^8}{256 r_{\rm dec}^4}
}\,.\label{eq:EqRip}
\end{align}
The argument of the square root is a polynomial of degree 8 in $\zeta_{\rm G}$ with branch points located as in the right panel of fig.\,\ref{fig:Radius} 
(the red dots, with $\zeta_{\rm G}$ promoted to be a complex variable) 
and the singularity closest to $\zeta_{\rm G} = 0$ sets the radius of convergence, as shown in the plot (notice that, since the polynomial 
has real coefficients, its 
roots are pairs of complex conjugate numbers). 

\begin{figure}[t]
\begin{center}
$$\includegraphics[width=.45\textwidth]{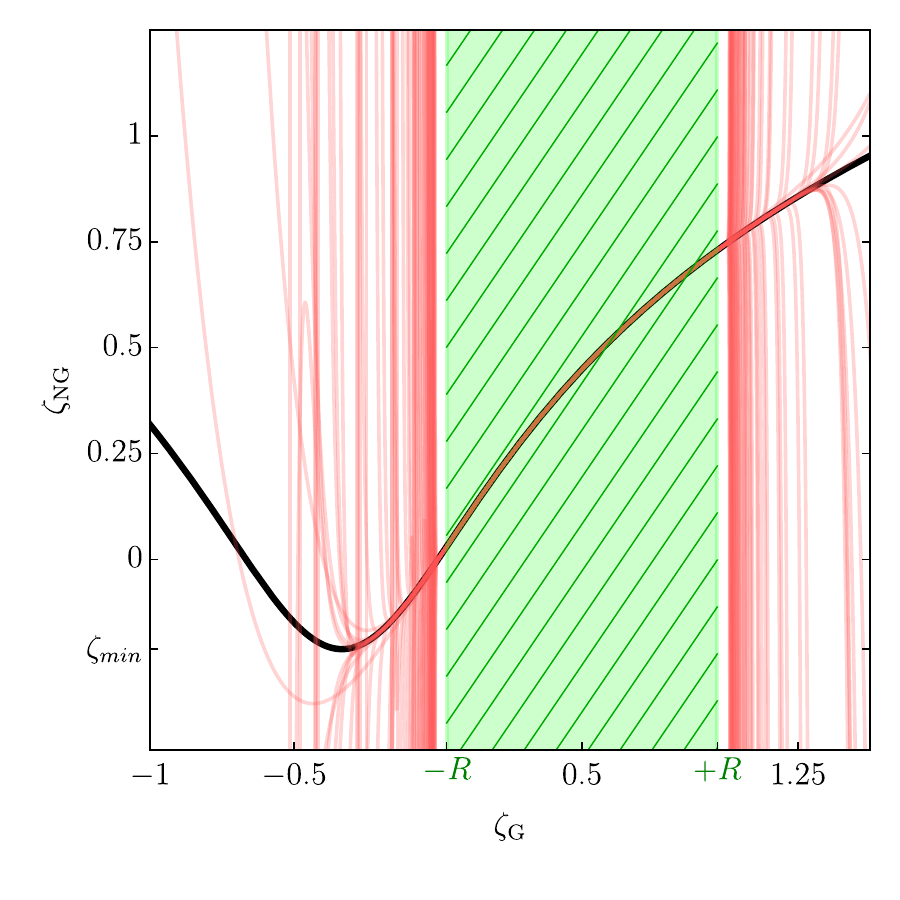}
\qquad\qquad\includegraphics[width=.45\textwidth]{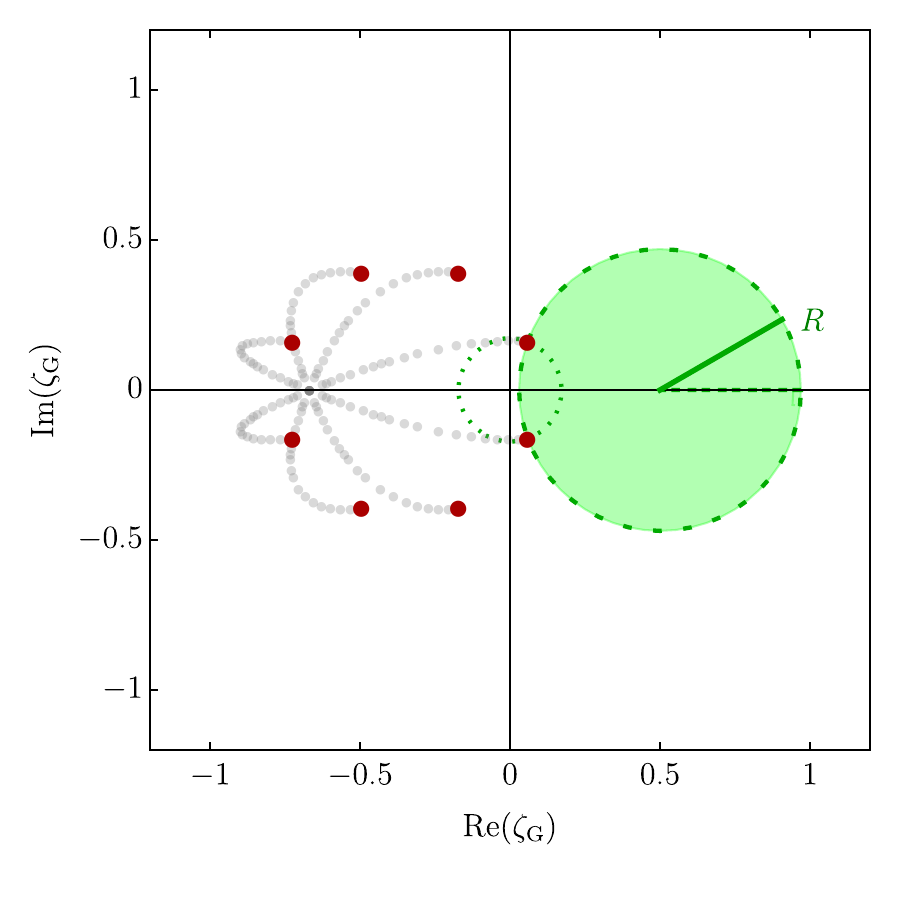}$$
\caption{\em  
Same as in fig.\,\ref{fig:Radius} but with the center of the power series expansion that is now 
set by $\zeta_{\rm G} = 0.5$ (see text for details).
 }\label{fig:RadiusOff}  
\end{center}
\end{figure} 
For completeness, we also show in light gray how these branch point singularities move as we increase the 
value of $r_{\rm dec}$ from $r_{\rm dec}=0.5$ towards $r_{\rm dec}=1$; as expected from our previous result, 
as $r_{\rm dec}\to 1$ the singularities eventually  collapse on the real singularity 
located at $\zeta_{\rm G} = -2/3$.  

As a side comment, notice that it is possible to tune the region of convergence if one takes a power-series
 expansion centered around a point that is not the origin.
This is shown in fig.\,\ref{fig:RadiusOff}. 
For illustration, we expand around the point $\zeta_{\rm G} = 0.5$. 
In the left panel of fig.\,\ref{fig:RadiusOff} we see that the region of convergence 
is way bigger than it was before, and we find $R\simeq 0.47$. 
In light of our previous discussion, this is now clear. 
If we consider the right panel of fig.\,\ref{fig:RadiusOff} we see that now the distance 
between the center of the expansion and the closest branch point singularity increased (since we moved the former 
towards the right while the latter obviously remained fixed) thus giving a wider region of convergence.

\bibliography{NonGaussian_Curvaton}

\end{document}